\newif\iffigs\figstrue
\documentclass[a4paper,11pt]{article}
\usepackage{latexsym,amssymb,lscape,graphics}
\usepackage{graphicx}        
\usepackage{longtable}
\usepackage{multirow}
\usepackage{color}
\usepackage{slashed,epsfig}
\usepackage{amsfonts}

\newtheorem{definizione}{Definition}[section]

\newcommand{\bd}{\begin{definizione}}
\newcommand{\ed}{\end{definizione}}

\def\IC{\relax\,\hbox{$\inbar\kern-.3em{\rm C}$}}
\def\IG{\relax\,\hbox{$\inbar\kern-.3em{\rm G}$}}
\def\IB{\relax{\rm I\kern-.18em B}}
\def\ID{\relax{\rm I\kern-.18em D}}
\def\IL{\relax{\rm I\kern-.18em L}}
\def\IF{\relax{\rm I\kern-.18em F}}
\def\IH{\relax{\rm I\kern-.18em H}}
\def\II{\relax{\rm I\kern-.17em I}}
\def\IN{\relax{\rm I\kern-.18em N}}
\def\IP{\relax{\rm I\kern-.18em P}}
\def\IQ{\relax\,\hbox{$\inbar\kern-.3em{\rm Q}$}}
\def\bfzero{\relax\,\hbox{$\inbar\kern-.3em{\rm 0}$}}
\def\IK{\relax{\rm I\kern-.18em K}}
\def\IG{\relax\,\hbox{$\inbar\kern-.3em{\rm G}$}}
 \font\cmss=cmss10 \font\cmsss=cmss10 at 7pt
\def\IR{\relax{\rm I\kern-.18em R}}
\def\ZZ{\relax\ifmmode\mathchoice
{\hbox{\cmss Z\kern-.4em Z}}{\hbox{\cmss Z\kern-.4em Z}}
{\lower.9pt\hbox{\cmsss Z\kern-.4em Z}} {\lower1.2pt\hbox{\cmsss
Z\kern-.4em Z}}\else{\cmss Z\kern-.4em Z}\fi}
\def\bfone{\relax{\rm 1\kern-.35em 1}}

\def\vf{\varphi}
\def\inbar{\vrule height1.5ex width.4pt depth0pt}
\def\bfzero{\relax{\rm I\kern-.18em 0}}
\def\bfone{\relax{\rm 1\kern-.35em 1}}

\DeclareFontFamily{U}{rsf}{} \DeclareFontShape{U}{rsf}{m}{n}{
  <5> <6> rsfs5 <7> <8> <9> rsfs7 <10-> rsfs10}{}
\DeclareMathAlphabet\Scr{U}{rsf}{m}{n}

 \def\cD{{\cal D}}

\newcommand{\ft}[2]{{\textstyle\frac{#1}{#2}}}
\def\tilde{\widetilde}

\def\1bar{1\hskip -.275cm -}
\def\2bar{2\hskip -.275cm -}
\def\3bar{3\hskip -.275cm -}

\newsavebox{\uuunit}
\sbox{\uuunit}
                 {\setlength{\unitlength}{0.825em}
                      \begin{picture}(0.6,0.7)
                                      \thinlines
                                      \put(0,0){\line(1,0){0.5}}
                                      \put(0.15,0){\line(0,1){0.7}}
                                      \put(0.35,0){\line(0,1){0.8}}
                                     \multiput(0.3,0.8)(-0.04,-0.02){10}{\rule{0.5pt}{0.5pt}}
                      \end {picture}}

\makeatletter \@addtoreset{equation}{section} \makeatother

\def\bfone{\relax{\rm 1\kern-.35em 1}}

\def\bfone{\relax{\rm 1\kern-.35em 1}}
\font\cmss=cmss10 \font\cmsss=cmss10 at 7pt

\newcommand{\sym}{\mathfrak{sp}}
\newcommand{\slal}{\mathfrak{sl}}

\def\IE{\relax{{\rm I\kern-.18em E}}}

\def\IGam{\relax{{\rm I}\kern-.18em \Gamma}}
\def\IGa{\IA}

\def\IA{\relax{\hbox{{\rm A}\kern-.82em {\rm A}}}}

\def\del{\partial}
\begin{document}
\begin{titlepage}
\vskip 0.2cm
\begin{center}
{\Large {\sc  Axial Symmetric  K\"ahler manifolds,\\[0.2cm] the $D$-map  of  Inflaton Potentials\\[0.2cm]
and the Picard-Fuchs Equation}}\\[1cm]
{ \large Pietro Fr\'e$^{a}$\footnote{Prof. Fr\'e is presently fulfilling the duties of Scientific Counselor of the Italian Embassy in the Russian Federation, Denezhnij pereulok, 5, 121002 Moscow, Russia.}, Alexander S. Sorin$^{b}$ }
{}~\\
{}~\\
\quad \\
{{\em $^{a}$  Dipartimento di Fisica, Universit\'a di Torino,}}
\\
{{\em $\&$ INFN - Sezione di Torino}}\\
{\em via P. Giuria 1, I-10125 Torino, Italy}~\quad\\
{\tt fre@to.infn.it}
{}~\\
\quad \\
{{\em $^{b}$\sl\small Bogoliubov Laboratory of Theoretical Physics}}\\
{{\em  {\tt and} Veksler and Baldin Laboratory of High Energy Physics,}}\\
{{\em Joint Institute for Nuclear Research,}}\\
{\em 141980 Dubna, Moscow Region, Russia}~\quad\\
\emph{e-mail:}\quad {\small {\tt sorin@theor.jinr.ru}}
\quad \\
\end{center}
~{}
\begin{abstract}
In this paper we provide a  definition of the $D$-map, namely of the mathematical construction implicitly utilized by supergravity that associates an axial symmetric K\"ahler surface to every positive definite potential function $V(\phi)$. The properties of the $D$-map are  discussed in general. Then the $D$-map is applied to the list of integrable cosmological potentials   classified by us in a previous publication with A. Sagnotti. Several interesting geometrical and analytical properties of the manifolds in the image of this $D$-map are discovered and illustrated. As a by-product of our analysis we demonstrate the existence of (integrable) Starobinsky--like potentials that can be embedded into supergravity. Some of them follow from constant curvature K\"ahler manifolds.  In the quest for a microscopic interpretation of inflaton dynamics we present the Ariadne's thread provided by a new mathematical concept that we introduce under the name of axial symmetric descendants of one dimensional special  K\"ahler manifolds. By means of this token we define a clearcut algorithm that to each potential function $V(\phi)$ associates a unique $4th$ order Picard-Fuchs equation of restricted type. Such an equation encodes information on the chiral ring of a superconformal field theory to be sought for, unveiling in this way a microscopic interpretation of the inflaton potential. We conjecture that the physical mechanism at the basis of the transition from a special manifold to its axial symmetric descendant is probably the construction of an Open String descendant of a Closed String model.
\end{abstract}
\end{titlepage}
\tableofcontents
\newpage
\section{Introduction}
The new data on the spectrum of the Cosmic Microwave Background radiation collected by the Planck mission \cite{Ade:2013uln},\cite{Ade:2013zuv}, that extend to a considereably higher precision those of the WMAP mission \cite{Hinshaw:2012aka}, have attracted a lot of interest and have been the stimulus for a new wave of extensive studies
on the theoretical models of  cosmological inflation.  Relying on the circumstantial evidence provided by these data that seem to favour one-scalar-field inflationary models, a lot of efforts have been devoted to the issue whether such one field inflationary models with a \textit{realistic potential} might be included into supergravity, obviously considering the latter as the low-energy effective theory of some more fundamental unified theory like superstring/brane theory \cite{johndimitri},\cite{Ketov:2010qz},\cite{Ketov:2012jt},\cite{Kallosh:2013hoa},\cite{Kallosh:2013lkr},\cite{Farakos:2013cqa}. In the course of the last year, pursuing a parallel line of investigation  in collaboration with A. Sagnotti, the two of us have developed a systematic study of the one-field scalar potentials that, upon insertion into Friedman equations, give rise to an integrable dynamical system. Such investigation resulted in a bestiary of \textit{integrable cosmological potentials} that was analysed in detail and presented in \cite{noicosmoitegr}. Some of these integrable potentials present phenomenologically interesting features and,  being mainly of the exponential type (linear or rational combinations of exponentials with different slopes), present the intriguing phenomenon of climbing scalars, first observed in
\cite{Dudasprimo,Dudas:2012vv}. According to the analysis of \cite{Sagnotti:2013ica}, climbing scalars might explain the observed distortion of the CMB spectrum at low angular momentum. In parallel with the compilation of the aforementioned bestiary the two of us, in collaboration with M. Trigiante, considered the problem of the possible inclusion of such integrable potentials into gauged supergravity, both extended and $\mathcal{N}=1$. The results of such investigations, that are quite extensive and already partially anticipated in the conclusions of \cite{noicosmoitegr}, will appear almost contemporarly with the present paper in \cite{nointegrable2}. As shown in \cite{nointegrable2}, although the class of potentials that includes the
 integrable ones (linear or rational combinations of exponentials) is quite ubiquitous among consistent truncations of gauged supergravities based on homogeneous scalar manifolds $\mathrm{G/H}$, the precise combinations requested by integrability are quite rare and difficult to be realized in the $\mathcal{N}$-extended context. Also in $\mathcal{N}=1$ supergravity, gauged by means of  a superpotential $W(\mathfrak{z})$, integrable scalar potentials, or other phenomenologically favoured potentials, meet serious obstacles in order to be embedded.
 \par
 A drastic shift of perspective on the supergravity embedding of cosmological scalar potentials occured in the last mid-summer with the publication of \cite{minimalsergioKLP}.  The authors of that paper advocated that  any positive definite potential can be embedded into $\mathcal{N}=1$ supergravity and in particular one can include such cosmological models as that known after the name of Starobinsky \cite{Starobinsky:1980te} which, at the  classical level, are equivalent to the supersymmetric coupling of higher curvature terms. Indeed relying on work of the end of the eighties on the superfield-superconformal approach to the construction of $\mathcal{N}=1$ supergravity \cite{Cecotti:1987qe},\cite{Cecotti:1987sa}, the authors of \cite{minimalsergioKLP} showed that, by using the new minimal set of auxiliary fields rather than the old one, the coupling of a higher curvature term $R^2$ to supergravity is equivalent to a standard supergravity model with second derivatives that includes a real scalar and a massive vector field \cite{massivetoine}. A posteriori, the resulting lagrangian  can be reinterpreted within  standard second order derivative supergravity as the Higgs phase of a model with a Wess-Zumino multiplet and a vector multiplet, the latter gauging a shift symmetry of the former. The axion field is eaten up by the gauge field which becomes massive. Instead, if one uses the old minimal set of auxiliary fields, supergravity coupled to an $R^2$ term produces a theory with two chiral multiplets \cite{Ferrara:2013wka}.
 \par
 Immediately after \cite{minimalsergioKLP}, the two of us published a paper \cite{primosashapietro}  where, relying on old papers on the rheonomic approach to supergravity and on the issue of the auxiliary fields \cite{contownsend}, \cite{D'Auria:1988qm}, we further emphasized that the mechanism of inclusion of arbitrary positive-definite one-scalar potentials into $\mathcal{N}=1$ supergravity is not necessarily linked to one better than to another scheme of construction of the Lagrangian (superfield, superconformal tensor calculus, or rheonomy) but simply to the shift of attention from the F-sector to the D-sector of the general scalar potential, as presented in the standard component form of the complete $\mathcal{N}=1$ supergravity Lagrangian, firstly constructed in \cite{standardN1}, later summarized in more geometrical terms in the book \cite{castadauriafre2} and   rephrased in our own recent paper \cite{primosashapietro}. A few new papers  \cite{Ferrara:2013wka}, \cite{Ferrara:2013kca}, \cite{Ferrara:2013pla} have contributed additional clarifications and additional examples to the scenario but the key point remains the same as already drafted in \cite{minimalsergioKLP} and emphasized by us in  \cite{primosashapietro}. The one-field potential is just the gauge contribution to the general $\mathcal{N}=1$ potential ensuing from the gauging of a translational or rotational symmetry of an underlying K\"ahler manifold and, by its own nature, it is the square of the momentum map of the gauged Killing vector. As firstly pointed out in \cite{minimalsergioKLP}, and already recalled, the standard Higgs mechanism implies that the axion field, partner of the inflaton, is eaten up by the vector field that gauges such a translation, or rotational symmetry, and by this token becomes massive.
 \par
 Such a universal picture perfectly fits into standard two-derivative supergravity and does not require any particular set of auxiliary fields to be implemented. To it we stick and from it we move on to analyse its profound consequences.
 \paragraph{The $D$-map.}
 As we already pointed out in \cite{primosashapietro}, this sort of Copernican revolution has the profound implication that each positive-definite potential $V(\phi)$ defines, through an algorithm motivated by supergravity, a K\"ahler manifold of complex dimension one, actually a two-dimensional surface, that enjoys the fundamental defining property of being axial symmetric. As already happened with other istances of geometrical maps like the time-honoroured $C$-map, also in this case, the geometrical construction utilized by supergravity can be isolated, purified of its physical non essential ingredients and formulated in pure mathematical terms. This is the first goal pursued by the present paper.  As a tribute to the supergravity origin of this new mathematical concept and imitating the successful practice initiated with the $C$-map, we name $D$-map the well defined algorithm that constructes an axial symmetric K\"ahler surface starting from the real data of a positive definite potential function $V(\phi)$.
 There are two immediate consequences of this approach. On one hand the geometrical structures attached to $V(\phi)$ by means of its $D$-map image provide a new arsenal of weapons to classify and analyse the nature of the various potential functions: curvature of the corresponding manifold, structure of its boundary, singularities if any, structure of the space of geodesics and so on. On the other hand and most prominently, the philosophical implication is that what comes first is not the potential, rather the geometry of the corresponding manifold, $V(\phi)$ being only the result of gauging and our anthropic perception through experiments and, for instance, measures of the CMB spectrum of an underlying internal geometry of microscopic origin. Relying on this viewpoint we have emphasized the reverse path from geometry to the potential (essentially the inverse of the $D$-map) and, analysing the general structure of axial symmetric euclidian metrics on two dimensional surfaces, we have retrieved the potential $V(\phi)$ as an yield of such metrics. We have also retrieved the process that leads from the field $\phi$ with canonical kinetic terms to the scalar field $C$ that is the natural partner of the gauged-away axion, as the integration of the standard geometrical equations defined by the complex structure $\mathfrak{J}$.
 \paragraph{The $D$-map images of the integrable potentials.} The next natural question raised by the $D$-map view-point concerns the geometrical status of the integrable potentials classified in our bestiary \cite{noicosmoitegr}. We have considered the axial symmetric K\"ahler surfaces associated with integrable potentials and we have analysed their properties, guided by the idea that the integrability of the ancestor should certainly reflect into special geometrical properties of the descendants. Indeed we have found that, at least for some of the infinite series of integrable potentials and for some of the sporadic ones, certain distingueshed miracles do happen. For instance the equation induced by the complex structure can be integrated in terms of classical special functions (hypergeometric and Appel functions) and the analytic form of the K\"ahler potential in terms of the resulting flat coordinate $C$ can be reconstructed in terms of power series that have  rational coefficients calculable to all orders. This is quite inspiring and reminiscent of what happens with the special K\"ahler geometries of Calabi Yau moduli spaces as calculated by means of mirror symmetry (for review and references we refer the reader to the book \cite{N2Wonder}).
 \paragraph{Constant curvature manifolds and Starobinsky--like models.} Another important fact discovered through our analysis of the $D$-map of integrable potentials is that certain very special instances of potentials lead to K\"ahler axial symmetric surfaces of constant curvature, namely to the standard coset manifold $\mathrm{SL(2,\mathbb{R})/O(2)}$ with some special value of its curvature. This has the relevant consequence that certain inflaton potentials can be simply obtained from quite conventional supergravity coupled to an $\mathrm{SL(2,\mathbb{R})/O(2)}$ Wess-Zumino multiplet, simply by choosing a suitable  parametrization of the coset  and gauging a compact or non compact one-dimensional subgroup of $\mathrm{SL(2,\mathbb{R})}$. Not only that. Since the momentum map is sensitive not only to the K\"ahler potential $\mathcal{K}$ but to the full norm of a holomorphic section of the Hodge bundle:
 \begin{equation}\label{normitalia}
  \log  \left|| W \right||^2 \, \equiv \, \mathcal{K} \, + \, \log \left| W \right|^2
 \end{equation}
 as long as a superpotential $W$ is invariant with respect to the gauged symmetry (and we will spell out the solution of such a constraint) we can add it to $ \mathcal{K}$  obtaining two indipendent modifications of an integrable scalar potential  associated with a constant curvature manifold. One modification is just inside the $D$-term and produces Starobinsky--like models, another one, more complicated, comes from  $F$-terms that are perfectly compatible with the gauged symmetry. In our opinion, these possible contributions to the potential function $V(\phi)$ of the inflaton, so far had  been overlooked. Furthermore it is also of some interest to stress that a Starobinsky--like potential is included in one of the integrable series.
 \paragraph{Microscopic Origin of Inflaton Dynamics.} Apart from these various results on integrable potentials and on their $D$-map images, that are interesting for their own sake, the main conceptual  challenge raised by the supergravity realization of  inflaton dynamics through the choice of appropriate variable curvature K\"ahler manifolds is the possibility of tracing back such dynamics to a fundamental microscopic theory.
 \par
 The low energy effective lagrangian of string theory, namely supergravity, already contains scalar fields in $D=10$, just the dilaton $\varphi(x)$ for type IIA or type I and even a primeval $\mathrm{SL(2,\mathbb{R})/O(2)}$ manifold for type IIB. Compactifying superstring theory on appropriate $6$-dimensional internal manifolds $\mathcal{M}_6$ one obtains many more scalars that are actually identified with the various possible moduli  that parameterize the geometry of the internal dimensions. As long as the compact internal dimensions are torii, orbifolds or orientifolds, the geometry of the scalar manifold which we finally observe in $4$-dimensional supergravity is that of suitable coset manifolds $\mathrm{G/H}$, namely of \textbf{constant curvature manifolds}. Any truncation thereof remains constant curvature. Therefore if we require non constant curvature K\"ahler manifolds, as it happens to be the case for the $D$-map of cosmological inflaton potentials (whether integrable or not integrable is not relevant at the level of the present discussion), there are only two possible sources of such scalar manifolds:
 \begin{description}
   \item[A)] Moduli spaces of smooth curved internal spaces $\mathcal{M}_6$, like for instance Calabi-Yau threefolds that break supersymmetry to $\mathcal{N}=2$ in type II theories and to $\mathcal{N}=1$ in type I or heterotic strings.
   \item[B)] Quantum corrections from target space instantons or from string loops.
 \end{description}
The second case is far away from being under control and not too much can be said. The first offers  instead a rich scenario of possibilities thanks to algebraic geometry and mirror symmetry. The K\"ahler geometry of the moduli space of complex structure deformations of CY threefolds and, via mirror symmetry, also that of K\"ahler structure deformations of the same threefold can be exactly calculated with various techniques of algebraic geometry, the most direct and prominent of which is provided by Picard-Fuchs equations.  The resulting K\"ahler manifolds are actually special K\"ahler manifolds and certainly, even in the case of one modulus \textbf{they are not axial symmetric}. Indeed they typically have no continuous isometry. Yet the fascination of special K\"ahler geometry is that hermitian data, like the metric and the K\"ahler potential, are derived from the more fundamental holomorphic data encoded in a section of the flat symplectic bundle whose existence defines special geometry. Such a section is typically identified with a solution of the Picard Fuchs equation which is a $4th$ order linear differential equation of restricted form. Holomorphicity allows to truncate  its solutions to the real or imaginary axis and hence to construct with it a K\"ahler potential that depends only on one coordinate, the remaining one being associated with axial symmetry.
\paragraph{Axial symmetric K\"ahler descendants.}  In this paper, in addition to the notion of $D$-map we introduce another mathematical notion, that of \textit{Axial symmetric descendants of one dimensional Special K\"ahler manifolds} which essentially coincides with  the above described  restriction to some real axis of the solutions to the Picard-Fuchs equations. Furthermore we show that from any positive-definite potential $V(\phi)$ not only we are able to construct its associated axial symmetric K\"ahler manifold but we can also construct the uniquely defined Picard Fuchs equation which permits to identify the latter as the axial symmetric descendant of an appropriate special K\"ahler manifold.
\par
The name descendant is utilized in this context since, in the back of our mind, we cherish the hypothesis that such  a procedure of restriction of a complex modulus to the real axis  can be traced back to Open or Heterotic Superstrings and is, in some sense, the outcome, at the level of supergravity, of the construction of Open String descendants of Closed Strings. Such a general idea has to be made circumstantial by means of appropriate examples and explicit constructions yet it is very attractive and is coherent with the general features of Superstring Theory.  We leave it for future investigations. We just stress that the notion of axial descendant is mathematically firmly established and the construction of a Picard Fuchs equation associated with any positive definite potential $V(\phi)$ is also completely algorithmic and gives a unique answer. Once this equation is established we come into possession of an identity card which is attached to a Calabi Yau threefold or, more abstractly, to the chiral ring of a $(2,2)$-superconformal field theory. Whether the $(2,2)$-superconformal field theory associated with phenomenologically favorite cosmological inflaton potentials or integrable ones is reasonable or not is a question that at the moment we are not able to answer but it is a fact that we have found a solid mathematical bridge that relates quite disparate structures and we opened a possible path for the microscopic explanation of inflaton dynamics.
\paragraph{Structure of the paper.}
Our paper is organized as follows. In the first two sections we analyse the structure of axial symmetric K\"ahler surfaces and we present  the mathematical definition of the $D$-map. In section 4 we reconsider in a detailed way how $\mathcal{N}=1$ supergravity realizes the $D$-map and we also present the usually disregarded $F$-type contributions to the inflaton potential. In sections 5,6,7 and 8 we analyse in full detail the K\"ahler manifolds that are in the $D$-map image of some of the the integrable potentials classified in  \cite{noicosmoitegr}. We focus in particular on the integrable series $I_1$, $I_2$ and $I_7$ that have specially nice analytically properties. Furthermore series $I_2$ that leads to Appel $F1$ functions hosts the $\gamma \, = \, - \, \ft 76$ best fit model that, according to the analysis of Sagnotti (\cite{Sagnotti:2013ica} and communication to the SQS13 workshop of Dubna) might better describe, among the integrable potentials, the distortion of the CMB spectrum at low angular momenta.
Section 9 is devoted to the search for a microscopic interpretation of inflaton potentials. There we introduce the concept of axial symmetric descendants of special K\"ahler manifold and we outline the algorithm for the construction of the Picard Fuchs equation associated to each positive definite potential. Section 10 contains our conclusions.
\section{The K\"ahler structure of Axial Symmetric two-dimensional Surfaces}
\label{axialsurf}
From a mathematical point of view, the $D$-map embedding of one-field inflaton models into $\mathcal{N}=1$ supergravity turns out to be nothing else but the uncovering of the  K\"ahler structure underlying every axial-symmetric two-dimensional surface together with the unique relation existing between the metric coefficients and the momentum-map of the Killing vector generating the axial symmetry. Supergravity simply takes advantage of this purely mathematical phenomena since the $D$-type contribution to the potential is simply  given by the square of the momentum map. For this reason we begin by summarizing some  simple mathematical results associated with axial symmetric two-dimensional surfaces.
\par
Let us consider a two-dimensional  Riemannian manifold $\Sigma$ whose points are labeled, in some open chart, by the two coordinates $x^\alpha \, = \, \{U,B\}$ and whose Riemannian metric has, in such a chart, the following form:
\begin{equation}\label{metraxia}
  ds^2_\Sigma \, = \, p(U) \, dU^2 \, + \, q(U) \, dB^2
\end{equation}
$p(U),q(U)$ being two positive definite functions of their argument. The manifold $\Sigma$ is an axial symmetric surface, since the metric (\ref{metraxia}) admits the Killing vector $\vec{k} \, = \, \partial_B$ which we can always interpret as the generator of infinitesimal rotations around a symmetry axis.
\par
As all other two-dimensional surfaces, $\Sigma$ has an underlying complex K\"ahlerian structure that we can systematically uncover (for details and conventions see section 2.4 of the book \cite{N2Wonder}). The first step is to determine the complex structure with respect to which the metric (\ref{metraxia}) is hermitian. By definition an almost complex structure is a tensor $\mathfrak{J}_\alpha^\beta$ which squares to minus the identity:
\begin{equation}\label{gormus}
  \mathfrak{J}_\alpha^\beta \,\mathfrak{ J}_\beta^\gamma \, = \, - \, \delta^\gamma_\alpha
\end{equation}
The almost complex structure $\mathfrak{J}_\alpha^\beta$ becomes a true complex structure if its Nienhuis tensor vanishes:
\begin{equation}\label{nienhuis}
  N_{\alpha\beta}^\gamma \, \equiv \, \partial_{[\alpha } \, \mathfrak{J}^\gamma_{\beta]} \, - \, \mathfrak{J}^\mu_\alpha \,
  \mathfrak{J}^\nu_\beta \,  \partial_{[\mu } \, \mathfrak{J}^\gamma_{\nu]} \, = \, 0
\end{equation}
Given a complex structure, a metric $g_{\alpha\beta}$ is hermitian with respect to it if the following identity is true:
\begin{equation}\label{hermitus}
  g_{\alpha\beta} \, = \, \mathfrak{ J}_\alpha^\gamma \, \mathfrak{ J}_\beta^\delta \, g_{\gamma\delta}
\end{equation}
Given the metric (\ref{metraxia}) there is a unique tensor $\mathfrak{ J}_\alpha^\beta$, which simulatenously satisfies eq.s(\ref{gormus},\ref{nienhuis},\ref{hermitus}) and it is the following:
\begin{equation}\label{forbito}
 \mathfrak{ J} \, = \, \left(\begin{array}{cc}
                               0 & \mathfrak{J}_B^U \\
                                \mathfrak{J}_U^B & 0
                             \end{array} \right ) \, = \, \left( \begin{array}{cc} 0 & \sqrt{\frac{p(U)}{q(U)}} \\
                              - \, \sqrt{\frac{q(U)}{p(U)}} & 0 \\
                     \end{array} \right)
\end{equation}
Next, according to theory, the K\"ahler 2-form is defined by:
\begin{eqnarray}
  \mathrm{K} &=& \mathrm{K}_{\alpha\beta} \, dx^\alpha \, \wedge \, dx^\beta \, = \, g_{\alpha \gamma} \, \mathfrak{J}^\gamma_\beta \, \, dx^\alpha \, \wedge \, dx^\beta \nonumber \\
  \null &=& \, - \, \sqrt{p(U) \, q(U) } \, dU \, \wedge \, dB \label{obrakalera}
\end{eqnarray}
and it is clearly closed. Hence the metric (\ref{metraxia}) is K\"ahlerian and necessarily admits a representation in terms of a complex coordinate $\zeta$  and a K\"ahler potential $\mathcal{K}(\zeta \, , \, \bar{\zeta})$. In terms of the complex coordinate:
\begin{equation}\label{zetosa}
  \zeta \, = \, \zeta(U,B)
\end{equation}
the K\"ahler 2-form $\mathrm{K}$ in  eq.(\ref{obrakalera}) should be rewritten as:
\begin{equation}\label{foxterry}
  \mathrm{K} \, = \, \partial \, \overline{\partial} \, \mathcal{K} \, = \, \partial_\zeta \, \partial_{\bar{\zeta}} \, \mathcal{K} \, d\zeta \, \wedge \, d\bar{\zeta}
\end{equation}
\subsection{Integrating the complex structure} The main question is: what is the complex coordinate $\zeta$? Given the complex structure  the complex coordinate is uniquely defined as a solution of the following differential equation\footnote{See eq.(2.4.27) of \cite{N2Wonder}.}:
\begin{equation}\label{golosina}
  \mathfrak{ J}_\alpha^\beta \, \partial_\beta \, \zeta \, = \, {\rm i} \partial_\alpha \, \zeta
\end{equation}
In our case the above equation becomes:
\begin{equation}\label{robedamatti}
  \sqrt{\frac{p(U)}{q(U)}} \, \partial_B \, \zeta(U,B) \, = \, {\rm i} \, \partial_U \, \zeta(U,B)
\end{equation}
which can be solved by means of the ansatz:
\begin{equation}\label{curlandia}
  \zeta \, = \, \rho(U) \, \exp\left[ {\rm i} B\right]
\end{equation}
 leading to
\begin{equation}\label{facilone}
  \frac{d}{dU} \, \log \, \rho(U) \, = \, \sqrt{\frac{p(U)}{q(U)} } \quad ; \quad \rho(U) \, = \, \exp\left[ C(U) \right] \quad ; \quad C(U) \, \equiv \, \int \,  \sqrt{\frac{p(U)}{q(U)} } \, dU
\end{equation}
Hence the function $C(U) \, \equiv \, \log |\zeta |$ admits the integrable representation spelled out in eq.(\ref{facilone}). For reasons that will become apparent in the sequel, we name the above solution $\zeta(U,B)$ of the differential equation (\ref{golosina}), based on the ansatz (\ref{curlandia}), the \textbf{Disk complex coordinate}.
\par
There is another ansatz which leads to another equally simple integration of the differential equation (\ref{golosina}) for the complex coordinate. If we replace
\begin{equation}\label{bartolomeo1}
  \zeta \,\mapsto \, t(U,B) \, \equiv \, {\rm i} \, C(U) \, - \, B
\end{equation}
then equation (\ref{golosina}) is satisfied by the same function $C(U)$ defined by the integral representation (\ref{facilone}). In this case
we have $C(U) \, = \, \mathrm{Im} t$. The alternative complex coordinate $t(U,B)$ is named the  \textbf{Plane complex coordinate}.
\par
Let us next consider the calculation of the K\"ahler metric in the complex set up.
\subsubsection{The K\"ahler metric in the Disk presentation}
Assuming that the K\"ahler potential $\mathcal{K}$ depends only on the modulus $\rho$ of the disk complex coordinate:
\begin{equation}\label{disculoJ}
  \mathcal{K} \, = \, J(\rho)
\end{equation}
the general expression for the K\"ahler metric
\begin{equation}\label{forniconeK}
    ds^2 \, \equiv \, \partial_\zeta \, \partial_{\bar{\zeta}} \, \mathcal{K} \, d\zeta \, d\bar{\zeta}
\end{equation}
takes the following explicit form:
\begin{eqnarray}
    ds^2  & = & \ft 14 \, A(U) \, \left(\rho^\prime(U)\right)^2 \, dU^2 \, + \, \ft 14 A(U) \left(\rho(U)\right)^2 \, dB^2 \nonumber\\
    A(U) &\equiv & \frac{d^2 J}{d\rho^2} \, + \, \frac{1}{\rho} \, \frac{d J}{d\rho}\label{garessio}
\end{eqnarray}
and recalling
\begin{equation}\label{corbezzolo}
    C(U) \, \equiv \, \log \,\rho(U)
\end{equation}
we easily verify that eq.(\ref{garessio}) is transformed into:
\begin{equation}\label{golodny}
    ds^2 \, = \, \ft 14 \left(\frac{dC}{dU}\right)^2 \, \frac{d^2 J}{dC^2} \, dU^2\, + \, \ft 14 \, \frac{d^2 J}{dC^2} \, dB^2
\end{equation}
The above form of the metric agrees with the original one (\ref{metraxia}) if:
\begin{eqnarray}
  \ft 14 \, \left(\frac{dC}{dU}\right)^2 \, \frac{d^2 J}{dC^2} &=&p(U) \label{pirla1} \\
 \ft 14 \, \frac{d^2 J}{dC^2} &=& q(U) \label{pirla2}
\end{eqnarray}
The two conditions (\ref{pirla1}) and (\ref{pirla2}) are consistent in virtue of the integral representation (\ref{facilone}) of the function $C(U)$. Suppose now that this latter can be analitically inverted:
\begin{equation}\label{caramba}
  U \, = \, \mathcal{U}(C)
\end{equation}
Inserting (\ref{caramba}) into $q(U)$ in eq.(\ref{pirla2}) we obtain:
\begin{equation}\label{fustagnovellutato}
  \ft 14 \, \frac{d^2 J}{dC^2} \, = \, \mathcal{Q}(C) \, \equiv \, q(\mathcal{U}(C)
\end{equation}
Hence, by means of a double integration we obtain the explicit form of the K\"ahler potential $J(C)$ and the complex K\"ahler structure of the  space defined by the metric (\ref{metraxia}) is uncovered.
\subsubsection{The K\"ahler metric in the Plane presentation}
Next we consider the alternative integration  (\ref{bartolomeo1}) of the complex structure equation (\ref{golosina}) which leads to the complex coordinate $t(U,B)$. Let us assume that the K\"ahler potential is a function, only of the imaginary part of $t$:
\begin{equation}\label{kalero2}
    \mathcal{K}\left(t,\bar{t}\right) \, = \, J \left(\mbox{Im} \,t \right) \, = \, J\left(C\right)
\end{equation}
Following the same steps as in the previous case we find that under such hypotheses the K\"ahler metric:
\begin{equation}\label{fornicone2}
    ds^2 \, \equiv \, \partial_t\, \partial_{\bar{t}} \, \mathcal{K} \, dt \, d\bar{t}
\end{equation}
takes the following explicit form:
\begin{eqnarray}
    ds^2  & = & \ft 14 \, J^{\prime\prime}(C) \, \left(C^\prime(U)\right)^2 \, dU^2 \, + \, \ft 14 J^{\prime\prime} \, dB^2 \label{garessio2}
\end{eqnarray}
As in the previous case, the above metric agrees with the original one (\ref{metraxia}) if   conditions (\ref{pirla1}) and (\ref{pirla2}) are satisfied.
\par
 It follows that for each axial symmetric metric we can reconstruct the complex K\"ahler structure in two  ways, either in the \textit{disk coordinate} or in the \textit{plane coordinate}. In both cases the K\"ahler potential is function only of the coordinate $C$ defined by the integral in eq.(\ref{facilone}) which we name the \textbf{flat coordinate}. It should be noted that the global structure of the manifold resulting from the disk and the plane presentation are inequivalent. In the disk case the coordinate $B$ is compact defined in the interval $[0,2\pi]$, while in the plane case it is non compact and it is defined over $\mathbb{R}$.
\subsubsection{The Canonical Coordinate $\phi$ and the $\mathcal{D}$-function}
Let us now go back to the general form (\ref{metraxia}) of the axial symmetric metric and let us introduce the following coordinate change. Define:
\begin{equation}\label{gonzallo}
  \phi \, = \, \phi(U) \, = \, \int \, 2 \,  \sqrt{p(U)} \, dU \quad ; \quad \ft 12 \, d\phi \, = \, \sqrt{p(U)} \, dU
\end{equation}
Assuming that the function $\phi(U)$ can be inverted
\begin{equation}\label{cursutu}
  U \, = \, U(\phi)
\end{equation}
and implementing the change of variable (\ref{gonzallo}), (\ref{cursutu}) in the metric (\ref{metraxia}) we obtain its new canonical form:
\begin{equation}\label{solarium}
  ds^2 \, = \, \ft 14 \, \left[ d\phi^2 \, + \, \left(\mathcal{ D}(\phi)\right)^2 \, dB^2 \right]
\end{equation}
where, by definition we have set:
\begin{equation}\label{finorco}
  \mathcal{ D}(\phi) \, = \, \sqrt{ q\left(U(\phi)\right)}
\end{equation}
The canonical form (\ref{solarium}) of the axial symmetric metric is just a particular case of the general one (\ref{metraxia}) where:
\begin{equation}\label{boraldo}
  p(\phi) \, = \,  \ft 14 \quad ; \quad q(\phi) \, = \, \ft 14  \left(\mathcal{ D}(\phi)\right)^2                                                                                                  \end{equation}
are the special forms taken in this gauge by  the defining functions.
\subsection{Interpretation of the $\mathcal{D}$-function}
First let us observe that in the canonical gauge, eq.s (\ref{pirla1}) and (\ref{pirla2}) become
\begin{equation}\label{critallica2}
   J^{\prime\prime}(C) \, \left(C^\prime(\phi)\right)^2 \, = \, 1 \quad ; \quad J^{\prime\prime}(C) \, = \,  \left(\mathcal{ D}(\phi)\right)^2
\end{equation}
Using such an information we can calculate the first derivative of the K\"ahler potential with respect to the flat coordinate $C$. We find:
\begin{eqnarray}\label{convezia}
  \frac{dJ}{dC}& = & \int \, \left(\mathcal{ D}(\phi)\right)^2 \, dC \, = \, \int \, \left(\mathcal{ D}(\phi)\right)^2 \,  \frac{dC}{d\phi} \, d\phi \nonumber\\
  & = & \int \, \mathcal{ D}(\phi)\, d\phi
\end{eqnarray}
What is the geometrical interpretation of $\frac{dJ}{dC}$? To answer this question, let us  recall the notion of momentum map for holomorphic Killing vectors of K\"ahler metrics. In the case of  one complex dimension if
\begin{equation}\label{sicomoro}
    \zeta \, \rightarrow \, \zeta \, + \, k^\zeta(\zeta) \quad ; \quad \bar{\zeta} \, \rightarrow \, \bar{\zeta} \, + \, k^{\bar{\zeta}}(\bar{\zeta})
\end{equation}
is an infinitesimal isometry of the metric (\ref{forniconeK}), there exist a real function $\mathcal{P}(\zeta,\bar{\zeta})$, such that
\begin{equation}\label{gospadi}
    k^\zeta (\zeta) \, = \, {\rm i} \, g^{\zeta\bar{\zeta}} \, \partial_{\bar{\zeta}} \, \mathcal{P} \quad ; \quad k^{\bar{\zeta}} (\bar{\zeta}) \, = \,
     - \, {\rm i}
     \, g^{\bar{\zeta}\zeta} \, \partial_\zeta \, \mathcal{P}
\end{equation}
In terms of the K\"ahler potential $\mathcal{K}$, supposedly invariant under the considered isometry ($k^\zeta \, \partial_\zeta \,\mathcal{K} \,  + \, k^{\bar{\zeta}}\,\partial_{\bar{\zeta}} \, \mathcal{K}\, = \, 0$), the momentum map of the Killing vector, satisfying the defining condition (\ref{gospadi}), is constructed through the following formula:
\begin{equation}\label{bozhemoi}
    \mathcal{P} \, = \, -\,{\rm i} \, \ft 12 \, \left(k^\zeta \, \partial_\zeta \,\mathcal{K} \,  - \, k^{\bar{\zeta}}\,\partial_{\bar{\zeta}} \, \mathcal{K} \right)
\end{equation}
For a K\"ahler metric which depends only from the modulus $\rho$ of $\zeta$ the infinitesimal shift of the phase $B$:
\begin{equation}\label{gossardo}
    B \, \rightarrow \, B \, + \, c
\end{equation}
is certainly an isometry and it is generated by the following killing vector:
\begin{equation}\label{compattokillino}
    k^\zeta \, = \, {\rm i} \, \zeta \quad ; \quad k^{\bar{\zeta}} \, = \, - \, {\rm i} \, \bar{\zeta}
\end{equation}
The corresponding momentum map is derived from eq. (\ref{bozhemoi}) and reads:
\begin{equation}\label{forzatutto}
    \mathcal{P} \, = \, \rho \, \frac{\mathrm{d}J}{\mathrm{d}\rho} \, = \, \frac{\mathrm{d}J}{\mathrm{d}C}
\end{equation}
Alternatively, if we consider the plane--type integration  (\ref{bartolomeo1}) of the complex structure equation (\ref{golosina}) and we  assume that the K\"ahler potential depends only on the imaginary part of the complex coordinate $t$,  then a  translation of its real part, namely $B\mapsto B+c$ is a non compact isometry of the corresponding K\"ahler metric. From the formula for the momentum-map (\ref{bozhemoi}) rewritten for the present case:
\begin{equation}\label{bozhemoi2}
    \mathcal{P} \, = \, -\,{\rm i} \, \ft 12 \, \left(k^t \, \partial_t \,\mathcal{K} \,  - \, k^{\bar{t}}\,\partial_{\bar{t}} \, \mathcal{K} \right)
\end{equation}
we obtain once again
\begin{equation}\label{fusecchio3}
    \mathcal{P}\, = \, J^{\prime}(C)
\end{equation}
This follows since  the killing vector of  translations is $k^t \, = \, 1$.
\par
From both interpretations we arrive at the conclusion that writing the axial symmetric metric in the canonical form (\ref{solarium}) the function  $\mathcal{D}(\phi)$ is the derivative with respect to $\phi$ of the momentum-map of the Killing vector which generates translations of the cyclic variable $B$:
\begin{equation}\label{coriandolino}
  \mathcal{D}(\phi) \, = \, \frac{d}{d\phi} \, \mathcal{P}(\phi) \, \equiv \, \mathcal{P}^\prime(\phi)
\end{equation}
The above identification (\ref{coriandolino}) is the basis for the $D$-map realized in supergravity.
\par
Before coming to that let us exemplify the constructions outlined above by means of the classical example of the hyperboloid and of the manifold of constant negative curvature.
\subsection{The classical hyperboloid constructions}
\begin{figure}[!hbt]
\begin{center}
\iffigs
\includegraphics[height=70mm]{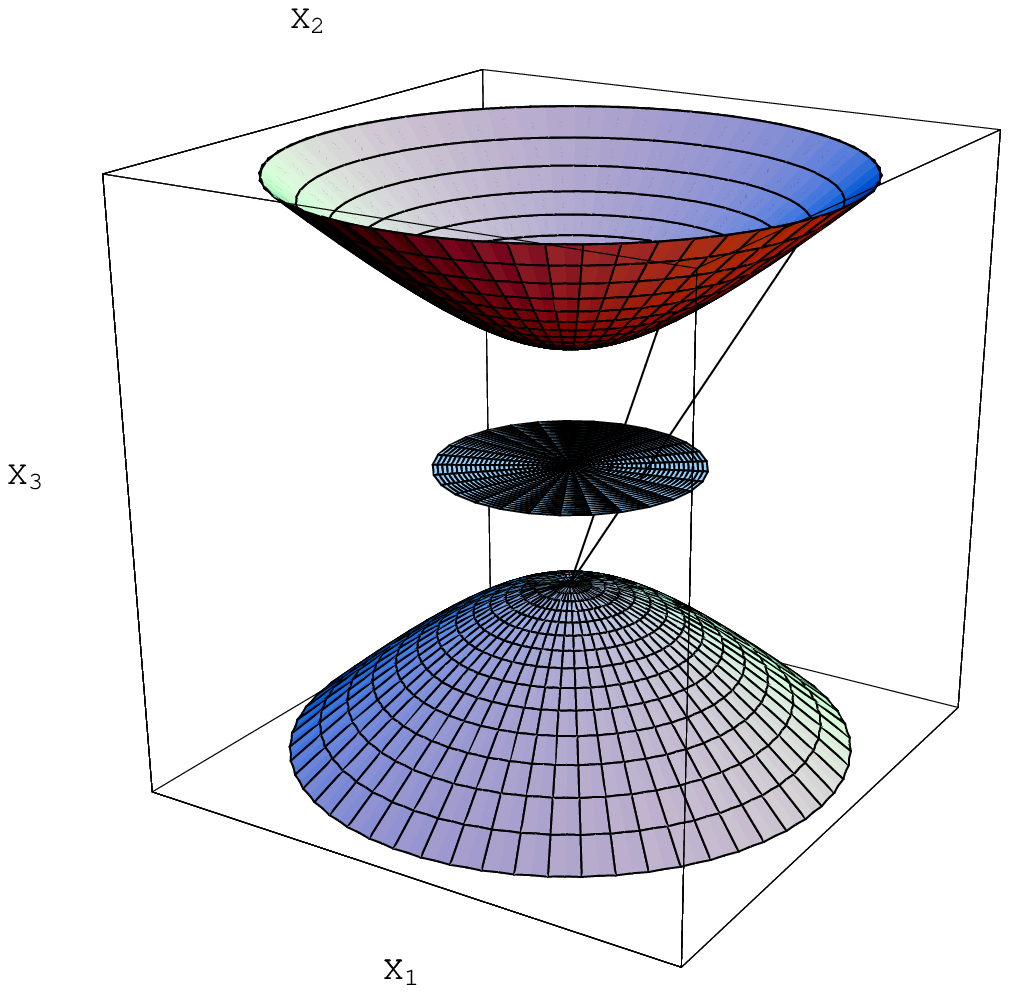}
 \includegraphics[height=70mm]{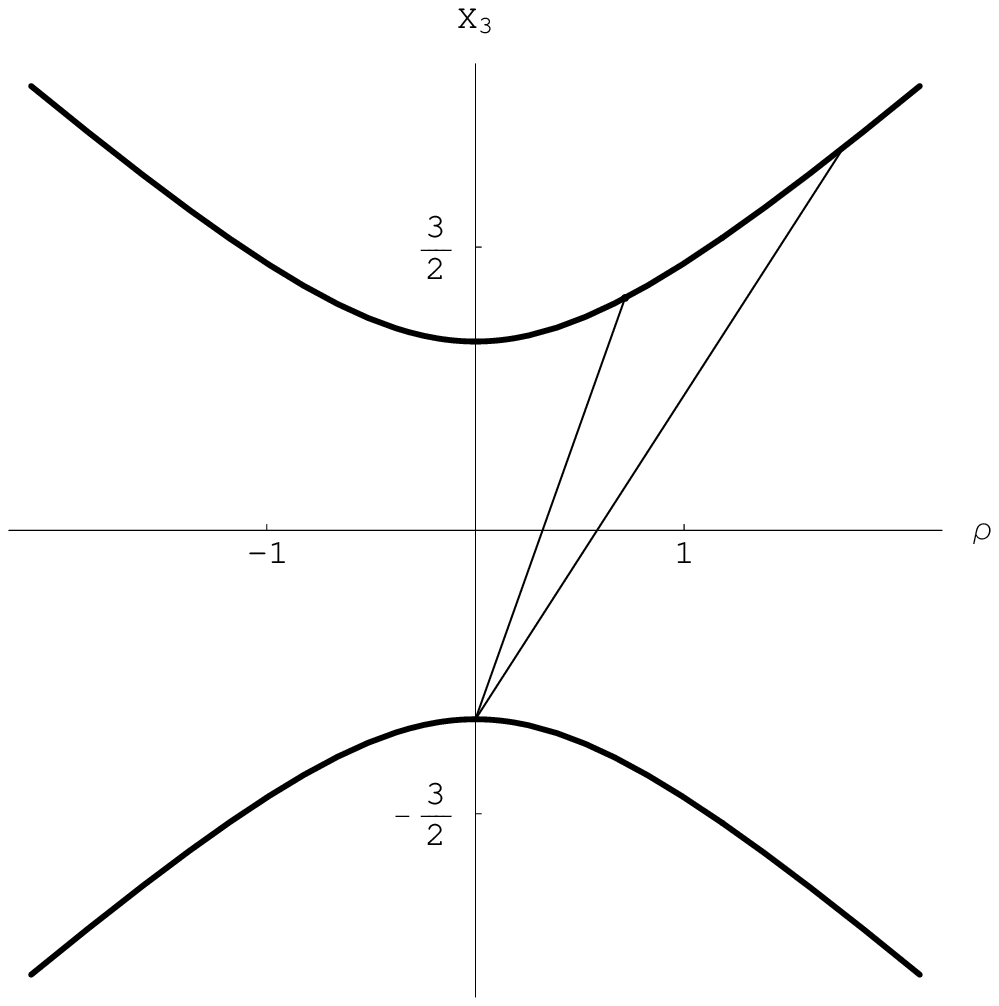}
\else
\end{center}
 \fi
\caption{\it Stereographic projection of the hyperboloid onto the unit disk of the complex plane. From any point of the upper branch of the hyperboloid we draw the line segment that links it to the vertex of the lower branch. This line segment intercepts the plane at $X_3=0$ in a point that necessarily lies inside the unit disk. The complex number associated with the latter point in the plane provides a complex coordinate system on the hyperboloid. The figure on the left is three-dimensional. The figure on the right corresponds to the intersection of the figure on the left with the plane $X_2=0$.}
\label{hyperboloide}
 \iffigs
 \hskip 1cm \unitlength=1.1mm
 \end{center}
  \fi
\end{figure}

Let us start from reviewing the classical constructions. If we name $\left\{X_1,X_2,X_3\right\}$ the coordinates of $\mathbb{R}^3$, the hyperboloid $\mathcal{C}_{\mathrm{Hyp}}$ (or the pseudo-sphere, as it was originally named by Beltrami) is the algebraic locus defined by the following algebraic quadratic equation:
\begin{equation}\label{geolocus1}
\vec{X}  \, \in \, \mathcal{C}_{\mathrm{Hyp}} \, \subset \, \mathbb{R}^3 \quad \Leftrightarrow \quad   X_3^2 \, - \, X_1^2 \, - \, X_1^2 \, = \, r^2
\end{equation}
The algebraic locus $\mathcal{C}_{\mathrm{Hyp}}$ has two branches that are visualized in fig.\ref{hyperboloide}.
Analytically $\mathcal{C}_{\mathrm{Hyp}}$ can be parameterized by means of two coordinates $\{\phi,B\}$ and by the following position that solves eq.(\ref{geolocus1}):
\begin{eqnarray}
  X_3 &=& \pm \, r \, \cosh \left( \, \omega \, \phi\right)\nonumber \\
 X_1&=& r \,\sinh \left( \, \omega \, \phi\right)\, \cos B\nonumber \\
  X_2&=& r \,\sinh \left( \, \omega \, \phi\right)\, \sin B \label{coshparamma}
\end{eqnarray}
the double sign choice in the parameterization of $X_3$ corresponding to the two branches. Considering next the standard Lorentz metric in $\mathbb{R}^3$ defined as it follows:
\begin{equation}\label{Lorenzo}
    ds^2_{L} \, \equiv \, - \, dX^2_3 \, + \,  dX^2_1 \, + \,  dX^2_2 \,
\end{equation}
if we restrict it to the $\mathcal{C}_{\mathrm{Hyp}}$ locus, by means of the parameterization  (\ref{coshparamma}) we obtain:
\begin{equation}\label{gospadin}
   \left. \, ds^2_{L} \right|_{\mathcal{C}_{\mathrm{Hyp}}} \, \equiv \, r^2 \, ds^2_{\mbox{Hyp}} \, = \,  r^2 \,\left[\omega^2 \, d\phi^2 \, + \, \left(\sinh \left( \, \omega \, \phi\right) \right)^2 \, dB^2 \right]
\end{equation}
There is a classical interpretation of the above metric in terms of the stereographic projection of the hyperboloid onto the unit disk inside the complex plane. Consider in $\mathbb{R}^3$ the straight line segment which goes from the point $(0,0,-r)$ to any point $P\, \in \, \mathcal{C}_{\mbox{Hyp}}$. This line intersects the plane $X_3\, = \, 0$ in a point $P^\prime$ which necessarily lies within the unit disk. In this way every point of the unit disk uniquely corresponds to a point of the upper branch of the hyperboloid and all points of the latter have a representative in the unit disk. The $x,y$ coordinates of these images in the plane $X_3=0$ are easily seen to be the following ones:
\begin{equation}\label{furtino}
    x \, \equiv \, \frac{X_1}{1+X_3} \, = \, \frac{\sinh \left( \, \omega \, \phi\right) \,\cos B}{1 \,+ \, \cosh\left( \, \omega \, \phi\right)} \quad ; \quad  y \, \equiv \, \frac{X_2}{1+X_3} \, = \,\frac{\sinh \left( \, \omega \, \phi\right) \,\sin B}{1 \,+ \, \cosh\left( \, \omega \, \phi\right)}
\end{equation}
and it follows that the complex coordinate defined below:
\begin{equation}\label{zetadef}
    \zeta  \, \equiv \, x \, + \, {\rm i} \, y \, = \, \frac{\sinh \left( \, \omega \, \phi\right) }{1 \,+ \, \cosh\left( \, \omega \, \phi\right)} \, \exp \left [\,{\rm i} \,B \right]
\end{equation}
lies in the unit disk
\begin{equation}\label{boundo}
    \left|\zeta\right|^2 \, < \, 1
\end{equation}
for all of its possible pre-images in the upper hyperboloid branch.
Introducing the following K\"ahler potential:
\begin{equation}\label{PoincKalDisk}
    \mathcal{K}\left(\zeta , \bar{\zeta} \right ) \, = \, 4 \, \log \, \left ( 1 \, - \, \left|\zeta\right|^2 \right)
\end{equation}
one verifies that the metric $ds^2_{\mbox{Hyp}}$ is indeed K\"ahlerian with respect to  it:
\begin{eqnarray}
    \left. ds^2_{\mbox{Hyp}}\right|_{r=1} & = & -\partial_\zeta \, \partial_{\bar{\zeta}} \, \mathcal{K} \, d\zeta \, d\bar{\zeta} \nonumber\\
    & = & \frac{4}{\left( 1 \, - \, \left|\zeta\right|^2\right)^2} \, \left|d\zeta\right|^2 \, = \, \omega^2 \, d\phi^2 \, + \, \left(\sinh \left( \, \omega \, \phi\right) \right)^2 \, dB^2 \label{ringo}
\end{eqnarray}
On the other hand, as it is well known, another equally good description of the geometry of the hyperboloid is obtained  mapping the unit disk into the upper complex half plane by means of the so called Cayley transformation which is recalled below:
\begin{eqnarray}
    t & \equiv & -\,{\rm i} \, \frac{\zeta \, + \, 1}{\zeta \, - \,  1}  \, \equiv \, {\rm i}\, C\left(\tilde{\phi}\right) \, + \, \tilde{B} \nonumber\\
    C\left(\tilde{\phi}\right) & = & \exp \left[ \omega \tilde{\phi}\right] \label{cayleus}
\end{eqnarray}
In terms of the new complex variable $t$ the hyperboloid metric takes the well known Poincar\'e form which is derived from the different K\"ahler potential mentioned below:
\begin{eqnarray}\label{fintallo}
  \left. ds^2_{\mbox{Hyp}}\right|_{r=1} & = & \frac{\left| dt\right|^2}{\left({\rm Im}\,t\right)^2} \, = \, \partial_{t}\,\partial_{\bar{t}} \, \widehat{\mathcal{K}} \, dt \, d\bar{t} \nonumber\\
   & = & \omega^2 \, d{\tilde{\phi}}^2 \, + \, \exp\left [- 2 \, \tilde{\phi}\right ]\,d{\tilde{B}}^2\label{formalinus}\\
 \widehat{\mathcal{K}} & = & -\, 4 \, \log \left[{\rm Im}\,t \right]
\end{eqnarray}
In view  of the generalization of the hyperboloid geometry that we are going to introduce and of the $D$-map that we are about to define, what is important for us to stress,  is that the metric (\ref{ringo}) and the metric (\ref{formalinus}) have exactly the same structure, namely:
\begin{equation}\label{grugno}
    ds^2_{\mbox{K\"ahler}}\, = \, \omega^2 \left( \, d\phi^2 \, + \, \mathcal{D}^2(\phi) \, dB^2\right)
\end{equation}
with two different choices of the function $\mathcal{D}(\phi)$:
\begin{equation}\label{gualdacosa}
    \mathcal{D}(\phi) \, = \, \left \{ \begin{array}{ccl}
                               \frac{1}{\omega} \, \sinh\left(\omega \, \phi\right) & ; & \mbox{unit disk} \\
                               \frac{1}{\omega} \,\exp\left[ \, -\, \omega \, \phi \right] & ; & \mbox{upper complex plane}
                             \end{array}
    \right.
\end{equation}
As we know from our construction, the two metrics (\ref{ringo}) and (\ref{formalinus}), notwithstanding the different form of $\mathcal{D}(\phi)$ are completely equivalent and describe exactly the same K\"ahler manifold. Indeed the coordinate transformation that maps one into the other is  immediately available from the Cayley transformation (\ref{cayleus}) and reads as follows
\begin{eqnarray}
  \phi &=& \frac{{\rm arccoth}\left(\frac{-\tilde{B}^2-e^{2
   \tilde{\phi}}-1}{\sqrt{\left(\tilde{B}^2+1\right)^2+
   e^{4 \tilde{\phi}}+2 e^{2 \tilde{\phi}}
   \left(\tilde{B}^2-1\right)}}\right)}{\omega } \nonumber \\
  B &=& -\arctan\left(\frac{2 \tilde{B}}{\tilde{B}^2+e^{2
   \tilde{\phi}}-1}\right) \label{baldone}
\end{eqnarray}
This is an important warning that the function $\mathcal{D}(\phi)$ has  not a unique correspondence with the K\"ahler manifold underlying eq.(\ref{grugno}). If two functions $D_1^2(\phi)$ and $\tilde{D}_2^2(\tilde{\phi})$ can be mapped one into the other by means of some appropriate coordinate transformation that generalizes that in (\ref{baldone}), than they correspond to the same K\"ahler manifold.

\section{Mathematical definition of the $D$-map}
\label{mathDmap}
From the mathematical point of view the $D$-map consists of a well-defined algorithm that associates a \textit{complex one-dimensional, non-compact axial symmetric K\"ahler manifold $\mathcal{M}_{J}$} to each \textit{positive definite potential $V(\phi)$}:
\begin{equation}\label{Dmappus}
\mbox{$D$-map} \quad : \quad    \underbrace{V(\phi)}_{\mbox{potential}} \, \mapsto \, \underbrace{\mathcal{M}_{J}}_{\mbox{K\"ahler manifold}}
\end{equation}
In the image of the $D$-map we find a class of manifolds that happen to be non algebraic deformations of the Poincar\'e, Klein, Beltrami, hyperboloid  model $\mathcal{M}_{PKB}$ \cite{PKB} of  Lobachevsky  geometry which we reviewed above, the pre-image of the latter manifold being either the pure exponential function $V_{PBK}(\phi) \, = \, g^2 \, \exp\left [\frac{\phi}{q}\right]$ or certain powers of the $Cosh\left[ \alpha \, \phi \right]$ function.  As we recalled, the classical manifold $\mathcal{M}_{PKB}$ is a quadratic algebraic surface in $\mbox{Mink}_{1,2}$, namely $\mathbb{R}^3$ endowed with the $\left(-,+,+\right)$ Lorentz metric.  Other positive definite potentials $V(\phi)$ map to manifolds that are non algebraic surfaces in the same space.
\par
Given any positive definite function $V(\phi)$ we identify its square root with the momentum map :
\begin{equation}\label{giocone}
   \mathcal{P}(\phi)\, \equiv \,\pm \,  \sqrt{V(\phi)}
\end{equation}
of the holomorphic Killing vector of the following axial symmetric real metric written in the canonical gauge:
\begin{equation}\label{metricozza}
    ds^2_{\mbox{K\"ahler}} \, = \, \ft 14 \, \mathrm{d}\phi^2 \, + \, \ft 14 \,\left(\mathcal{P}^\prime(\phi)\right)^2 \, dB^2
\end{equation}
As we explained in  section \ref{axialsurf} there are two distinct paths that lead to the interpretation of (\ref{metricozza}) as the K\"ahler metric of a suitable one dimensional complex K\"ahler manifold. In the first interpretation, namely the disk one, the isometry associated with the momentum map function $\mathcal{P}(\phi)$  is a compact rotation, while in the second interpretation, the plane-one, such isometry is a non-compact translation. Correspondingly we introduce the following definitions.
\begin{definizione}
Given a positive definite potential function $V(\phi)$ we name \textbf{Disk type $D$-map} of the latter the axial symmetric K\"ahler manifold whose K\"ahler metric has the form (\ref{metricozza}), where the square root of the potential function $\sqrt{V(\phi)} \, \equiv \, \mathcal{P}(\phi)$ is interpreted as the momentum-map of a \textbf{compact isometry} $B\, \rightarrow \, B+c$. The K\"ahler potential of such a manifold depends only on the modulus $\rho(\phi)$ of the complex coordinate $\zeta \, = \, \rho \, \exp[ {\rm i} B]$ and, as a consequence of the assumed interpretation, we have:
\begin{description}
  \item[a)] Modulus of $\zeta$
  \begin{equation}\label{disko1}
    \rho(\phi) \, \equiv \, \left| \zeta \right| \, = \, \exp\left [C(\phi)\right] \quad ; \quad C(\phi) \, \equiv \, \int \, \frac{d\phi}{\mathcal{P}^\prime(\phi)} \, = \, \int \, \frac{d\phi}{\partial_\phi \,\sqrt{V(\phi)}}
  \end{equation}
  \item[b)] K\"ahler potential
  \begin{eqnarray}
    \mathcal{K}(\zeta,\zeta)&=& J(C) \, = \, J(\rho) \, = \, J(C(\phi)) \, = \, J(\phi) \, = \, \int \, \frac{\mathcal{P}(\phi)}{\mathcal{P}^\prime(\phi)} \, d\phi \, = \,2 \, \int \, \frac{V(\phi)}{V^\prime(\phi)} \, d\phi \label{celerus1}
  \end{eqnarray}
\end{description}
\end{definizione}
Alternatively we have:
\begin{definizione}
Given a positive definite potential function $V(\phi)$ we name \textbf{Plane-type $D$-map} of the latter the K\"ahler manifold whose K\"ahler metric has the form (\ref{metricozza}), where the square root of the potential function $\sqrt{V(\phi)} \, \equiv \, \mathcal{P}(\phi)$ is interpreted as the momentum-map of a \textbf{non compact isometry} $B\, \rightarrow \, B+c$. The K\"ahler potential of such a manifold depends only on the imaginary part $C(\phi)$ of the complex coordinate $t \, = \, {\rm i} \, C \, + \, B$ and, as a consequence, of the assumed interpretation we have:
\begin{description}
  \item[a)] Imaginary part of $t$
  \begin{equation}\label{plane1}
    \mathrm{Im} \, t(\phi) \, \equiv\,  C(\phi) \, \equiv \, \int \, \frac{d\phi}{\mathcal{P}^\prime(\phi)} \,  = \, \int \, \frac{d\phi}{\partial_\phi\sqrt{V(\phi)}} \,
  \end{equation}
  \item[b)] K\"ahler potential
  \begin{eqnarray}
    \mathcal{K}(t,t)&=& J(\mathrm{Im} \, t) \, = \, J(C) \, = \,  J(C(\phi)) \, = \, J(\phi) \, = \, \int \, \frac{\mathcal{P}(\phi)}{\mathcal{P}^\prime(\phi)} \, d\phi \, = \,2 \, \int \, \frac{V(\phi)}{V^\prime(\phi)} \, d\phi \label{celerus2}
  \end{eqnarray}
\end{description}
\end{definizione}
\subsection{Curvature of the K\"ahler manifolds in the image of the $D$-map}
In order to grasp the nature of the K\"ahler manifolds in the image of the $D$-map, the first and most fundamental step is that of calculating their intrinsic curvature. To this purpose we introduce the vielbein formalism.
We name $E^0 \, , \, E^1$ the zweibein and we  set:
\begin{eqnarray}
  E^1 &=& \ft 12 \, \mathrm{d}\phi \nonumber \\
  E^2 &=&  \ft 12 \, \mathcal{P}^\prime(\phi) \, \mathrm{d}B \label{zweibein}
\end{eqnarray}
Accordingly the Levi-Civita  spin connection is calculated by means of the standard structural equations:
\begin{eqnarray}
  \mathrm{d}E^1 \, + \, \omega \, \wedge \, E^2&=& 0 \\
  \mathrm{d}E^2 \, - \, \omega \, \wedge \, E^1 &=& 0 \label{levicivspincon}
\end{eqnarray}
Eq.s (\ref{levicivspincon}) are immediately solved by
\begin{equation}\label{cristubulu}
    \omega \, = \, - \, \mathcal{P}^{\prime\prime}(\phi) \, \mathrm{d}B
\end{equation}
From this result it follows the curvature two form:
\begin{eqnarray}
    \mathfrak{R} & \equiv & \mathrm{d}\omega \, \equiv \, R(\phi) \, E^1 \, \wedge \, E^2 \nonumber\\
    R(\phi) & = & - 4 \, \frac{\mathcal{P}^{\prime\prime\prime}(\phi)}{\mathcal{P}^{\prime}(\phi)}\label{garducci}
\end{eqnarray}
In terms of the scalar potential $V(\phi)$ we obtain the following relation:
\begin{equation}\label{giunone}
     R(\phi) \, = \, -\, 4 \, \left ( \frac{V^{\prime\prime\prime}}{V^{\prime}} \, - \, \ft 32 \, \frac{V^{\prime\prime}}{V} \, - \, \ft  34 \, \left( \frac{V^\prime}{V}\right)^2 \right)
\end{equation}
The underlying K\"ahler manifold is a homogeneous space $\frac{\mathrm{SU(1,1)}}{\mathrm{U(1)}}$ only when the curvature $R(\phi)$ turns out to be a constant. This can happen also with non trivial potentials $V(\phi)$ which therefore simply emerge from an astute change of coordinates.
\par
From eq.(\ref{garducci}) we easily derive the condition on the potential $V(\phi)$ in order to get constant curvature. Setting:
\begin{equation}\label{flatcondizia}
    R(\phi) \, = \, - \, 4 \, \omega \quad ; \quad Q(\phi) \, \equiv \, \mathcal{P}^{\prime}(\phi)
\end{equation}
we immediately derive the differential equation:
\begin{equation}\label{barlucco}
    Q^{\prime\prime}(\phi) \, - \, \omega \, Q(\phi) \, = \,0
\end{equation}
whose general integral leads to the following form of the potential corresponding to constant curvature K\"ahler manifolds:
\begin{equation}\label{pastrone}
    V(\phi) \, = \, \left( \alpha \, \cosh \left[\sqrt{\omega} \,\phi\right] \, + \, \frac{\beta}{\sqrt{\omega}} \, \sinh \left[\sqrt{\omega} \,\phi\right] \, + \, \gamma\right)^2 \quad ; \quad \alpha,\, \beta,\,\gamma \, \in \, \mathbb{R}
\end{equation}
When $R(\phi)$ is not constant the manifold is necessarily non homogeneous and one can try to study its properties from the behavior of its curvature applying case by case standard techniques of differential geometry. Understanding the intrinsic geometric properties of the K\"ahler manifold associated with each considered inflaton potential is the preliminary and main step in order to understand its possible dynamical origin in string theory.
\section{Supergravity realization of the $D$-map}
\label{sugraDmap}
Our starting point is a generic theory of $\mathcal{N}=1$ supergravity with $n+1$ Wess-Zumino multiplets and $m$ vector multiplets.
The complex scalar fields forming the bosonic sector of the Wess-Zumino multiplets are named $\mathfrak{z}^\alpha$ ($\alpha \, = \, 1,\dots,n+1$), while the gauge fields, forming the bosonic sector of the vector multiplets are named $A^\Lambda_\mu$.
According to classical results  \cite{standardN1} reviewed and rederived, within the rheonomic framework in \cite{primosashapietro},
the general form of the scalar potential of gauged $\mathcal{N}=1$ supergravity takes the following form:
\begin{eqnarray}
  V &=& \underbrace{\ft 14 \, g_{\alpha\beta^\star} \, \mathcal{H}^\alpha \,\mathcal{H}^{\beta^\star}}_{\mbox{chiralinos contr.}} \, - \, \underbrace{3 \, S\, S^\star}_{\mbox{gravitino contr.}} \, + \,  \underbrace{\ft 13 \, \left(\mbox{Im} \,\mathcal{N}^{-1}\right)_{\Lambda\Sigma}\, D^\Lambda \, D^\Sigma }_{\mbox{gaugino contr.}} \nonumber \\
   &=& e^2 \exp\left[\widehat{\mathcal{K}}( \mathfrak{z}, \bar{\mathfrak{z}})\right]\left(g^{\alpha\beta^\star} \, \mathcal{D}_\alpha\,W(\mathfrak{z}) \, \mathcal{D}_{\beta^\star} \overline{W}(\bar{\mathfrak{z}})
   \, -\, 3\left| W(\mathfrak{z}) \right|^2 \right) \, +\,
   \, \ft 13 \, g^2 \, \delta_{\Lambda\Sigma}\, \mathcal{P}^\Lambda( \mathfrak{z}, \bar{\mathfrak{z}}) \, \mathcal{P}^\Sigma ( \mathfrak{z}, \bar{\mathfrak{z}}) \label{masterformulone}
\end{eqnarray}
where the origin of the various contributions from the different auxiliary fields of the theory is mentioned explicitly. There are two independent coupling constants in the above scalar potential that correspond to switching on and off the two types of available gaugings.
The coupling constant $e$ relates to the $F$-part of the gauging and can be considered as an overall constant in front of the superpotential $W(\mathfrak{z})$:
\begin{equation}\label{superpottolone}
  \widehat{W}(\mathfrak{z}) \, = \, e \, {W}(\mathfrak{z})
\end{equation}
which, by definition, is a holomorphic local function of the $n+1$ complex scalar fields $\mathfrak{z}^\alpha$ (actually a section of the Hodge bundle over the underlying K\"ahler manifold). The second independent coupling constant is $g$ which corresponds to the gauging of isometries of the K\"ahler manifold. In the kinetic part of the bosonic lagrangian $g$ appears in the covariant derivative of the scalar fields:
\begin{eqnarray}\label{gothicozeta}
  &&\mathcal{L}_{kinetic}^{SUGRA} \, \supset \, 2 \, g_{\alpha\beta^\star} \, \nabla_\mu \mathfrak{z}^\alpha \,  \nabla_\nu \bar{ \mathfrak{z}}^{\beta^\star} \, g^{\mu\nu} \nonumber\\
  &&\nabla_\mu \mathfrak{z}^\alpha \, \equiv \, \partial_\mu \, \mathfrak{z}^\alpha \, + \,g \, A^\Lambda_\mu \, k^\alpha_\Lambda(\mathfrak{z})
\end{eqnarray}
where $k^\alpha_\Lambda(\mathfrak{z})$ are the holomorphic killing vectors. In the scalar potential $g$ appears in front of the positive definite term obtained by squaring the momentum maps $\mathcal{P}^\Lambda(\mathfrak{z},\bar{\mathfrak{z}})$ of the same Killing vectors.
\par
 We consider the case where the K\"ahler manifold has the following direct product structure:
\begin{equation}\label{direttoprodo}
    \mathcal{M}_{\mbox{K\"ahler}} \, = \, \mathcal{M}_{J} \, \otimes \, \mathcal{M}_{\mathcal{K}}
\end{equation}
the submanifold $\mathcal{M}_{J} $ being an axial symmetric K\"ahler manifold with complex dimension one while the manifold $\mathcal{M}_{\mathcal{K}}$ has complex dimension $n$.
\par
According to the discussion of previous sections for the axial symmetric $\mathcal{M}_J$  we can use either the disk or the plane integration of the complex structure equations, leading to either one of the following relation between the flat coordinate $C$ and
the utilized complex coordinate:
\begin{equation}\label{guberno}
  \left \{ \begin{array}{cccl}
             \zeta  & = &  \exp[C] \, \exp[{\rm i}B] & \mbox{disk} \\
             t & = & {\rm i} \, C \, - \, B & \mbox{plane}
           \end{array}\right.
\end{equation}
Axial symmetry of the K\"ahler manifold $\mathcal{M}_J$ implies that in both interpretations the group of translations $B\rightarrow B+c$ is a group of isometries for the K\"ahler metric. This  amounts to assume the following structure for the complete K\"ahler potential of the manifold (\ref{direttoprodo}).
\begin{equation}\label{fustacchio}
   \widehat{\mathcal{K}}\left( \mathfrak{z},\bar{ \mathfrak{z}}\right) \, = \, J \left(C\right) \, + \, \mathcal{K}(z,\bar{z})
\end{equation}
where $J \left(C\right)$ is any real  function of its argument while $\mathcal{K}(z,\bar{z})$  is a generic K\"ahler potential for the manifold $\mathcal{M}_{\mathcal{K}}$.
\par
The K\"ahler metric that defines the kinetic terms of the scalars in the lagrangian takes the form:
\begin{equation}\label{menopausa}
    ds^2_{\mbox{K\"ahler}} \, = \, \ft 14 \, J^{\prime\prime}\left(C\right) \left( dC^2 \, + dB^2\right) \, + \, g_{ij^\star}\, dz^i \, d\bar{z}^{j^\star}
\end{equation}
and the $G$-function (the logarithm of the  invariant norm of  a holomorphic section of the Hodge-bundle) is the following one:
\begin{equation}\label{ficino}
    G \, \equiv \,  \log \left|| W(\mathfrak{z}) \right||^2 \, = \,  J\left(C\right) \, + \, \mathcal{K}(z,z) \, + \, \log \left| W\left(\left\{\begin{array}{c}
              \zeta\\
              t
            \end{array}\right\},
    \,z\right)\right|^2
\end{equation}
The mechanism that allows to generate an inflaton potential is based on the gauging of the translation isometry $B \to B+ c$ with $c\in \mathbb{R}$. In order for this gauging to be feasible it is necessary that such an isometry of the K\"ahler metric extends to a \textit{bona fide} global symmetry of the entire supergravity lagrangian coupled to the $n+1$ Wess-Zumino multiplets. This happens if and only if the function $G$ is invariant under the translation $B\to B + c$. Such invariance implies that the holomorphic superpotential $W(\mathfrak{z})$
should depend on the field $\left\{\begin{array}{c}
              \zeta\\
              t
            \end{array}\right\}$ in the  following  uniquely permissible way (where $N\in \mathbb{R}$):
\begin{equation}\label{decuppiu}
  W(\mathfrak{z}) \, = \, \left \{\begin{array}{ccc}
                                    \zeta^N \, \mathcal{W}(z) & ; & \mbox{disk} \\
                                    \exp\left[-{\rm i} \, N \, t \right] \, \mathcal{W}(z) & ; & \mbox{plane}
                                  \end{array} \right.
    \end{equation}
Under these conditions the Wess-Zumino part of the complete scalar potential takes the following form:
\begin{equation}\label{gozzo}
    V_{WZ} \, = \, \exp\left[\hat{J}\, \right] \,
    \left[\underbrace{\exp\left[\mathcal{K}\right]\left(g^{ij^\star} \, \mathcal{D}_i\,\mathcal{W} \, D_{j^\star} \overline{\mathcal{W}}\, -\, 3\left| \mathcal{W} \right|^2 \right)}_{\mbox{potential of the $n$ multiplets}}\, +\,
    \exp\left[ \mathcal{K}\right] \frac{(\hat{J}^\prime)^2}{\hat{J}^{\prime\prime} } \, \left|\mathcal{W}\right|^2 \right]
\end{equation}
where we have redefined the K\"ahler potential of the $\mathcal{M}_J$ factor by reabsorbing the contribution of  either $\zeta$ or $t$ to the superpotential, namely we have set:
\begin{equation}\label{riassorbo}
  \hat{J}(C) \, \equiv \, J(C) \, + \, 2 \, N \, C \, = \, \left \{\begin{array}{lcl}
                                                                     J\left(\log \, |\zeta|\right) \, + \,  \log \left| \zeta^N \right|^2 & ; & \mbox{disk}\\
                                                                      J\left(\mathrm{Im} \, t\right) \, + \, \log \left|\exp\left[ - \, {\rm i} \, N \, t\right] \right|^2 & ;& \mbox{plane}
                                                                   \end{array}
   \right.
\end{equation}
In addition, the gauging of the translation symmetry $B \, \rightarrow \, B+c$ produces a contribution to the full potential
\begin{equation}\label{fullopotto}
    V \, = \, V_{WZ}  \, + \, V_{YM}
\end{equation}
of the $D$-type, that according to the general formula (\ref{masterformulone}) has the following structure:
\begin{equation}\label{additional}
    V_{YM} \, = \, g^2  \,  \mathcal{P}^2
\end{equation}
the function $\mathcal{P}$ being the momentum map of the Killing vector that generates the aforementioned translation isometry. According to the discussions presented in the previous mathematical sections the momentum map $\mathcal{P}$ is  in both cases:
\begin{eqnarray}\label{fanciulla}
    \mathcal{P} & = & \frac{\mathrm{d}\hat{J}}{\mathrm{d}C} \, \equiv \,\hat{J}^\prime \nonumber\\
    V_{YM} & = & g^2 \, \left(\hat{J}^\prime(C) \right)^2
\end{eqnarray}
The complete potential (\ref{fullopotto}) can be reduced to a function of the single field $C$ if the other moduli fields
$z^i$ can be stabilized in a $C$-independent way. A very simple calculation, shows that this occurs generically if the superpotential $\mathcal{W}(z)$ has a critical point, namely if a set of values $z_0^i$ does exist such that:
\begin{equation}
 \left.   \mathcal{D}_i \,\mathcal{W} \right|_{z^i=z^i_0} \, = \, 0 \quad ; \quad \mathcal{W} \left (  z_0 \right ) \, = \, \mathcal{W}_0 \, \in \, \mathbb{C} \label{modstab}
\end{equation}
Indeed taking the derivative of the full $WZ$-part (\ref{gozzo}) of the full potential (\ref{fullopotto}) with respect to the fields $z^i$ and using covariance in the Hodge bundle we immediately see that all terms of such a derivative are proportional to $\mathcal{D}_i \mathcal{W}$ or to its complex conjugate, so that a critical point of the superpotential is always an extremum of the full potential.
After stabilization the residual one-field potential reduces to\footnote{The modified form of the potential due to the insertion of an $F$-term that appears in eq.(\ref{VdiC}) is identical with that  presented earlier in eq.(5.26) of \cite{Ferrara:2013kca}.} :
\begin{eqnarray}\label{VdiC}
    V(C) & = & g^2 \, \times  \, \underbrace{\left({ \hat{J}}^\prime(C) \right)^2}_{D-\mbox{potential}} \, + \, \underbrace{e^2 \, e^{K_0} \,\left |\mathcal{W}_0\right |^2 \,}_{\bar{e}^2} \times \, \underbrace{\exp \left [ \hat{J}(C) \right] \, \left( \frac{\left( \hat{J}^\prime\right)^2}{\hat{J}^{\prime\prime}} \, - \, 3 \right)}_{F-\mbox{potential}}\nonumber\\
    &=& g^2 \, \times  \, \left({ \hat{J}}^\prime(C) \right)^2 \, + \,\bar{e}^2  \, \exp \left [ \hat{J}(C) \right] \, \left( \frac{\left( \hat{ J}^\prime\right)^2}{\hat{J}^{\prime\prime}} \, - \, 3 \right)
\end{eqnarray}
which depends on two coupling constants $g$ due to gauging and $\bar{e}$ due to the original superpotential. The very important point to stress is that the effective coupling constant $\bar{e}$ is proportional to $\left |\mathcal{W}_0\right |^2$. If the other moduli are stabilized at a critical point of the superpotential $\mathcal{W}(z)$ where this latter vanishes, then the effective potential of the residual inflaton field is positive definite and it is purely of the $D$-type. In case the critical point of $\mathcal{W}(z)$ is not a zero of the same $\mathcal{W}(z_0)\, \ne \, 0$, then the inflaton potential has also an $F$-type part, whose structure is displayed in eq. (\ref{VdiC}).
\par
As we know, in terms of the flat coordinate $C$, the scalar kinetic term is not canonical. After neglecting the $B$ field that is eaten up by the vector field in the Higgs mechanism, the relevant scalar part of the Lagrangian reduces to :
\begin{equation}\label{guterlat}
    \mathcal{L}_{scalar} \, = \, 2 \, \left ( \ft 14  \, \hat{J}^{\prime\prime} (C) \, \partial^\mu C \, \partial_\mu C \, - \, V(C)\right)
\end{equation}
What remains to be done in order to make contact with cosmological inflaton models is just to choose the canonical gauge defined by equation (\ref{critallica2}). Using the field redefinition implied by such equation:
\begin{equation}\label{balucco}
    C \, = \, C(\phi)
\end{equation}
the kinetic term reduces to its canonical form $\ft 12 \partial^\mu \, \phi \, \partial_\mu \, \phi$, while the potential is read-off from the following substitution:
\begin{equation}\label{conglomerato}
    V(\phi) \, = \, V\left(C(\phi)\right)
\end{equation}


The intriguing question is how to work out the corresponding K\"ahler potential or, formulating it in a more intrinsic way, independently from the used coordinates, \textit{what is the geometry of the underlying one-dimensional K\"ahler manifold?}
\section{The $D$-map of some integrable cosmological potentials}
\label{IntegDmap}
In a recent paper \cite{noicosmoitegr} A. Sagnotti and the two of us have presented a bestiary  of one-field potentials that lead to integrable cosmological models, discussing also several properties of the ensuing exact solutions.
The bestiary consists of $10$ families\footnote{The infinite series $I_{10}$ appearing in Table \ref{tab:families}, had been overlooked in \cite{noicosmoitegr} and can be obtained from an appropriate analytic continuation of the series $I_9$.} of potentials depending on one or more parameters and of $28$ sporadic cases.
The $10$ families are recalled for the reader's convenience in Table \ref{tab:families}, while a subset of the $28$ sporadic potentials is displayed in Table \ref{Sporadic}.
\begin{table}[h!]
\centering
\begin{tabular}{|l|c|}
\hline
\null & Infinite families of Integrable Potentials \\
\hline
\hline
\null&\null\\
$I_1$ & $\! C_{11} \, e^{\,\vf} \, + \, 2\, C_{12} \, + \, C_{22} \, e^{\, - \vf}$  \\
\null&\null\\
\hline
\null&\null\\
$I_2$ & $\! C_1 \, e^{\,2\,\gamma \,\varphi}\, +\, C_2e^{\,(\gamma+1)\, \varphi} \ \ \ $  \\
\null&\null\\
\hline
\null&\null\\
$I_3$ & $\! C_1 \, e^{\, 2\, \varphi} \ + \ C_2$    \\
\null&\null\\
\hline
\null&\null\\
$I_4$ & $\! C_1 \, \varphi \,  e^{\, 2\, \varphi} $    \\
\null&\null\\
\hline
\null&\null\\
$I_5$ & $\! C_1 \, \log \left( \coth [\varphi]\right) \, + \,C_2 $    \\
\null&\null\\
\hline
\null&\null\\
$I_6$ & $\! C_1 \arctan\left(e^{-\,2\,\varphi}\right)\, + \, C_2 $    \\
\null&\null\\
\hline
\null&\null\\
$I_7$ & $\! C_1 \, \Big(\cosh\,\gamma\,\varphi \Big)^{\frac{2}{\gamma} \, - \, 2}\, + \, C_2 \Big( \sinh\,\gamma\,\varphi \Big)^{\frac{2}{\gamma} \, - \, 2}$  \\
\null&\null\\
\hline
\null&\null\\
$I_8$ & $\! C_1 \left(\cosh [2 \, \gamma \, \varphi] \right)^{\ft 1 \gamma -1}\,\cos\left[\left(\ft 1 \gamma -1\right)\, \arccos\left(\tanh[2\,\gamma\, \varphi]\,+\,C_2\right)\right]$ \\
\null&\null\\
\hline
\null&\null\\
$I_9$ & $\! C_1 \ e^{2\,\gamma\,\varphi} \  + \ C_2 \
e^{\frac{2}{\gamma}\,\varphi}\ \ $ \\
\null&\null\\
\hline
\null&\null\\
$I_{10}$ & $\! C_1 \ e^{2\,\gamma\,\varphi} \ \cos\left( 2\, \varphi \, \sqrt{1-\gamma^2} \ + \, C_2 \right) \ \ $ \\
\null&\null\\
\hline
\hline
\end{tabular}
\caption{The families of integrable potential classified in \cite{noicosmoitegr}. In all   cases  $C_i$ should be real parameters and $\gamma \in \mathbb{Q}$ should just be a rational number. Furthermore the field $\varphi$ is identified with a field $\phi$ admitting a canonical normalization in $D=4$ by means of the universal relation $\varphi \, = \, \sqrt{3} \,\phi$. The infinite series $I_{10}$ appearing above, had been overlooked in \cite{noicosmoitegr} and can be obtained from an appropriate analytic continuation of the series $I_9$. }
\label{tab:families}
\end{table}
\begin{table}[h!]
\centering
\begin{tabular}{|lc|}
\hline
\null & \null \\
\null & Sporadic Integrable Potentials \\
\null & \null \\
 \null & $\begin{array}{lcr}\mathcal{V}_{Ia}(\varphi)
   & = & \frac{\lambda}{4} \left[(a+b)
   \cosh\left(\frac{6}{5}\varphi\right)+(3 a-b)
   \cosh\left(\frac{2}{5}\varphi\right)\right] \end{array}$ \\
   \null & \null \\
   \null & $\begin{array}{lcr}\mathcal{V}_{Ib}(\varphi) & = & \frac{\lambda}{4} \left[(a+b)
   \sinh\left(\frac{6}{5}\varphi\right)-(3 a-b)
   \sinh\left(\frac{2}{5}\varphi\right)\right] \end{array} $\\
   \null & \null \\
   where & $ \left\{a,b\right\} \, = \, \left\{
\begin{array}{cc}
 1  & -3  \\
 1  & -\frac{1}{2} \\
 1  & -\frac{3}{16}
\end{array}
\right\} $\\
\null & \null \\
\hline
\null & \null \\
\null & $\begin{array}{lcr}
\mathcal{V}_{II}(\varphi)
   & = & \frac{\lambda}{8} \left[3 a+3 b- c+4( a- b)
   \cosh\left(\frac{2}{3}\varphi \right)+(a+b+c) \cosh\left(\frac{4}{3}\varphi \right)\right]\nonumber\ ,
\end{array}$\\
\null & \null \\
where  & $\left\{a,b,c\right\} \, = \, \left\{
\begin{array}{ccc}
 1  & 1  & -2
   \\
 1  & 1  & -6
   \\
 1  & 8  & -6
   \\
 1  & 16  & -12
    \\
 1  & \frac{1}{8} &
   -\frac{3}{4} \\
 1  & \frac{1}{16} &
   -\frac{3}{4}
\end{array}
\right\} $\\
\null & \null \\
\hline
\null & \null \\
\null & $\begin{array}{lcr}
\mathcal{V}_{IIIa}(\varphi) & = & \frac{\lambda}{16} \left[\left(1-\frac{1}{3
   \sqrt{3}}\right) e^{-6 \varphi
   /5}+\left(7+\frac{1}{\sqrt{3}}\right)
   e^{-2 \varphi
   /5} \right. \\
   && \left. +\left(7-\frac{1}{\sqrt{3}}\right)
   e^{2 \varphi /5}+\left(1+\frac{1}{3
   \sqrt{3}}\right) e^{6 \varphi
   /5}\right]\ .
   \end{array}$\\
\null&\null\\
\hline
\null & \null \\
\null &$\begin{array}{lcr}
  \mathcal{V}_{IIIb}(\varphi) &=& \frac{\lambda}{16} \left[\left(2-18
   \sqrt{3}\right) e^{-6 \varphi
   /5}+\left(6+30 \sqrt{3}\right) e^{-2
   \varphi /5}\right.\\
   &&\left. +\left(6-30
   \sqrt{3}\right) e^{2 \varphi
   /5}+\left(2+18 \sqrt{3}\right) e^{6
   \varphi /5}\right]
   \end{array}
 $ \\
 \null&\null\\
\hline
\end{tabular}
\caption{In this Table of the $28$ sporadic integrable potentials classified in \cite{noicosmoitegr} we have selected those that are pure linear combinations of exponentials. We plan to consider sporadic potentials more in depth elsewhere.}
\label{Sporadic}
\end{table}
It was pointed out that some of these models display phenomenologically attractive features, in some cases yielding a graceful exit from inflation. In a couple of separate publications  Sagnotti has also shown that the phenomenon of climbing scalars, displayed by all of the integrable models we were able to classify, has the potential ability to explain the  oscillations in the low angular momentum part of the CMB  spectrum, apparently observed by PLANCK. In his recent talk given at the Dubna SQS2013 workshop, our coauthor has also shown a best fit to the PLANCK data for the low $\ell$ part of the spectrum, by using the series of integrable potentials $I_2$\footnote{In comparing the following equation with the Table of paper \cite{noicosmoitegr}, please note the coefficient $\sqrt{3}$ appearing in the exponents that has been introduced to convert the unconventional normalization of the field $\varphi$ used there to the canonical normalization of the field $\phi$ used here.}
\begin{equation}\label{gammaserie}
    V(\phi) \, = \, a \, \exp\left[ 2\, \sqrt{3} \, \gamma \, \phi\right] + b \, \exp\left[  \sqrt{3} \, (\gamma +1)\, \phi\right]
\end{equation}
This best fit selects the particularly nice value $\gamma \, = \, -\ft 76$.
\par
In paper \cite{noicosmoitegr} we posed the question whether integrable potentials can be fitted into supergravity and in a forthcoming publication \cite{nointegrable2} we show that, although their type is very natural in gauged extended supergravities, the precise combinations implied by integrability are hard to be met. For instance, by means of the classification of all the gaugings of the  $STU$ model with $\mathcal{N}=2$ supersymmetry, in  \cite{nointegrable2} we are able to exclude the presence of integrable potentials in such an  environment. However, although hard, the task is not impossible and, within $\mathcal{N}=1$ supergravity, gauged by superpotentials of the type that appear in flux compactifications, we were able to single out a pair of integrable cases (they will be presented in \cite{nointegrable2}). According to the nomenclature put forward in the introduction we refer to this embedding of cosmological models in supergravity via the superpotential as to the \textit{F-type embedding}.
\par
On the other hand, adopting the point of view of the $D$-map,  arising  from the ideas put forward in \cite{minimalsergioKLP}, that was firstly advocated in \cite{primosashapietro} and was thoroughly discussed in the previous section \ref{Dmappus}, the embedding of integrable potentials into supergravity becomes feasible for all the cases where the potential is positive definite. In particular the condition of D-type embedding  is certainly fulfilled by the best fit case of eq.(\ref{gammaserie}) when $a >0,b>0$ and by almost all instances in our bestiary of integrable potentials.
\par
In this framework, the effort to understand the Physics underlying the emergence of such integrable potentials changes gear. Instead of looking for the mechanisms that determine suitable superpotentials, the focus is shifted on trying to understand the nature of the corresponding one-dimensional K\"ahler geometry. Along the road towards the solution of this problem the formula (\ref{giunone}) that relates the potential to the curvature  of the K\"ahler manifold constitutes a first illuminating step. This was alreday pointed out in \cite{primosashapietro}. Here according to the mathematical definition we provided in section \ref{mathDmap} we explore the properties of the K\"ahler manifolds in the $D$-map image of some of the series of integrable potentials listed in Tables \ref{tab:families} and \ref{Sporadic}.
\section{The $D$-map of the $I_1$ series}
The first series of integrable potentials $I_1$ is the simplest one and, as we showed in \cite{noicosmoitegr}, allows for the reduction of Friedman equations to linear equations in standard cosmic time. From the present point of view of the $D$-map, the distinct models encompassed in this series are actually two. The first model correspond to the choice $C_{11} \ne 0 , C_{22} \ne 0$ in which case by means of a shift of the variable $\phi$ we can always equalize the two coefficients and reduce the candidate momentum map to the following form:
\begin{equation}\label{garlando1}
\mathcal{P}(\phi) \, = \, \sqrt{\epsilon +\cosh \left(\sqrt{3} \phi \right)+1}
\end{equation}
The remaining coefficient $C_{12}$ has been written as $1+\epsilon$ because at $\epsilon \, = \, 0$ the $D$-image of the potential becomes a constant negative curvature manifold, namely  $\mathrm{SU(1,1)}/\mathrm{U(1)}$. For different values of $\epsilon$ we get a deformation of such a classical manifold, which is again recovered for a second value $\epsilon \, = \, 2$.
\par
The second model corresponds to the case where one of the two coefficients $C_{11}$, or $ C_{22}$ vanishes. Which one vanishes does not matter since we can always change the sign of $\phi$, if necessary, and reduce ourselves to $C_{22} \, = \,  0$. Furthermore the remaining coefficient $C_{11}$, if it does not vanish, can be  reduced to $1$ by a means of a shift in $\phi$ plus a rescaling of the entire potential.
The case where $C_{12} \, = \, C_{22} \, = \, 0$ corresponds to  pure exponential potential and hence to a constant curvature coset manifold $\mathrm{SL(2,R)}/\mathrm{O(2)}$.
Hence the only other non trivial model encompassed in series $I_1$ is provided by the following momentum-map
\begin{equation}\label{garlando2}
  \mathcal{P}(\phi) \, = \, \sqrt{1+e^{\sqrt{3} \phi }}
\end{equation}
\subsection{Analysis of the $D$-map of models $I_1A$}
We name $I_1A$ the K\"ahler manifolds defined by the momentum-map (\ref{garlando1}) and we proceed to a survey of their properties.
\par
The first important item in our constructions is the derivation of the flat coordinate, namely the integration of the complex-structure equation. By explicit evalutation of the integral in eq.(\ref{disko1}) or (\ref{plane1}), we obtain:
\begin{eqnarray}
  C(\phi) &=& \int \frac{d\phi}{\mathcal{P}^\prime(\phi)} \, = \, \int \,\frac{2 \sqrt{\mu +\cosh
   \left(\sqrt{3} \phi \right)}
   \mathrm{csch}\left(\sqrt{3}
   \phi \right)}{\sqrt{3}} \, d\phi \nonumber \\
  \null &=&\frac{1}{3}
   \left(\sqrt{\epsilon } \log
   \left(-\frac{2 \epsilon
   +\cosh \left(\sqrt{3} \phi
   \right)+2 \sqrt{\epsilon
   \left(\epsilon +\cosh
   \left(\sqrt{3} \phi
   \right)+1\right)}+1}{\cosh
   \left(\sqrt{3} \phi
   \right)+1}\right)\right.\nonumber\\
   &&\left.-\sqrt{\epsilon +2} \log
   \left(\frac{\sqrt{\frac{\epsilon +\cosh \left(\sqrt{3}
   \phi \right)+1}{\epsilon
   +2}}+1}{1-\sqrt{\frac{\epsilon +\cosh \left(\sqrt{3}
   \phi \right)+1}{\epsilon
   +2}}}\right)\right)
\end{eqnarray}
which is explicit in terms of elementary transcendental functions but cannot inverted since this requires the solution of transcendental equations. For the two special values of $\epsilon$ corresponding to constant curvature manifolds such inversion becomes instead possible.
\paragraph{Curvature.}
Applying the general formula (\ref{garducci}) for the curvature to the present case we obtain the following simple result:
\begin{equation}\label{coronello}
  R(\phi) \, = \, -\frac{9 \epsilon  (\epsilon
   +2)}{\left(\epsilon +\cosh  \left(\sqrt{3} \phi  \right)+1\right)^2}-3
\end{equation}
which confirms what we already anticipated, namely that for $\epsilon \, = \, \{0,-2\}$ we retrieve a constant curvature manifold.
Furthermore we see that for $\phi \to \pm\infty$ the curvature approaches the universal value $-3$, while at $\phi \, = \, 0$ it attains an $\epsilon$ dependent value.
\begin{figure}[!hbt]
\begin{center}
\iffigs
\includegraphics[height=70mm]{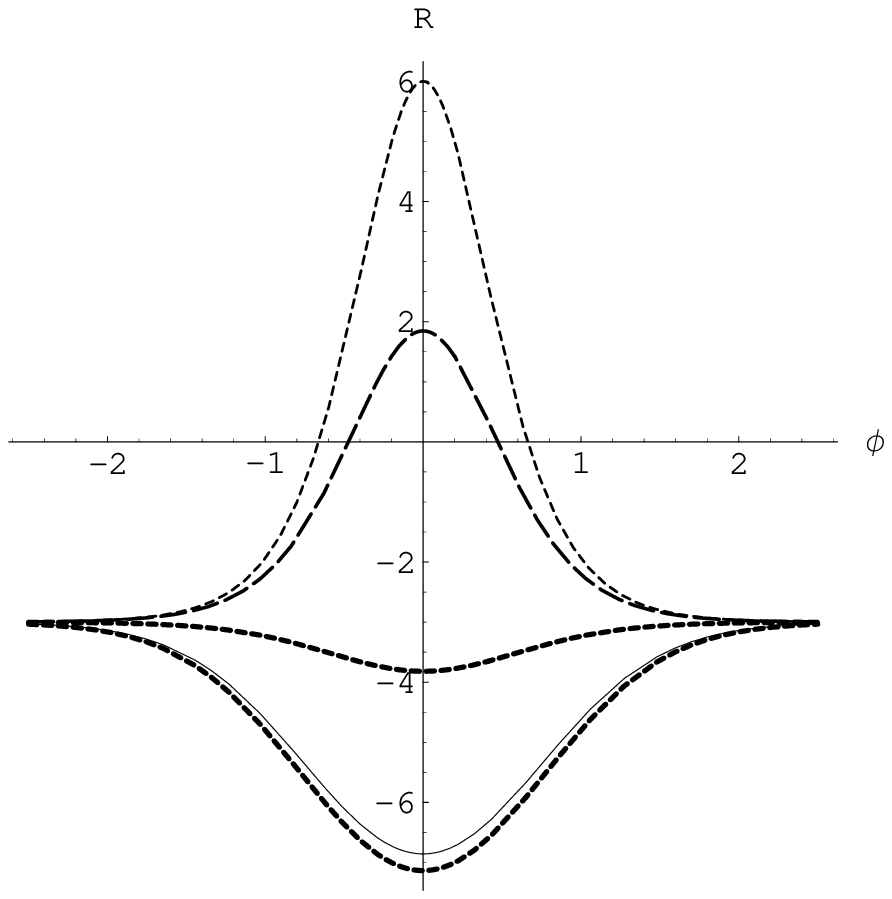}
 \includegraphics[height=70mm]{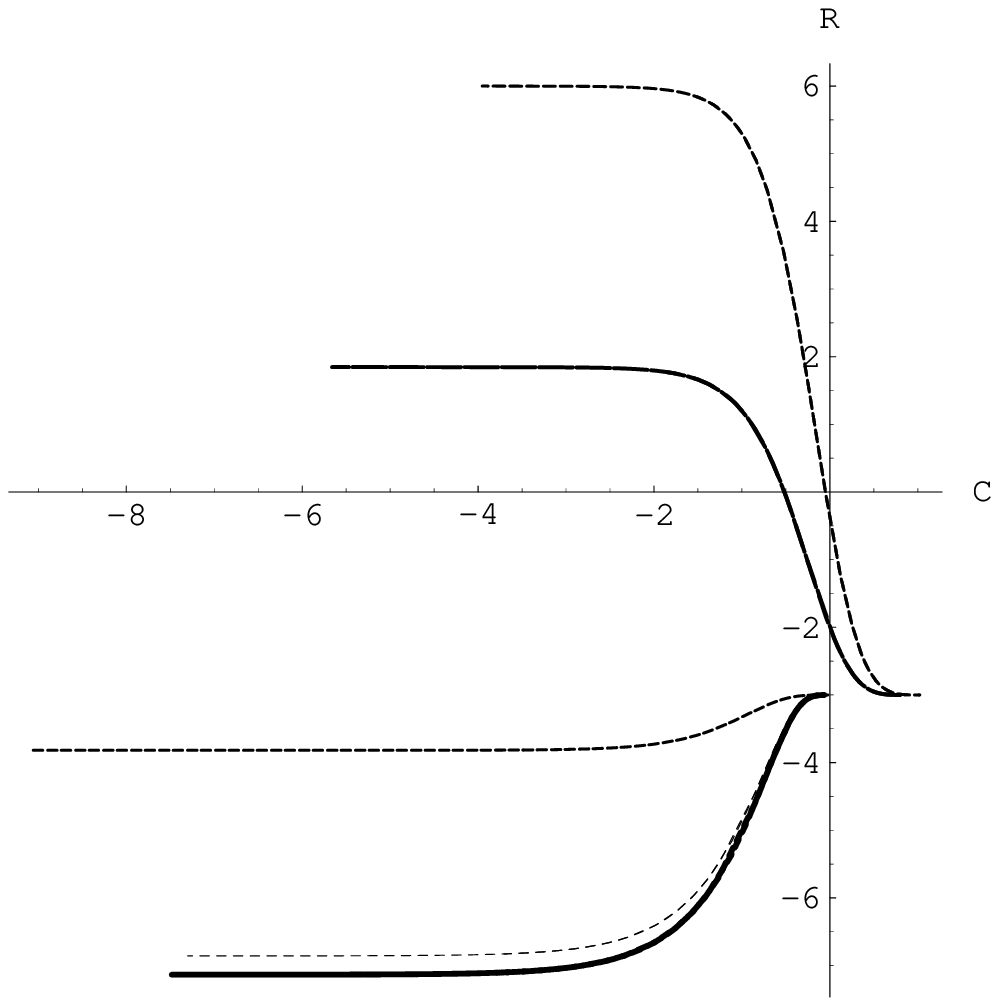}
\else
\end{center}
 \fi
\caption{\it This picture displays, for values values of the parameter $\epsilon$, the plots of the curvature for the K\"ahler manifolds in the image of the $D$-map of the integrable series $I_1A$. The figure on the left displays the plots of $R$ as a function of the original value $\phi$, while the figure on the right gives the plot of $R$ as a function of the flat coordinate $C$. As one sees the range of $C$ is from $C\, = \, -\infty$ that corresponds to $\phi = 0$ to $C=0$ that corresponds $\phi \, = \, \pm \infty$. The curvature is some $\epsilon$-dependent number for large negative $C$ while it is a universal $-3$ for $C\to 0$}
\label{CurveSerieIA}
 \iffigs
 \hskip 1cm \unitlength=1.1mm
 \end{center}
  \fi
\end{figure}
The plots of the curvature displayed in fig.\ref{CurveSerieIA} reveal, first of all, that the range of the flat coordinate $C$ is from $C\, = \, -\infty$ to $C=0$. This suggests that the correct interpretation of the $D$-map is the disk one, where $C \, \equiv \, \log |\zeta|$ has precisely such a range since $|\zeta | \, < \, 1$. Hence for $\epsilon$ different from the critical values the K\"ahler geometry defined by these model is a deformation of the classical K\"ahler geometry in the unit disk. Furthermore the analysis of fig.\ref{CurveSerieIA} shows that at small radii (for  $|\zeta | \,\ll \, 1 $) the curvature is almost constant but has a different value depending on $\epsilon$. On the other hand the traveller that  approaches the boundary of the manifold at $\zeta \simeq 1$, all of sudden feels a sharp change in the curvature that smoothly but very quickly reaches the universal value $-3$.
\par
For this model  we do not dwell on the power series reconstruction of the K\"ahler potential, as we plan to do in other cases, yet we can predict from the above considerations that it should be of the  form discussed later in eq.(\ref{seriosa}).
\subsection{Analysis of the $D$-map of models $I_1B$}
\label{serieUnoB}
We name $I_1B$ the K\"ahler manifolds defined by the momentum-map (\ref{garlando2}) whose properties we analyse in the sequel of this section.
\par
\begin{figure}[!hbt]
\begin{center}
\iffigs
\includegraphics[height=70mm]{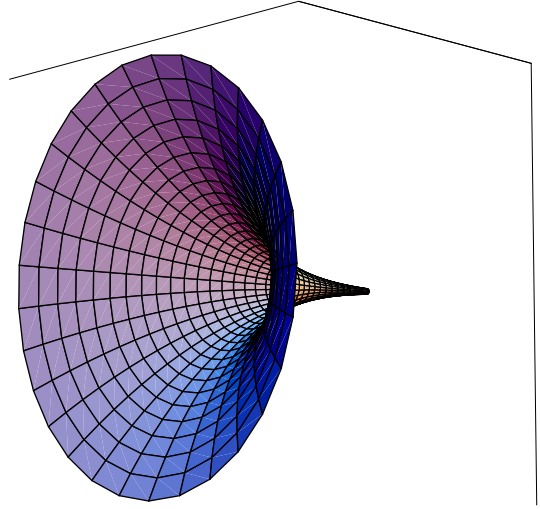}
\else
\end{center}
 \fi
\caption{\it This figure shows an intrinsic picture of the K\"ahler manifold in the $D$-map image of the integrable potential of series $I_1B$. The graph is obtained as surface of revolution of the plot of the function: $\frac{\sqrt{3} e^{\sqrt{3}\phi }}{2
   \sqrt{1+e^{\sqrt{3} \phi }}}$ that is the derivative of the momentum map in eq. (\ref{garlando2})  }
\label{trombettaExp}
 \iffigs
 \hskip 1cm \unitlength=1.1mm
 \end{center}
  \fi
\end{figure}
\begin{figure}[!hbt]
\begin{center}
\iffigs
\includegraphics[height=70mm]{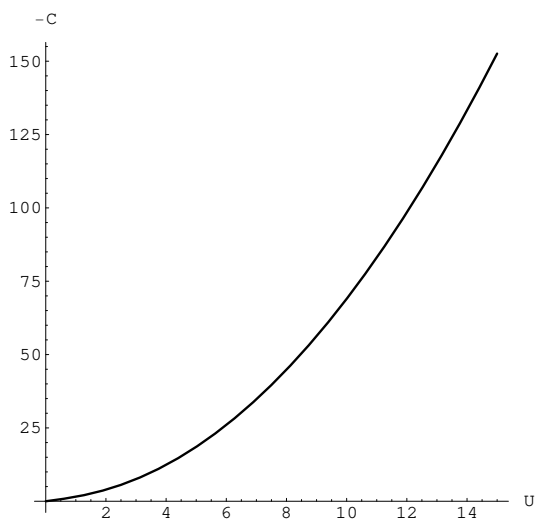}
\else
\end{center}
 \fi
\caption{\it Plot of the flat coordinate $-C$ versus the coordinate $U$  for the $D$-map image of the integrable potential $I_1B$.}
\label{palotto}
 \iffigs
 \hskip 1cm \unitlength=1.1mm
 \end{center}
  \fi
\end{figure}

An intrinsic picture of this manifold is displayed in fig.\ref{trombettaExp}.
Direct integration of the complex structure equation provides the form of the flat coordinate $C$ that is the following one:
\begin{equation}\label{furgonepiatto}
  C(\phi) \, = \,\frac{1}{3} \left(-\log \left(e^{-\sqrt{3} \phi }
   \left(2 \sqrt{1+e^{\sqrt{3}
   \phi }}+e^{\sqrt{3} \phi }+2\right)\right)-2 \, e^{-\sqrt{3} \phi } \, \sqrt{1+e^{\sqrt{3} \phi }}\right)
\end{equation}
It is convenient to introduce the following change of coordinate:
\begin{equation}\label{ciangio}
 \phi \,\to\, -\frac{2 \log (U)}{\sqrt{3}},\quad\quad \mbox{d$\phi$}\,\to \, -\frac{2
   \mbox{d$U$}}{\sqrt{3} U}
\end{equation}
which reduces the metric to the following form:
\begin{equation}\label{matrozzaIB}
  ds^2_{\mbox{K\"ahler IB}} \, = \, \frac{9\, \mbox{d$B$}^2\,+\,16\,\mbox{d$U$}^2 \,\left(U^2+1\right)}{48 \, U^2\,\left(U^2+1\right)}
\end{equation}
and the flat coordinate to:
\begin{equation}\label{gornaialavanda}
  C(U) \, = \, \frac{1}{3} \left(\, -\, 2\,
   \sqrt{U^2+1}\, U\, - \,\log \left(2 \,U \,\left(U+\sqrt{U^2+1}\right)
  \, +\, 1\right)\right)
\end{equation}
Calculating the curvature by means of equation (\ref{garducci}) and then substituting the coordinate $\phi$ with the coordinate $U$ we obtain:
\begin{equation}\label{siciliano}
 R(U) \, = \, -3-\frac{9}{\left(1+\frac{1}{U^2}\right)^2}
\end{equation}
The plot of the curvature with respect to the coordinate $U$ and to the flat coordinate $C$ is presented in fig.\ref{CurvatureIB}.
\begin{figure}[!hbt]
\begin{center}
\iffigs
\includegraphics[height=70mm]{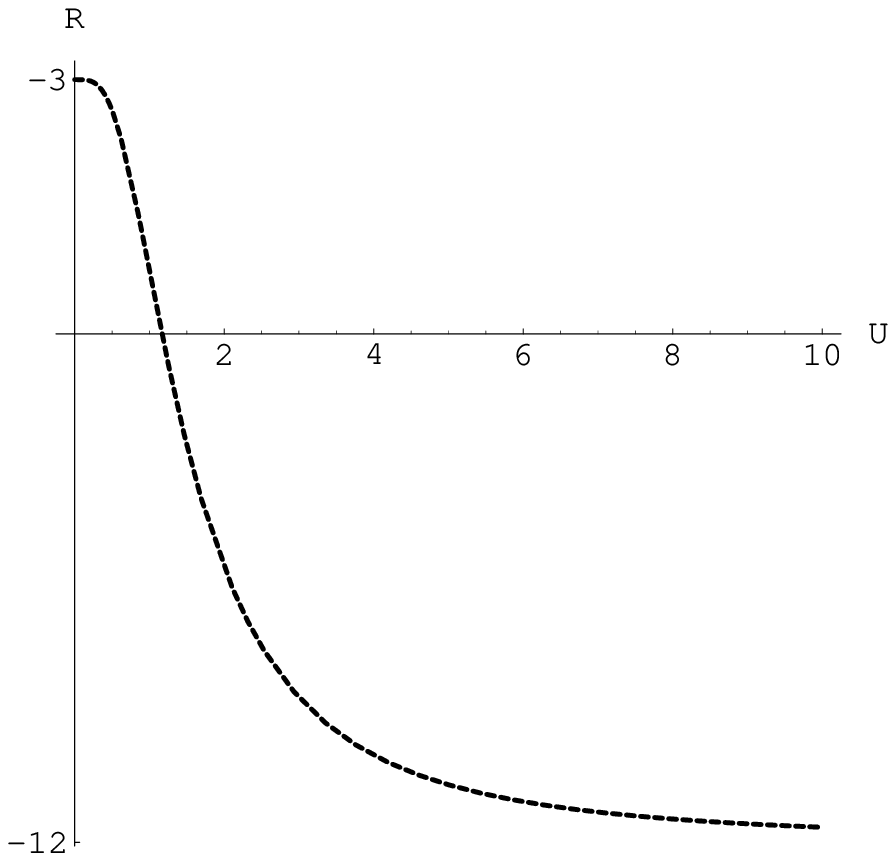}
\includegraphics[height=70mm]{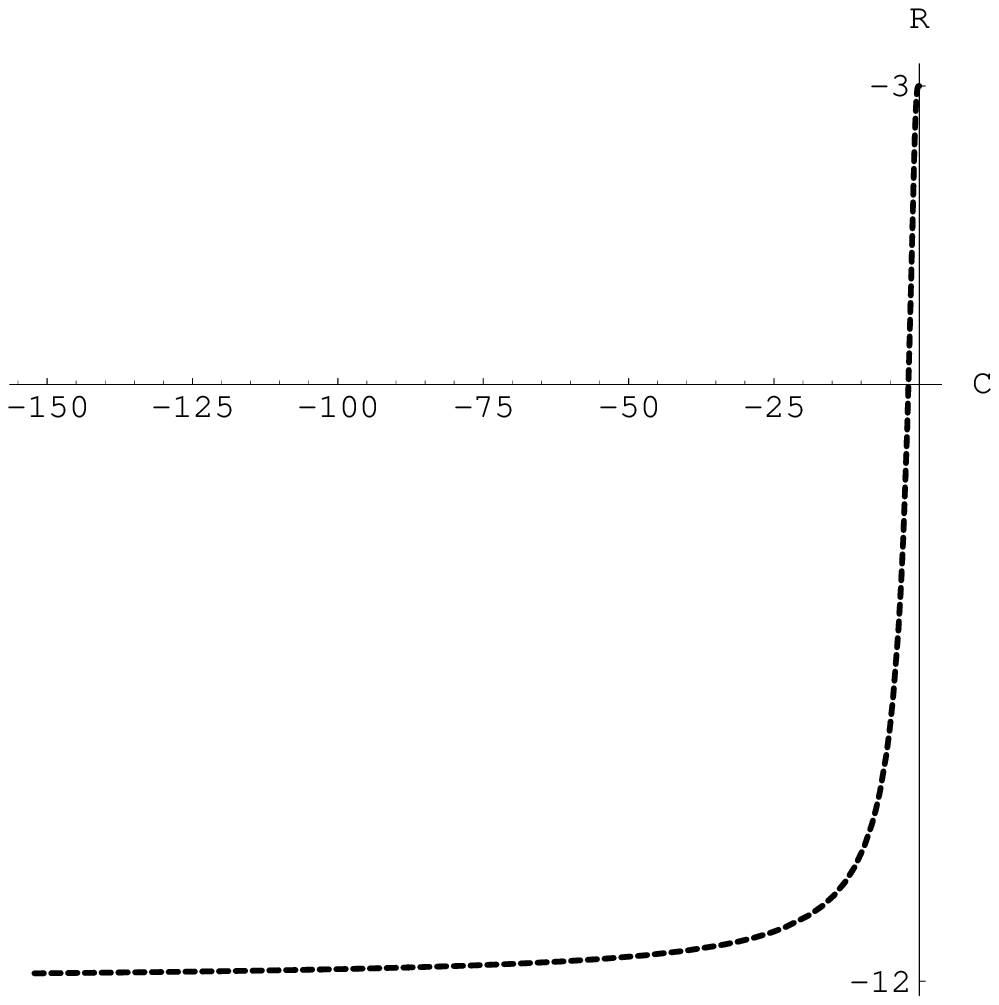}
\else
\end{center}
 \fi
\caption{\it This figure shows the plot of the curvature for the K\"ahler manifold in the $D$-map image of the integrable potential $I_1B$. The figure on the left shows the plot of $R$ with respect to the coordinate $U$, while that on the right shows the plot of the curvature with respect to the flat coordinate $C$.}
\label{CurvatureIB}
 \iffigs
 \hskip 1cm \unitlength=1.1mm
 \end{center}
  \fi
\end{figure}
\paragraph{K\"ahler potential.} Calculating the K\"ahler potential from eq.(\ref{celerus1}) or (\ref{celerus2}) we find:
\begin{equation}\label{zardo}
 J(\phi) \, = \, \frac{2 \, \phi }{\sqrt{3}}\, - \, \frac{2}{3} \, e^{-\sqrt{3} \phi }
\end{equation}
which transformed to the $U$-variable gives:
\begin{equation}\label{gecolona}
J(U) \, = \,  -\frac{2 \, U^2}{3}-\frac{4 \, \log (U)}{3}
\end{equation}
\begin{figure}[!hbt]
\begin{center}
\iffigs
\includegraphics[height=70mm]{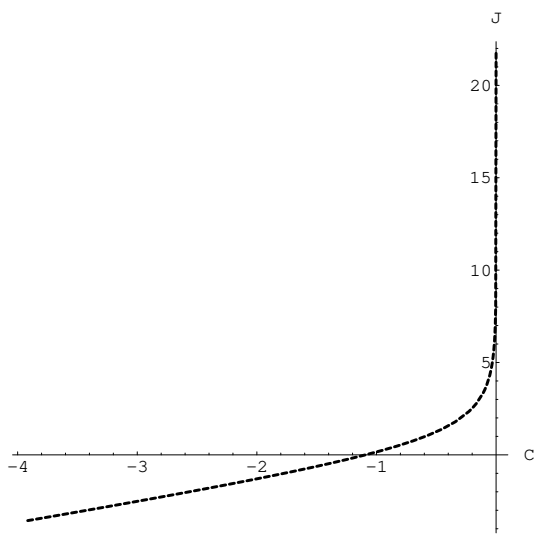}
\else
\end{center}
 \fi
\caption{\it This figure shows the plot of the K\"ahler potential for the K\"ahler manifold in the $D$-map image of the integrable potential $I_1B$. The K\"ahler potential is plotted against the flat coordinate $C$.}
\label{CurvatureIB}
 \iffigs
 \hskip 1cm \unitlength=1.1mm
 \end{center}
  \fi
\end{figure}
Next we try to determine the form of the exact K\"ahler potential according to the following series expansion:
\begin{equation}\label{gartoccio}
J(C)\, = \,   \beta +\alpha  \log (-\,C)+\sum _{i=1}^{\infty }(- C)^i p_i
\end{equation}
This development of the $J(C)$ function is appropriate to the plane interpretation of the $D$-map which is supported by the following three arguments:
\begin{enumerate}
  \item The range of the flat coordinate $ - \, C$ is from zero to infinity (upper complex plane) as it is evident from the plot in figure \ref{palotto}.
  \item The leading singularity of the K\"ahler potential $J(U)$ is $\log(U)$ in a variable $U$ in which the flat coordinate $C$ is analytic and admits a Taylor series development around $U=0$.
  \item The original model defined by eq.(\ref{garlando2}) for the momentum-map is a deformation:
   \begin{equation}\label{sinovio}
   \mathcal{P}_{\mathrm{undeformed}}(\phi) \, = \, \exp\left[ \frac{\sqrt{3}}{2} \, \phi\right] \, \equiv\, \sqrt{e^{\sqrt{3} \phi} } \quad \Rightarrow \quad  \mathcal{P}_{\mathrm{deformed}}(\phi) \, = \,  \sqrt{e^{\sqrt{3} \phi} \, + \, \underbrace{\lambda}_{\mbox{deform. parameter}} }
 \end{equation}
 of the model defined by $\mathcal{P} \, = \, \exp\left[ \frac{\sqrt{3}}{2} \, \phi\right]$ which corresponds to the standard presentation of the $\mathrm{SL(2,\mathbb{R}})/\mathrm{O(2)}$ coset by means of the Poincar\'e metric in the upper complex plane.
\end{enumerate}
The Taylor series development of the function $-C(U)$ is the following one:
\begin{eqnarray}\label{rapido}
  - \, C(U) & = & \frac{4 \, U}{3}+\frac{2
   \, U^3}{9}-\frac{\, U^5}{30}+\frac{\, U^7}{84}-\frac{5\, U^9}{864}
   +\frac{7\, U^{11}}{2112}-\frac{7\, U^{13}}{3328}+\frac{11\, U^{15}}{7680}+O\left(\, U^{16}\right)
\end{eqnarray}
Inserting eq.(\ref{rapido}) into eq.(\ref{gartoccio}) and comparing the series development in $U$ of the result with eq.(\ref{gecolona}) we obtain a triangular system of linear equations for the coefficients $p_i$ that can be solved at any desired order. The result is:
\begin{eqnarray}\label{furtaccio}
J(C) & = & \frac{4}{3} \mathrm{Log}\left[\frac{4}{3}\right]\, -\, \frac{4 }{3}\, \mathrm{Log}[-C] -\frac{C^2}{4}+\frac{39
  \, C^4}{1280}-\frac{87
  \, C^6}{8960}+\frac{98739
 \,  C^8}{22937600}\nonumber\\
   &&-\frac{285687
  \, C^{10}}{126156800}+\frac{19447413849
  \, C^{12}}{14694744064000}-\frac{244110
   0969
  \, C^{14}}{2938948812800} \, + \, \mathcal{O} \left(C^{16}\right)
\end{eqnarray}
As we see odd order coefficients vanish $p_{2\i +1} \, = \, 0$, while the even ones $p_{2\, i}$ are all rational numbers.
\par
\begin{figure}[!hbt]
\begin{center}
\iffigs
\includegraphics[height=70mm]{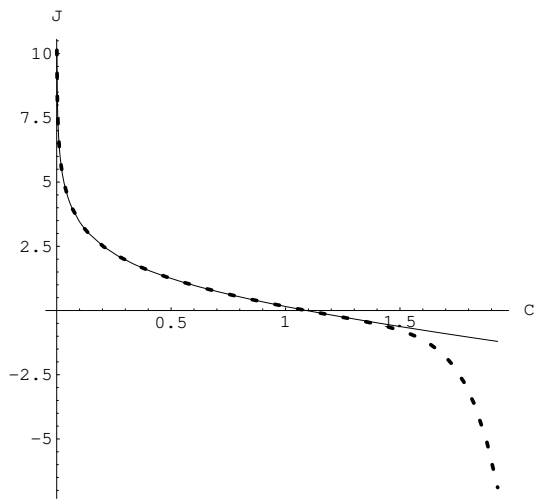}
\else
\end{center}
 \fi
\caption{\it This figure shows the plot of the K\"ahler potential for the K\"ahler manifold in the $D$-map image of the integrable potential $I_1B$. The K\"ahler potential is plotted against the flat coordinate $C$, both for the exact function and for the truncation to the 14th order of the series expansion. The solid line is the exact function, while the dashed line is the truncated series. As we see the approximation is excellent within the convergence radius that is approximately $C\, = \,\ft 32$. For larger values of $C$ the series diverges and another representation should be found. }
\label{paratto}
 \iffigs
 \hskip 1cm \unitlength=1.1mm
 \end{center}
  \fi
\end{figure}

In fig.\ref{paratto} the plot of the series development (\ref{furtaccio}) is compared with the graph of the true function $J(C)$: as one sees the convergence radius is approximate $\ft 32$.
\section{The $D$-map of the $I_7$  series of integrable potentials }
The integrable potentials of the series 7 in Table \ref{tab:families} are particularly  interesting since when we put either $C_1 = 0$ or $C_2 =0$ they take the form of a perfect square, as suggested by their momentum map interpretation. Here we consider the case $C_2 \, = \, 0$ for exemplification.
A second reason of interest in this series is provided by the fact that for two values of the index $\gamma$, specifically $\gamma \, = \, \ft 12$ and $\gamma \, = \, \ft 13$ the corresponding K\"ahler manifold in the image of the $D$-map has constant curvature and reduces to the homogeneous space $\frac{\mathrm{SU(1,1)}}{\mathrm{U(1)}}$. This fact offers a unique possibility of better understanding the mechanism of generation of a scalar potential by means of the gauging. As clearly exemplified by these cases the inverse of the $D$-map is not unique since  the potential depends not only on the K\"ahler metric but also  on the section of the Hodge bundle that is included in the definition of the K\"ahler potential.
\par
A third reason of interest in the $I_7$ series stems from the fact that for the special value $\gamma \, = \, \ft 15$, although the corresponding K\"ahler manifold is not of costant curvature, yet the flat coordinate  $C(\phi)$ defined by eq. (\ref{disko1}) can be inverted in terms of known special functions $\phi \, = \, \phi (C)$,  leading thus to an explicit analytic form of the K\"ahler potential
(\ref{celerus2}) in terms of its argument $C\, = \, \log |\zeta| $. This allows to study the generic form of these geometries that is encoded in the following series expansion:
\begin{equation}\label{seriosa}
    \mathcal{K}\left(\zeta , \, \zeta\right) \, = \, J(C) \,  = \,  \alpha \, C \, + \, \beta \, \sum_{k=1}^\infty \, c_{k} \, \exp\left[ 2\, k \, C\right] \quad ; \quad C \, \equiv \, \log |\zeta |
\end{equation}
When the underlying space is just $\frac{\mathrm{SU(1,1)}}{\mathrm{U(1)}}$, the explicit form of the K\"ahler potential (\ref{seriosa}) is indeed
\begin{equation}\label{funambolo}
  \mathcal{K}\left(\zeta , \, \zeta\right) \, = \,  \alpha \,  \log\,|\zeta| \, + \, \beta \, \log\,\left( 1 \, - \, |\zeta|^2\right)
\end{equation}
corresponding to the stereographic projection of the hyperboloid (see fig. \ref{hyperboloide} and eq.(\ref{PoincKalDisk})) and the coefficients
$c_k$ of the series expansion (\ref{seriosa}) are easily predicted:
\begin{equation}\label{cikappa}
  c_k \, = \, - \, \frac{1}{k}
\end{equation}
In the case $\gamma \, = \, \ft 15$ the coefficients $c_k$ can be explicitly calculated to any order. For other integrable potentials they can also be evaluated but only by inverting a triangular infinite system of linear equations.
\par
Let us then examine the general properties of the $D$-map of the $I_7$ integrable series.
\par
Using the appropriate conversion of normalizations we have:
\begin{eqnarray}
    V(\phi) &= &  \, \left( \mathcal{P}(\phi) \right)^2 \nonumber\\
    \mathcal{P}(\phi) & = & g \cosh
   ^{\frac{1}{\gamma
   }-1}\left(\sqrt{3} \gamma
   \phi \right) \label{goodwishes}
\end{eqnarray}
where $g$ is a coupling constant that we will normalize to convenient values depending on the case.
\paragraph{\bf The curvature.}
Using this information in the curvature formula eq.(\ref{garducci}) we find:
\begin{equation}\label{fluorescenza}
    R_{\gamma|\cosh}(\phi) \, = \, -12 \left((\gamma  (6 \gamma -5)+1)
   \tanh ^2\left(\sqrt{3} \gamma
   \phi \right)+(3-5 \gamma ) \gamma
   \right)
\end{equation}
\begin{figure}[!hbt]
\begin{center}
\iffigs
\includegraphics[height=70mm]{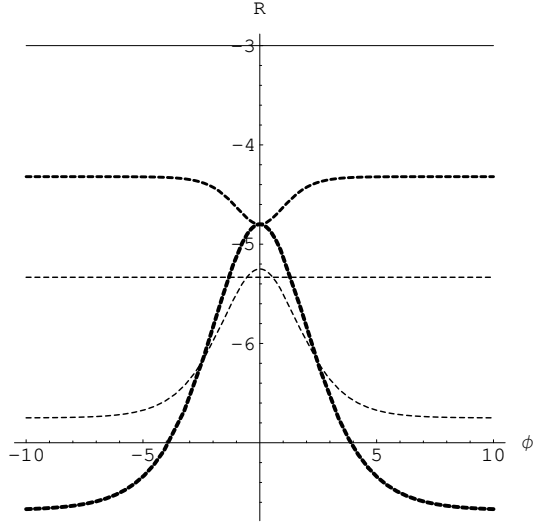}
\else
\end{center}
 \fi
\caption{\it Plots of the curvature for the K\"ahler manifolds in the $D$-map image of the $I_7$ series. The dashed thick line with a maximum corresponds to the case $\gamma \, =\, \ft 15$. The dashed thick line with a minimum corresponds to the case $\gamma \, = \, \ft 25$. The thinner dashed line with a maximum corresponds to $\gamma \, = \, \ft 14$. The thin  dashed straight line corresponds to $\gamma \, = \, \ft 13$. Finally the straight solid line corresponds to $\gamma \, = \, \ft 12$ }
\label{coshgammaCurv}
 \iffigs
 \hskip 1cm \unitlength=1.1mm
 \end{center}
  \fi
\end{figure}
The asymptotic value of the curvature is the same at $\phi\, = \, \pm \infty$ and we have:
\begin{equation}\label{fortepiano}
    R_{\gamma|\cosh}(\pm \infty) \, = \, -12 \, (1\, -\, \gamma)^2
\end{equation}
As we already mentioned a very intriguing fact is that for the two values $\gamma= \ft 12$ and $\gamma = \ft 13$ the dependence on $\phi$ of the curvature disappears. This means that in these two cases the K\"ahler manifold is just the homogeneous space $\mathrm{SU(1,1)}/\mathrm{U(1)} \, \sim \, \mathrm{\mathrm{SL(2,\mathbb{R})}}/\mathrm{O(2)} $ and that the potential is created by gauging some appropriate subgroup of $\mathrm{SL(2,\mathbb{R})}$.
 The plots of the K\"ahler curvature for some different values of $\gamma$ are displayed in fig.\ref{coshgammaCurv}.
\par
For all the  $\gamma$.s different from the two critical values the plots shows that also in this case two asymptotically homogeneous K\"ahler manifolds are smoothly connected.
\paragraph{\bf The $C$-function.}
A very nice property of this series is that equation (\ref{disko1}) can be integrated for all values of $\gamma$ and we obtain:
\begin{equation}\label{goodyear}
 C_\gamma(\phi) \, = \,    \frac{\cosh ^{\frac{\gamma -1}{\gamma
   }}\left(\sqrt{3} \gamma  \phi
   \right) \left(\,
   _2F_1\left(1,\frac{\gamma -1}{2
   \gamma };\frac{3}{2}-\frac{1}{2
   \gamma };\cosh ^2\left(\sqrt{3}
   \gamma  \phi
   \right)\right)-1\right)}{3 g
   (\gamma -1)^2}
\end{equation}
where $_2F_1\left(\dots\right)$ denotes an hypergeometric function of the specified arguments. The explicit result of the integration implies that the function $C_\gamma(\phi)$ should satisfy a second order linear differential equation whose second independent solution might be of relevance in searching for the origin of these geometries. Indeed performing the transformation:
\begin{equation}\label{goliardo}
    \phi \, = \, \frac{\mathrm{ArcCosh}(T)}{\sqrt{3} \gamma
   }
\end{equation}
we get
\begin{equation}\label{olograf}
    C_\gamma(T) \, = \, \frac{T^{\frac{\gamma -1}{\gamma }}
   \left(\, _2F_1\left(1,\frac{\gamma
   -1}{2 \gamma
   };\frac{3}{2}-\frac{1}{2 \gamma
   };T^2\right)-1\right)}{3 g (\gamma -1)^2}
\end{equation}
and the form taken by the hypergeometric equation satisfied by $C_\gamma(T) $ is the following one:
\begin{equation}\label{curiale1}
    \left (\, \left (2 T \,+ \, 3\right) \gamma \, -\, 1\right ) C_\gamma'(T)\, +\, 2 \,T\, \left(T\,+\,1\right)
   \, \gamma  \, C_\gamma''(T)\, \, = \, 0
\end{equation}
Since the function $C_\gamma(T)$ appears in (\ref{curiale1}) only under derivatives the effective degree of the equation is the first order and the second solution is just trivial being a constant. Hence up to an additive constant the function $C_\gamma(\phi)$ is uniquely defined as in eq.(\ref{goodyear})).
\paragraph{\bf The K\"ahler potential.} Also eq.(\ref{celerus2}) is easily integrated and we get:
\begin{eqnarray}\label{CalleroCosh}
  J(\phi) & = &   \frac{\log \left(\sinh\left(\sqrt{3} \gamma  \phi\right)\right)}{3 (1-\gamma) \gamma } \nonumber\\
    J(T)      & = & \frac{\log (T-1)+\log (T+1)}{6
   \gamma -6 \gamma ^2}
\end{eqnarray}

\begin{figure}[!hbt]
\begin{center}
\iffigs
\includegraphics[height=70mm]{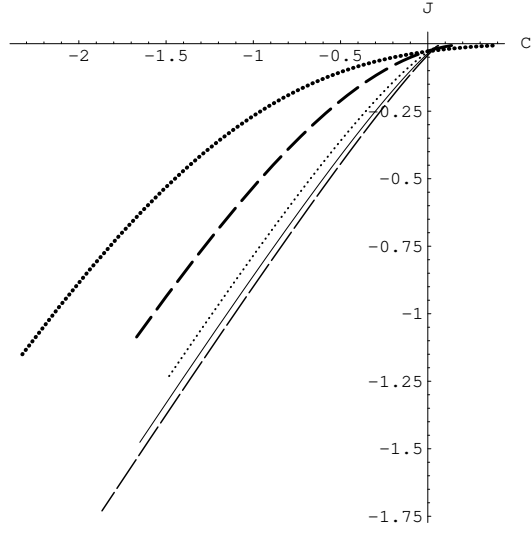}
\else
\end{center}
 \fi
\caption{\it In this picture  we see plots of the  the K\"ahler potential $J(C)$ with respect to the coordinate $C$ for $I_7$ series.
The line with big dots corresponds to the case $\gamma \, = \, \ft 25$. The dashed line with fat long dashes
corresponds to the case $\gamma \, = \, \ft 12$. The line with small thin dots corresponds to the case $\gamma \, = \, \ft 23$. The solid line corresponds to $\gamma \, = \, \ft 34$. Finally the line with very long thin dashes corresponds to
$\gamma \, = \, \ft 45$. }
\label{coshgammaKallo}
 \iffigs
 \hskip 1cm \unitlength=1.1mm
 \end{center}
  \fi
\end{figure}

The variable $T$ being defined by eq.(\ref{goliardo}). For generic $\gamma$ we cannot invert the function $C_\gamma(T)$ in order to obtain $T=T(C)$ to be substituted back into eq.(\ref{CalleroCosh}). Hence the analytic form of the K\"ahler potential is not reachable. Yet we can grasp the nature of these spaces by performing a parametric plot of the K\"ahler potential versus the $C(T)$ function. Such plots for various values of the parameter $ùgamma$ are displayed in fig.\ref{coshgammaKallo}.
As it clearly appears from such pictures, the K\"ahler potentials produced by the $D$-map of the integrable series $I_7$ are relatively small deformations of the logarithmic K\"ahler potential corresponding to the constant curvature Poincar\'e plane equivalent to the hyperboloid model. It is very instructive to consider in detail the two cases where the integrable potential leads exactly to standard hyperbolic geometry. From these examples we better understand the mechanism of deformation where we depart from pure constant curvature.
\subsection{The K\"ahler manifold in the image of the $\gamma \, = \, \ft 12$ potential}
\begin{figure}[!hbt]
\begin{center}
\iffigs
\includegraphics[height=70mm]{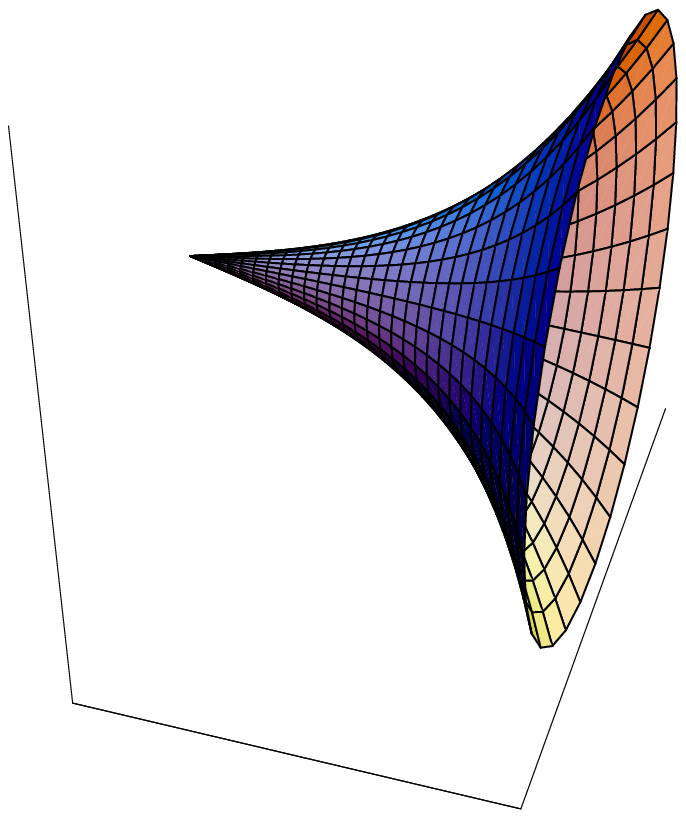}
\else
\end{center}
 \fi
\caption{\it Intrinsic view of the costant negative curvature surface whose metric is given in eq.(\ref{gospadin}). The surface is obtained
as surface of revolution of the plot of the function $\sinh\left[\frac{\sqrt{3} \, \phi}{2}\right]$ ($\phi >0$). This is the model of the K\"ahler manifold sitting in the image of the $D$-map for the $\gamma \, = \, \ft 12$ case of the $I_7$ integrable series. }
\label{trombola}
 \iffigs
 \hskip 1cm \unitlength=1.1mm
 \end{center}
  \fi
\end{figure}

In the case $\gamma \, = \, \ft 12$ the curvature of the K\"ahler manifold is constant equal to $-3$ and we expect to reconstruct the geometry of the coset manifold $\frac{\mathrm{SU(1,1)}}{\mathrm{U(1)}}$ in a particular parameterization.
The first signal of the specialty of this case arises from the fact that for this value of $\gamma$  the hypergeometric function
$\,_2F_1\left(1,-\frac{1}{2};\frac{1}{2};z^2\right)$ appearing in eq. (\ref{olograf}) is actually an elementary transcendental function $1 - z \,\mathrm{ArcTanh}[z]$.
It is therefore convenient to normalize the momentum map in a way that will be useful to simplify final expressions and to perform a separate direct integration in order to obtain the simplest form of the $C$-function. Specifically we set:
\begin{equation}\label{specPgam12}
  \mathcal{P}_{\ft 12} \, = \, -\frac{4}{3} \cosh \left(\frac{\sqrt{3} \phi}{2}\right) \quad ; \quad g \, = \, -\ft 34
\end{equation}
and from the integration formula (\ref{disko1}) we obtain:
\begin{equation}\label{garazde1}
    C_{\ft 12}(\phi) \, = \, \log \left(\coth \left(\frac{\sqrt{3}
   \phi }{4}\right)\right)
\end{equation}
The above relation can be easily inverted and we get:
\begin{equation}\label{goladno1}
    \phi(C) \, = \, \frac{4 \mathrm{ArcCoth}\left(e^C\right)}{\sqrt{3}}
\end{equation}
Inserting (\ref{goladno1}) in the formula (\ref{CalleroCosh}) for the K\"ahler potential (with $\gamma \, = \,\ft 12$) we get:
\begin{equation}\label{fascollo}
  J(C) \, = \,   \frac{4 }{3}\, C\, -\frac{4}{3} \log
   \left(1-e^{2 C}\right)\, + \,\ft 23 \,{\rm i} \,\pi\,+ \, c_0
\end{equation}
where $c_0$ is an arbitrary integration constant that we can use to dispose of the irrelevant constant imaginary contribution $\ft 23 \,{\rm i} \,\pi\,$ due the change of sign in the logarithm. Comparing eq. (\ref{fascollo}) with eq.(\ref{funambolo}) we see that they agree by setting $\alpha \, = \, \ft 43 \, = \, -\, \beta$  and that the  correct interpretation of the result is provided by the Disk-type $D$-map. Setting:
\begin{equation}\label{foratto}
    \zeta \, = \, \exp\left [ C\right ] \, \exp\left [ {\rm i} B\right] \quad ;
    \quad \bar{\zeta} \, = \, \exp\left [ C\right ] \,\exp\left [ {\rm i} B\right]
\end{equation}
we get the identification
\begin{eqnarray}\label{fortissimo1}
 J(C) \, = \,    G\left(\zeta \, , \, \bar{\zeta}\right) & \equiv & \log \left|| W\right||^2 \, = \, \mathcal{K}\left(\zeta \, , \, \bar{\zeta}\right) \, + \, \log \left| W\right|^2 \nonumber\\
    & = &  \log \,\left(\zeta^{\ft 23} \,\bar{\zeta}^{\ft 23}\right ) \, - \,
    \log \left( 1 \, - \,\left| \zeta\right|^2\right)^{\ft 43}
\end{eqnarray}
Hence the invariant function $G\left(\zeta \, , \, \bar{\zeta}\right)$ emerges from the standard K\"ahler potential yielding the metric of the hyperboloid in the Disk-projection (see eq.s(\ref{PoincKalDisk}) and (\ref{ringo}) (with an overall normalization $\ft 43$) and a superpotential $W(\zeta) \, = \, \zeta^{\ft 23}$. Indeed the form of the $\frac{\mathrm{SU(1,1)}}{\mathrm{U(1)}}$ invariant metric realized by the $D$-map of the $\gamma \, = \, \ft 12$ integrable $I_7$ potential happens to be that introduced in eq.(\ref{gospadin})  obtained from the stereographic projection of the hyperboloid (fig.\ref{hyperboloide}). An intrinsic view of this two-dimensional surface is provided in fig.\ref{trombola}.
\par
This reconstruction suggests that the case of the integrable potential under consideration is just an instance in the more general family that can be extracted inverting  the $D$-map procedure from the invariant function $G_N$ that follows:
 \begin{eqnarray}
    J_N(C) & = &    G_N\left(\zeta \, , \, \bar{\zeta}\right) \, \equiv \, \log \left|| W_N\right||^2  \nonumber\\
    & = &  \log \,\left(\zeta^{\ft {2\,N}{3} }\,\bar{\zeta}^{\ft {2\,N}{3}}\right ) \, - \,
    \log \left( 1 \, - \,\left| \zeta\right|^2\right)^{\ft 43}\nonumber\\
   &=& N \,\frac{4 }{3}\,C \,-\,\frac{4}{3} \log
   \left(1-e^{2 C}\right)\, + \,\ft 23 \, + \, e^2 \label{carasco}
 \end{eqnarray}
where we have deleted the imaginary constant $\ft 23 \,{\rm i} \,\pi$ and replaced $c_0$ with the parameter $e$. This latter is the coupling constant that switches on and off the $F$-gauging. At $e=0$ the scalar potential is only of $D$-type. At $g=0$ it is only of $F$-type. In general it contains both contributions.
\par
Following the reverse path, from the function $J_N(C)$ we obtain the momentum map, as function  of $C$:
\begin{equation}\label{gollum}
    \mathcal{P}_N(C) \, = \, \frac{\mathrm{d} J_N}{\mathrm{d}C} \, = \,- \,\left(\frac{1}{3} (-4 N+4 \coth
   (C)+4)\right)
\end{equation}
The field $\phi$ whose kinetic term is canonical is immediately retrieved ( as function of $C$) from the following calculation:
\begin{eqnarray}\label{corsettino}
    \phi(C) & = & \int \, \mathrm{d}C \, \sqrt{\frac{\mathrm{d}^2 J_N}{\mathrm{d}C^2}} \, = \, \int \, \mathrm{d}C \,\frac{2 \mathrm{csch}(C)}{\sqrt{3}} \nonumber\\
    & = & \frac{2 \log \left(\tanh
   \left(\frac{C}{2}\right)\right)}{\sqrt{3}}
\end{eqnarray}
This relation is easily inverted and we obtain:
\begin{equation}\label{curnabislo}
    C(\phi) \, = \,  2 \, \mathrm{ArcTanh} \, \left(e^{\frac{\sqrt{3} \phi}{2}}\right)
\end{equation}
which is just another way of writing eq.(\ref{goladno1}). Inserting this result into eq.(\ref{gollum}) we finally obtain the expression of the momentum map and of the $D$-part of the potential in terms of the field $\phi$:
\begin{equation}\label{YMpartus}
    \mathcal{P}_N(\phi) \, = \, \left(\frac{4}{3} \left(-N+\cosh
   \left(\frac{\sqrt{3} \phi
   }{2}\right)+1\right)\right) \quad ; \quad V_{YM} \, = \, g^2 \, \left(\mathcal{P}_N(\phi)\right)^2
\end{equation}
The integrable potential we originally considered corresponds to the choice $N=1$, but the entire family of potentials displayed in (\ref{YMpartus}) are all associated with the gauging of a compact $U(1)$ isometry of the standard Poincar\'e homogeneous K\"ahler manifold.
\par
\begin{figure}[!hbt]
\begin{center}
\iffigs
\includegraphics[height=70mm]{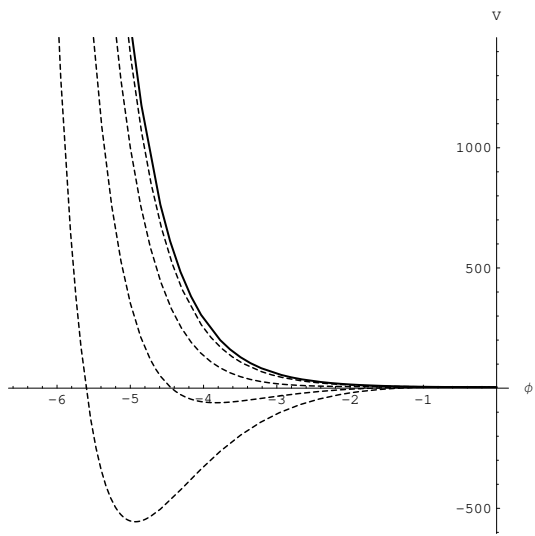}
\else
\end{center}
 \fi
\caption{\it Plots of the potentials obtained combining $D$-type and $F$-type of gaugings in the case of the standard $\mathrm{SU(1,1)/U(1)}$ K\"ahler manifold represented in the unit disk. The various curves correspond to different values of $\lambda \, = \, \frac{e}{g}$. The solid thick line corresponds to $\lambda \, = \, 0$, namely to the pure $D$-type, positive definite potential. As $\lambda$ increases the potential starts deforming to the right and after some critical value of $\lambda$ it becomes negative in a certain region where it develops a minimum. The  $\lambda$ values of the plotted curves are $\lambda \, = \, 4,3,2,1$. The deeper the minimum the highest the value of $\lambda$. }
\label{capziosa}
 \iffigs
 \hskip 1cm \unitlength=1.1mm
 \end{center}
  \fi
\end{figure}

Consider next the case $N=0$ and let us switch on also the $F$-gauging by letting $ e \, \ne \, 0$.
The $V_{WZ}$ part of the potential is obtained from eq.s (\ref{conglomerato},\ref{VdiC}) first calculating the necessary derivatives of $J_1(C)\,=\,G(C)$ and then substituting eq.(\ref{curnabislo}) in the resulting expression. The result is the following potential function:
\begin{equation}\label{acciderba}
  V_{WZ} \, = \,  e^2 \, \frac{\left(-9+e^{4\, \mathrm{ArcTanh} \,\left(e^{\frac{\sqrt{3} \phi
   }{2}}\right)}\right) }{3
   \sqrt[3]{\left(1-e^{4 \, \mathrm{ArcTanh} \,\left(e^{\frac{\sqrt{3} \phi
   }{2}}\right)}\right)^4}}
\end{equation}
Defining $\lambda \, = \, \frac{e}{g}$ the ratio of the $F$-coupling constant with respect to the $D$-one, we get (up to the an overall scale in front) a family of potentials whose plot is displayed in fig.\ref{capziosa} for some values of $\lambda$.

\subsection{The K\"ahler manifold in the image of the $\gamma \, = \, \ft 13$ potential}
As in the previous case, we choose a normalization of the potential and of the momentum map that will prove handy in subsequent calculations. We we set:
\begin{equation}\label{fuccino}
\mathcal{P}_{\ft 13} \, = \,  -\frac{3}{2} \cosh ^2\left(\frac{\phi }{\sqrt{3}}\right) \, = \, - \, \sqrt{V(\phi)}
\end{equation}
Correspondingly the metric of the space under consideration is the following one:
\begin{equation}\label{babbutmammut}
  ds^2_{\gamma \, = \, \ft 13} \, = \,\frac{1}{4} \left(\mathrm{d}\phi ^2
  +\frac{3}{4} \mathrm{d}B^2 \sinh^2\left(\frac{2 \phi}{\sqrt{3}}\right)\right)
\end{equation}
and the picture of the underlying surface is the same as for the case $\gamma \, = \, \ft 12$ already displayed in fig. \ref{trombola}.
\par
As in the previous case of $\gamma = \ft 12$, also for $\gamma \, = \, \ft 13$ the general formula (\ref{olograf}) for the coordinate $C$ in terms of the coordinate $\phi$ simplifies and the hypergeometric function becomes an elementary transcendental function. Explicitly either by direct integration of eq.(\ref{disko1}) or by limit $\gamma \to \ft 13$ in (\ref{olograf}), we obtain:
\begin{equation}\label{custoza1}
C_{\ft 13}(\phi) \, = \,  \log \left(\coth \left(\frac{\phi }{\sqrt{3}}\right)\right)
\end{equation}
which can be easily inverted yielding:
\begin{equation}\label{phiinC13}
 \phi \, = \,  \sqrt{3} \mathrm{ArcCoth}\,\left(e^C\right)
\end{equation}
\par
Inserting the appropriate value $\gamma \, = \, \ft 13$ in the general formula (\ref{CalleroCosh}) for the K\"ahler potential we find:
\begin{equation}\label{fardellotto}
 J_{\ft 13}(\phi) \, = \, \frac{3}{2} \log \left(\sinh \left(\frac{\phi  }{\sqrt{3}}\right)\right)
\end{equation}
and upon use of  (\ref{custoza1}) we get:
\begin{equation}\label{gongolini}
  J_{\ft 13}(\phi)\, = \, -\frac{3}{4} \left(2 C+\log\left(1-e^{-2 C}\right)\right)
\end{equation}
which agrees with eq.(\ref{funambolo}) setting $\zeta \, = \, \exp[\, - \, C] \, \exp[{\rm i} \, B]$ and $\alpha \, = \, \ft 34 \, = \, -\, \beta$.
Also in this case, as in the previous one we might generalize the corresponding supergravity model in two ways keeping fixed the K\"ahler metric and hence the ungauged part of the lagrangian. One deformation corresponds to changing the value of $\alpha$ in the K\"ahler potential and results into the addition of a constant term to the scalar potential. The other deformation is obtained exactly as in the previous case by switching on the $F$-gauge as well. \textit{Mutatis mutandis}, namely changing the powers and the coefficients of the field $\phi$ inside the hyperbolic functions
the additional $V_{WZ}$ potential will be of the form (\ref{acciderba}). We do not dwell on this since it would just be repetitive. We rather address the question of what is the coset parametrization that leads to the  form (\ref{babbutmammut}) of the metric.
 Let us introduce the standard generators of the $\slal(2,\mathbb{R})$ Lie algebra:
 \begin{equation}\label{sl2geni}
   L_0 \, = \,\left(
\begin{array}{ll}
 \frac{1}{2} & 0 \\
 0 & -\frac{1}{2}
\end{array}
\right) \quad ; \quad L_1 \, = \, \left(
\begin{array}{ll}
 0 & 1 \\
 0 & 0
\end{array}
\right) \quad ; \quad L_{-1} \, = \, \left(
\begin{array}{ll}
 0 & 0 \\
 1 & 0
\end{array}
\right)
 \end{equation}
 and construct the following family of group  elements of $\mathrm{SL(2,\mathbb{R})}$ depending on the two parameters $(\phi,B)$:
 \begin{eqnarray}\label{fantasioso}
  \mathbb{ L}\left(\phi,\, B\right) & =&  \exp\left[ B \, \ft 12 \left(L_1 \, - \, L_{-1}\right)\right] \, \cdot \, \exp\left[ \phi \, \frac{1}{\sqrt{3}} \left(L_1 \,+ \, L_{-1}\right)\right] \nonumber\\
  & = &\left(
\begin{array}{ll}
 \cos \left(\frac{B}{4}\right) \cosh
   \left(\frac{\phi
   }{\sqrt{3}}\right)+\sin
   \left(\frac{B}{4}\right) \sinh
   \left(\frac{\phi
   }{\sqrt{3}}\right) & \cosh
   \left(\frac{\phi
   }{\sqrt{3}}\right) \sin
   \left(\frac{B}{4}\right)+\cos
   \left(\frac{B}{4}\right) \sinh
   \left(\frac{\phi
   }{\sqrt{3}}\right) \\
 \cos \left(\frac{B}{4}\right) \sinh
   \left(\frac{\phi
   }{\sqrt{3}}\right)-\cosh
   \left(\frac{\phi
   }{\sqrt{3}}\right) \sin
   \left(\frac{B}{4}\right) & \cos
   \left(\frac{B}{4}\right) \cosh
   \left(\frac{\phi
   }{\sqrt{3}}\right)-\sin
   \left(\frac{B}{4}\right) \sinh
   \left(\frac{\phi
   }{\sqrt{3}}\right)
\end{array}
\right) \nonumber\\
 \end{eqnarray}
 The matrices $\mathbb{L}(\phi,B)$ constitute the required parametrization of the coset. Calculating the left-invariant one-form:
 \begin{equation}\label{sinistrainvaria}
  \Omega\left( \phi, \, B\right) \, \equiv \, \mathbb{L}^{-1}(\phi,B) \, \partial_\phi \, \mathbb{L}(\phi,B) \, d\phi \,  + \,
   \mathbb{L}^{-1}(\phi,B) \, \partial_B \, \mathbb{L}(\phi,B) \, dB
 \end{equation}
 and calculating the zweibein:
 \begin{equation}\label{zwibinno}
   E^1 \, \equiv \,  2 \, \mbox{Tr} \left( L_0 \, \Omega \right ) \quad ; \quad  E^2 \, \equiv \,  \ft 12 \,  \, \mbox{Tr} \left(\left(L_1 \, + L_{-1} \right)\, \Omega \right )
 \end{equation}
 we find:
 \begin{eqnarray}
 E^1 & = & \frac{1}{2} \,{\mathrm{d}B} \sinh
   \left(\frac{2 \phi }{\sqrt{3}}\right) \nonumber\\
   E^2 & = & \frac{\mathrm{d}\phi }{\sqrt{3}}
    \label{custasi}
 \end{eqnarray}
 and the bilinear form:
 \begin{equation}\label{ciulifischio}
   ds^2 \, = \, \ft 34 \,\left (  \left ( E^1\right)^2 \, + \,  \left ( E^2\right) \right)
 \end{equation}
 reproduces the metric (\ref{babbutmammut}).
\subsection{The K\"ahler manifold in the image of the $\gamma \, = \, \ft 15$ potential}
In this case the momentum map takes the following form:
\begin{equation}\label{coarse}
  P_{\ft 15}(\phi) \, = \,\frac{25}{12} \cosh
   ^4\left(\frac{\sqrt{3} \phi
   }{5}\right)
\end{equation}
where, once again we have chosen a normalization that will prove useful in the sequel.
The corresponding metric is the following one.
\begin{equation}\label{filiut}
  ds^2_{\gamma \, = \, \ft 15} \,  = \, \ft 14 \left( d\phi^2 \, + \, \frac{25}{3} \cosh
   ^6\left(\frac{\sqrt{3} \phi
   }{5}\right) \sinh
   ^2\left(\frac{\sqrt{3} \phi
   }{5}\right) \, dB^2 \right)
\end{equation}
\begin{figure}[!hbt]
\begin{center}
\iffigs
\includegraphics[height=90mm]{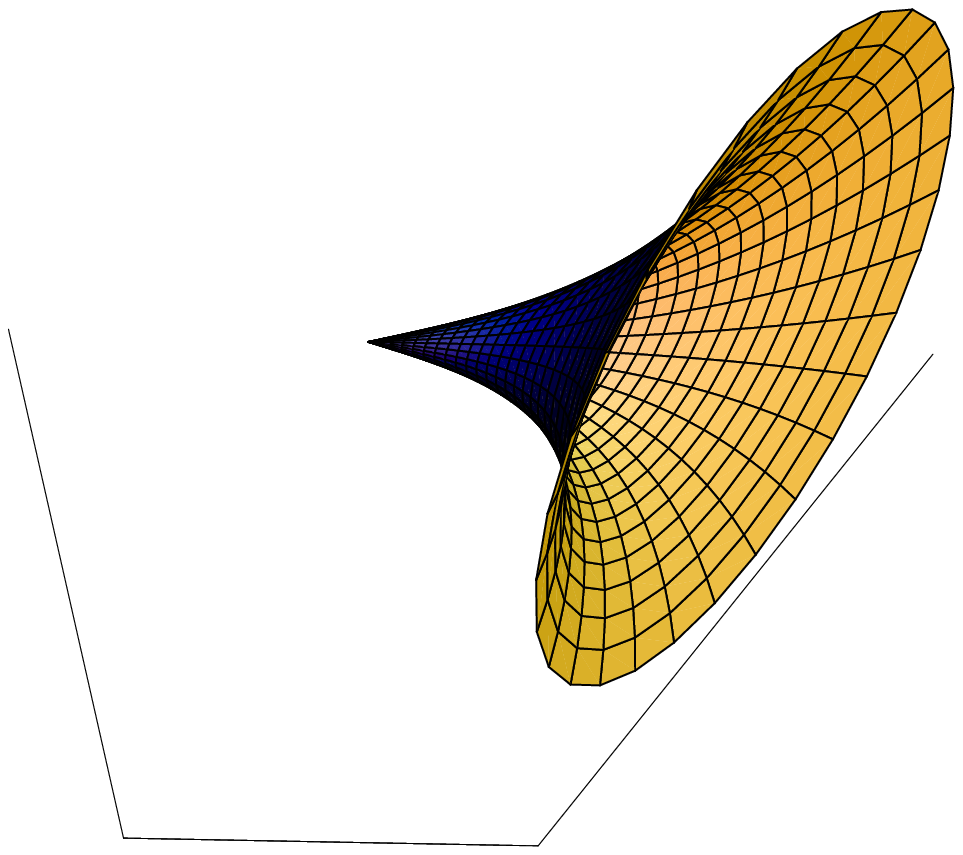}
\else
\end{center}
 \fi
\caption{\it Intrinsic view of the costant negative curvature surface whose metric is given in eq.(\ref{filiut}). The surface is obtained
as surface of revolution of the plot of the function $\frac{5 \cosh ^3\left(\frac{\sqrt{3}
   \phi }{5}\right) \sinh
   \left(\frac{\sqrt{3} \phi
   }{5}\right)}{\sqrt{3}}$. This is the model of the K\"ahler manifold sitting in the image of the $D$-map for the $\gamma \, = \, \ft 15$ case of the $I_7$ integrable series. }
\label{trombolata5}
 \iffigs
 \hskip 1cm \unitlength=1.1mm
 \end{center}
  \fi
\end{figure}
A picture of the corresponding two dimensional surface is displayed in fig.\ref{trombolata5}.
\par
Performing the integral (\ref{disko1}), we obtain:
\begin{equation}\label{Cfunzia15}
 C_{\ft 15}(\phi) \, = \, \alpha +\frac{1}{2}  \left(\mathrm{sech}^2\left(\frac{\sqrt{3} \phi }{5}\right)\, - \, 2
    \log \left(\cosh \left(\frac{\sqrt{3} \phi }{5}\right)\right)+\log \left(\cosh^2\left(\frac{\sqrt{3} \phi }{5}\right)-1\right)\right)
\end{equation}
where $\alpha$ is an arbitrary integration constant.
If we fix such integration constant to the value $\alpha \, = \, - \, \ft{1}{2}$ eq.(\ref{Cfunzia15}) can be easily inverted and one obtains:
\begin{equation}
  \phi \, = \,  \frac{5 \mathrm{ArcSech}\,\left(\sqrt{W\left(-e ^{2 \, C}\right)+1}\right)}{\sqrt{3}}
  \label{lambertus}
\end{equation}
where $W(x)$ denotes the Lambert omega function.  Since $W(x)$ is real only for $x>0$, it is convenient to change the sign of its argument. This is easily done by redefining the coordinate $C(\phi)$ which, being an integral, is defined up to an additive constant,  by means of the shift:
\begin{equation}\label{sciftus}
  C(\phi) \, \rightarrow  \,  C(\phi) \, + \, {\rm i} \, \frac{1}{2} \, \pi
\end{equation}
After this redefinition, substituting (\ref{lambertus}) in the expression  (\ref{fluorescenza}) for the curvature and in the expression (\ref{CalleroCosh}) for the K\"ahler potential we obtain:
\begin{eqnarray}
 R_{\ft 15} &=&\frac{24}{25} \left(3 W\left(e^{2 \, C}\right)\, - \, 5\right) \label{cova1}\\
  J_{\ft 15} &=&\log \left(W\left(e^{2 C}\right)\right)\, - \, \frac{25}{24} \log \left(W\left(e^{2 \, C}\right)+1\right) \label{fanta1}
\end{eqnarray}
Plots of the curvature and of the K\"ahler potential are shown in fig.\ref{RJfutto}:
\begin{figure}[!hbt]
\begin{center}
\iffigs
\includegraphics[height=70mm]{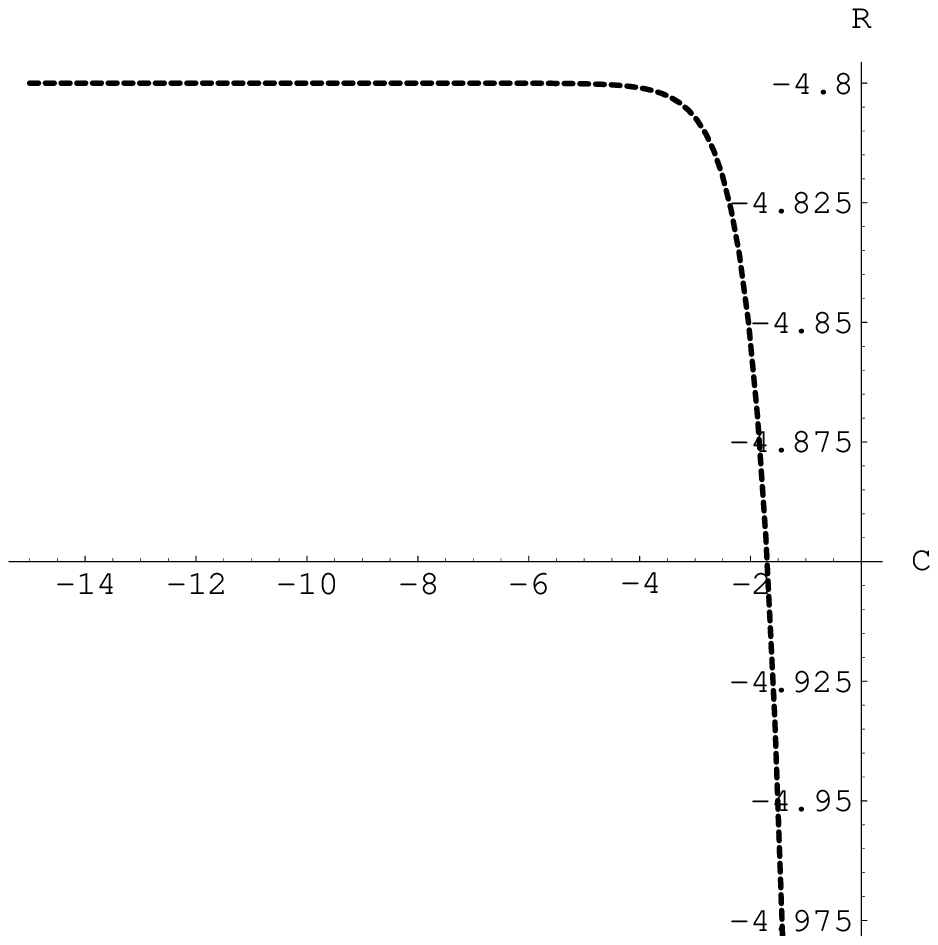}
\includegraphics[height=70mm]{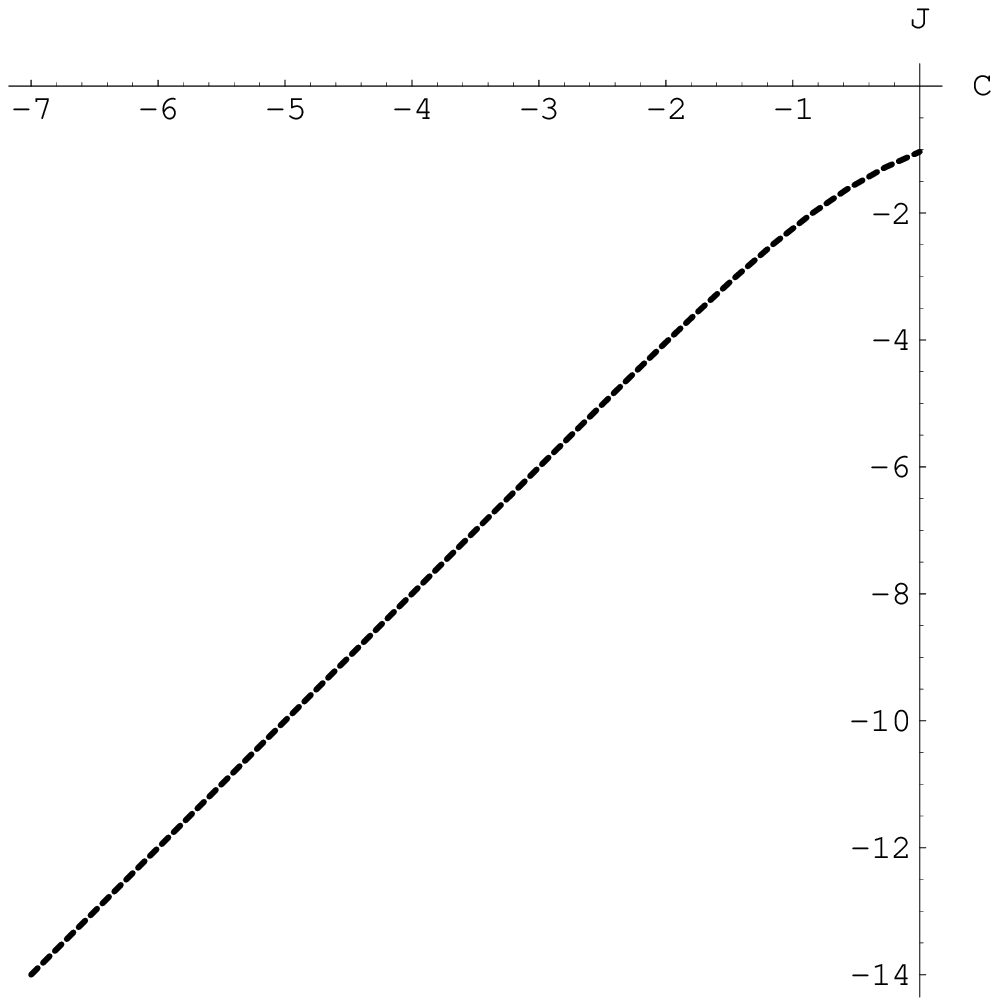}
\else
\end{center}
 \fi
\caption{\it  Plots of the curvature $R$ and of the K\"ahler potential $J$ for the K\"ahler manifold in the $D$-map image of the $\gamma \, = \, \ft 15$ potential of the integrable series $I_7$. The curvature is presented in the plot on the left, the K\"ahler potential in the plot on the right. The permitted range of $C$ where both functions are real is $C<0$ negative. This is consistent with the interpretation of $C$ as the logarithm of the norm of the complex coordinate $\zeta$. Indeed, as in the case of the Poincar\'e metric, the modulus of the coordinate $\zeta$ has to remain inside the unit disk.}
\label{RJfutto}
 \iffigs
 \hskip 1cm \unitlength=1.1mm
 \end{center}
  \fi
\end{figure}
Having an explicit representation of the K\"ahler potential $J(C)$ in terms of known functions we can calculate its power series development which is to be compared with equation (\ref{seriosa}).
Explicitly we find:
\begin{eqnarray}
  J_{\ft 15}(C) &=& 2 C\, - \, \frac{49 \, e^{2 C}}{24} \, + \, \frac{41 \, e^{4 C}}{16}\, - \, \frac{641 \, e^{6 C}}{144}\nonumber\\
 && +\frac{2543 \, e^{8 \, C}}{288}\, -\, \frac{1205 \, e^{10 \, C}}{64}\, + \, \frac{30251 \, e^{12 \, C}}{720}
 \, - \, \frac{11700601 \, e^{14 \, C}}{120960} \nonumber\\
 &&+\frac{656143 \, e^{16 \, C}}{2880}\, - \, \frac{528316681 \, e^{18 \, C}}{967680} \, +\, \frac{36095435 \, e^{20\, C}}{27216} \label{omnibus}
\end{eqnarray}
As a final general remark we can conclude that whenever the function $J(C)$ has a power series expansion of the form (\ref{seriosa}) the appropriate interpretation is that of disk-type $D$-map and the underlying geometry is a deformation  of the classical Poincar\'e geometry in the unit circle obtained by stereographic projection of the hyperboloid.
\section{K\"ahler manifolds in the $D$-map image of  $I_2$  integrable potentials}
The integrable potentials of the series $I_2$ in Table \ref{tab:families} are those mentioned in eq.(\ref{gammaserie}). Applying eq.(\ref{giunone}) to these potentials we find the following expression for the
K\"ahler curvature:
\begin{eqnarray}
  R_\gamma(\phi) &=& \, - \, 4 \, \frac{\mathrm{N}(\phi)}{\mathrm{D}(\phi)} \nonumber\\
  \mathrm{N}(\phi) &=& 3 \left(8 a^3 e^{6 \sqrt{3}
   \gamma  \phi } \gamma ^3+b^3
   e^{3 \sqrt{3} (\gamma +1)
   \phi } (\gamma +1)^3+4 a^2 b
   e^{\sqrt{3} (5 \gamma +1)
   \phi } \left(5 \gamma
   ^3+1\right)\right.\nonumber\\
   &&\left.+2 a b^2 e^{2
   \sqrt{3} (2 \gamma +1) \phi }
   \left(8 \gamma ^3-3 \gamma
   ^2+6 \gamma +1\right)\right) \nonumber\\
  \mathrm{D}(\phi)&=& 4 \left(e^{2 \sqrt{3} \gamma
   \phi } a+b e^{\sqrt{3}
   (\gamma +1) \phi }\right)^2
   \left(2 a e^{2 \sqrt{3}
   \gamma  \phi } \gamma +b
   e^{\sqrt{3} (\gamma +1) \phi
   } (\gamma +1)\right)\label{gammaserioso}
\end{eqnarray}
Let us now go a little bit deeper in the the analysis of such models and in particular let us discuss  appropriate normalizations in order to make precise contact with the K\"ahler geometry of the Lobachevsky-Poincar\'e plane of which they happen to be a deformation. In this case we expect to have a plane interpretation of the $D$-map and the expansion of the $J(C)$ function should be of the type (\ref{gartoccio}) already met in the case of the $I_1B$ series, the flat coordinate $C$ being the imaginary part of the coordinate $t$ in the plane. The geometry of an $\mathrm{SL(2,\mathbb{R})}/\mathrm{O(2)}$ manifold is  encoded in the following   K\"ahler potential :
\begin{equation}\label{su11callo}
    \mathcal{K}_{\slal(2,\mathbb{R})}^{(q)} (t,\bar{t}) \, = \, - \, \log \, \left [(t \, - \, \bar{t})^q \right]
\end{equation}
that leads to the Poincar\'e metric:
\begin{equation}\label{gorlo}
    ds^2_{\mbox{Poincar\'e}} \, = \, \frac{q}{4} \, \frac{\mathrm{d}t \, \mathrm{d}\bar{t}}{\left(\mbox{Im}  \,t\right)^2}
\end{equation}
The number $q$ is the only parameter, named the weight which is simply related to the constant curvature of the manifold as it follows:
\begin{equation}\label{Rqrela}
    R_{\slal(2,\mathbb{R})}^{(q)} \, =\, - \, \frac{4}{q}
\end{equation}
The relation between the standard description (\ref{su11callo}) of the Poincar\'e plane and that  in terms of the coordinate $\phi$, utilized above is easily established by setting
\begin{equation}\label{faccina}
    t\, = \, {\rm i}\, \underbrace{\exp\left[\frac{1}{\sqrt{q}} \,\phi\right]}_{C(\phi)} \, + \, B
\end{equation}
Indeed the $\mathrm{SU(1,1)}/\mathrm{U(1)}$ metric (\ref{gorlo}) is turned into the form (\ref{metricozza}) if we set:
\begin{equation}\label{marsilio}
    \mathcal{P}^\prime (\phi)\, = \, - \, \sqrt{\frac{q}{4}} \, \exp\left [ - \,\frac{1}{\sqrt{q}} \,\phi\right]
\end{equation}
which upon use of eq.(\ref{garducci}) yields the result (\ref{Rqrela}). This provides an additional check of the position $C(\phi)\, = \, \exp\left[\frac{1}{\sqrt{q}} \,\phi\right]$ since in the standard formulation (\ref{su11callo}), the momentum map
 for a translation Killing vector $k \, = \, 1$ is:
\begin{equation}\label{dillo}
    \mathcal{P}\, = \, \frac{q}{2} \, \frac{1}{C}\, = \, \frac{q}{2} \, \exp\left [ - \,\frac{1}{\sqrt{q}} \,\phi\right]
\end{equation}
which is consistent with (\ref{marsilio}).
\par
Hence we observe that when we interpret the square root of the potential (\ref{gammaserie}) as a momentum map, at $b=0$ we match the case of an $\slal(2,\mathbb{R})$ homogeneous space with:
\begin{equation}\label{costaneto}
    q \, = \, \frac{1}{3 \gamma ^2}
\end{equation}
To match the correct normalization of such a momentum map we have also to fix the value of the parameter $a$ as it follows:
\begin{equation}\label{fistola}
    a \, = \, \frac{4}{9 \, \gamma^4}
\end{equation}
and in view of this we also define:
\begin{equation}\label{costolettadimaiale}
    b \,= \, \frac{4}{9 \, \gamma^4}\, \lambda
\end{equation}
so that the constant curvature case is retrieved by setting $\lambda=0$. Finally we recall that, since the kinetic term of the field $\phi$ is canonical, namely the K\"ahler metric has the form (\ref{metricozza}), we can perform a redefinition of  $\phi$ by means of a constant shift. Setting:
\begin{equation}\label{glipsa}
    \phi \, \rightarrow \, \phi \, + \, \frac{1}{\sqrt{3}(\gamma\, - \, 1)}
\end{equation}
together with (\ref{fistola}), (\ref{costolettadimaiale}), we obtain that the momentum map is:
\begin{equation}\label{firitone}
  \mathcal{P}_\gamma \, = \,  -\frac{\sqrt{e^{2 \sqrt{3}
   \gamma  \phi }+e^{\sqrt{3}
   (\gamma +1) \phi }} \lambda
   ^{\frac{\gamma }{\gamma
   -1}}}{3 \gamma ^2}
\end{equation}
\subsection{Integration of the complex structure equation}
Inserting the form (\ref{firitone}) of the momentum map into eq. (\ref{plane1}) and performing the integral we obtain the following remarkable exact result for the flat coordinate $C$:
\begin{equation}\label{fattuccio}
  C_\gamma(\phi) \, = \, e^{-\sqrt{3} \gamma  \phi }\, F_1\left(\frac{\gamma }{\gamma-1};-\frac{1}{2},1;2+\frac{1}{\gamma -1};-e^{-\sqrt{3} (\gamma -1) \, \phi },-\frac{e^{-\sqrt{3} (\gamma -1) \phi } (\gamma +1)}{2 \, \gamma }\right)
\end{equation}
where $F_1(\dots)$ denotes an Appel function F1 which is defined by the following series development in two variables:
\begin{equation}\label{appellofunzio}
    F_1(a,b_1,b_2,c;x,y) \, = \, \sum_{m,n=0}^\infty \, \frac{(a)_{m+n} \, (b_1)_m \, (b_2)_{n}}{(c)_{m+n} \, m! \, n!} \, x^m \,y^n
\end{equation}
having denoted:
\begin{equation}\label{risingfactorial}
    (r)_n \, \equiv \, \frac{\Gamma(r+n)}{\Gamma(n)}
\end{equation}
Implementing the following coordinate transformation:
\begin{equation}\label{gorcizza}
\phi \, = \, -\frac{\log (T)}{\sqrt{3} \gamma }
\end{equation}
the expression for the flat coordinate $C$ becomes:
\begin{equation}\label{fischiettando}
  C_\gamma(T) \, = \, T F_1\left(-\frac{\gamma }{1-\gamma};-\frac{1}{2},1;\frac{1-2 \gamma
   }{1-\gamma
   };-T^{1-\frac{1}{\gamma
   }},\frac{T^{1-\frac{1}{\gamma }}
   (-\gamma -1)}{2 \gamma }\right)
\end{equation}
and with the further position:
\begin{equation}\label{ruttolando}
  U \, = \, T^{1-\frac{1}{\gamma }} \, \rightarrow \, T \, = \, U^{\frac{\gamma}{\gamma-1}}
\end{equation}
we obtain:
\begin{equation}
 C_\gamma(U) \, = \, U^{\frac{\gamma}{\gamma-1}}\, F_1\left(-\frac{\gamma }{1-\gamma};-\frac{1}{2},1;\frac{1-2 \gamma
   }{1-\gamma
   };-U,- \,\frac{\gamma \, + \,1 }{2\, \gamma} \,U\right)
 \end{equation}
 In order to discuss $C_\gamma(U)$ we need to consider the series expansion in $U$ of Appel functions of the following type:
\begin{equation}\label{tritone}
   \mathcal{F}(U) \, = \,  F_1(a;b,w;c;-U,-p U)
\end{equation}
our case corresponding to:
\begin{equation}\label{valliparami}
    a \, = \, \frac{\gamma}{\gamma \, -\, 1} \quad ; \quad b \, = \, - \, \frac{1}{2}\quad ; \quad w \, = \, 1\quad ; \quad c \, = \, \frac{1\, - \, 2 \, \gamma}{1\, - \, \gamma} \quad ; \quad p \, = \, - \,\frac{\gamma \, + \,1 }{2\, \gamma}
\end{equation}
From (\ref{appellofunzio}) it follows that:
\begin{equation}\label{targovy}
    \mathcal{F}(U) \, = \, \sum_{k=0}^\infty \, c_k \, U^k
\end{equation}
where
\begin{eqnarray}\label{gingle}
c_k & = &   \frac{(-1)^k \Gamma (a+k)}{\Gamma
   (a)} \, \sum _{n=0}^k
   \frac{p^n (\Gamma (b+k-n) \Gamma
   (n+w))}{n! (\Gamma (b) \Gamma (w)
   (k-n)!)} \nonumber\\
   & = & \frac{(-1)^k \Gamma (c) \Gamma (a+k)
   \Gamma (b+k) \,
   _2F_1(-k,w;-b-k+1;p)}{\Gamma (a)
   \Gamma (b) \Gamma (k+1) \Gamma
   (c+k)}
\end{eqnarray}
\subsection{The case  $\gamma \, = \, -\ft 76 $ }
\begin{figure}[!hbt]
\begin{center}
\iffigs
 \includegraphics[height=60mm]{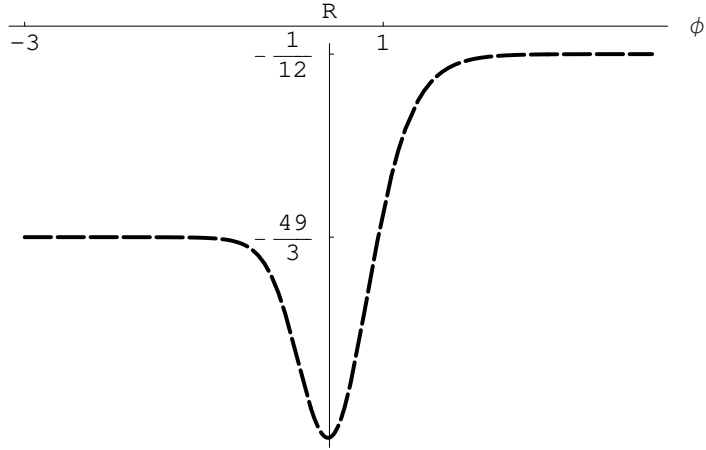}
\else
\end{center}
 \fi
\caption{\it
Plot of the K\"ahler curvature, versus coordinate $\phi$ for the best fit model $\gamma \, = \, -\ft 76$. The plots for various value of $\lambda$ are obtained from the present one simply by shifting the coordinate $\phi$}
\label{settesestiR}
 \iffigs
 \hskip 1cm \unitlength=1.1mm
 \end{center}
  \fi
\end{figure}

As an illustration of reconstruction of the K\"ahler potential in  the series (\ref{gammaserie}), we utilize the best fit model $\gamma \, = \, - \ft 76$ proposed by Sagnotti.
 Inserting the  value $\gamma \, = \, - \ft 76$ in eq.(\ref{gammaserioso})  and furthermore  redefining the parameters as in equations (\ref{fistola}), (\ref{costolettadimaiale}), (\ref{glipsa}), we obtain:
\begin{eqnarray}
  R_{-\ft 76}(\phi) &=&\, - \, \frac{2744+5996 e^{\frac{13
   \phi }{2 \sqrt{3}}}+9844
   e^{\frac{13 \phi
   }{\sqrt{3}}}+e^{\frac{13
   \sqrt{3} \phi }{2}}}{12
   \left(1+e^{\frac{13 \phi
   }{2 \sqrt{3}}}\right)^2
   \left(14+e^{\frac{13 \phi
   }{2 \sqrt{3}}}\right)}
\end{eqnarray}
where the overall scale $a$ and the parameter $\lambda$ cancel. The function $R_{-\ft 76}(\phi) $ has the property:
\begin{equation}\label{giocattolo}
    R_{-\ft 76}(-\infty) \, = \, \frac{49}{12} \quad ; \quad R_{-\ft 76}(\infty) \, = \, \frac{1}{48}
\end{equation}
The structure of the curvature function, whose plot is displayed  in fig.\ref{settesestiR}, reveals that this space (as all the elements of the $I_2$ series) realizes a smooth transition from an $\mathrm{SL(2,\mathbb{R})}/\mathrm{O(2)}$ with one value of the curvature $R_{-\infty}$ to another one with a different value $R_{\infty}$. The shape of such a transition is that of a kink.
 In conclusion the best fit K\"ahler manifold is a kink connecting the two $q$-indices:
\begin{equation}\label{figliomio}
    q_{\infty} \, = \, 192 \quad ;\quad q_{-\infty} \, = \, \frac{48}{49}
\end{equation}
These rational numbers are likely to hide some profound meaning in terms of brane-wrapping or similar higher dimensional mechanisms as suggested by the brane interpretation of the best fit model put forward by Sagnotti and briefly summarized in \cite{noicosmoitegr}.
\par
\begin{figure}[!hbt]
\begin{center}
\iffigs
 \includegraphics[height=60mm]{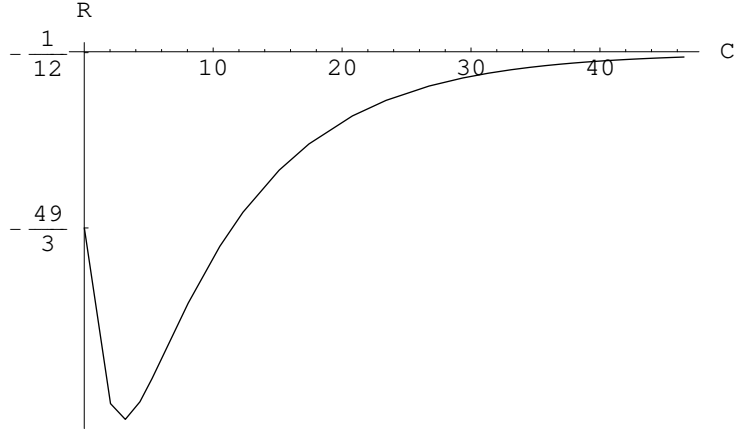}
\else
\end{center}
 \fi
\caption{\it
Dependence of the curvature on the flat coordinate $C$, namely on the imaginary part  of the complex coordinate $t$, for the $\gamma \, = \, - \ft 76$ model. The parameter $\lambda$ was set to one. All other values of $\lambda$ simply result in a rescaling of $C$.}
\label{R76fromC}
 \iffigs
 \hskip 1cm \unitlength=1.1mm
 \end{center}
  \fi
\end{figure}

In the case we want to consider the variable $U$ introduced above (see eq.s (\ref{gorcizza}) and (\ref{ruttolando})) takes the following explicit form:
\begin{equation}\label{gonone}
    U \, = \, e^{\frac{13 \phi }{2
   \sqrt{3}}}
\end{equation}
whose range is from $0$ to $+\infty$. In terms of this new coordinate we have (setting $\lambda=1$):
\begin{eqnarray}
  C_{-\ft 76}(U) &=& U^{7/13}
   F_1\left(\frac{7}{13};-\frac{1}{2},1;\frac{20}{13};-U,
   -\frac{U}{14}\right) \label{C76funzia}\\
  R_{-\ft 76}(U) &=& -\frac{910}{U+14}-\frac{1}{12}+\frac
   {91}{U+1}-\frac{169}{4 (U+1)^2}\label{R76funzia}
\end{eqnarray}
The dependence of the curvature from the imaginary part $C$ of the complex coordinate $t$ is displayed in fig. \ref{R76fromC}.
\par
Let us next derive the form of the K\"ahler potential of this manifold first as a function of the field $\phi$ and then as a function of the flat coordinate $C$.
\par
\begin{figure}[!hbt]
\begin{center}
\iffigs
 \includegraphics[height=60mm]{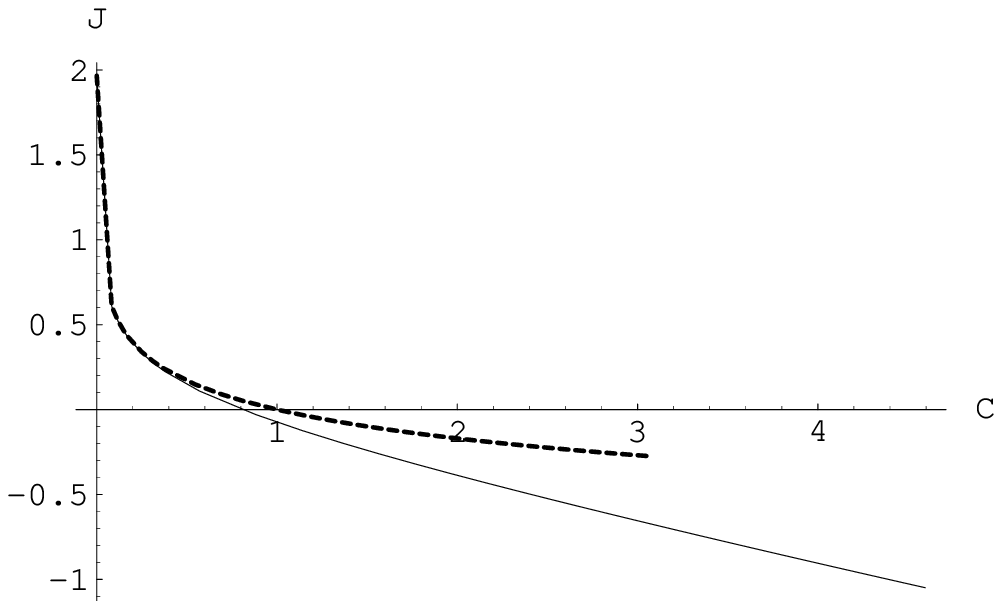}
 \includegraphics[height=60mm]{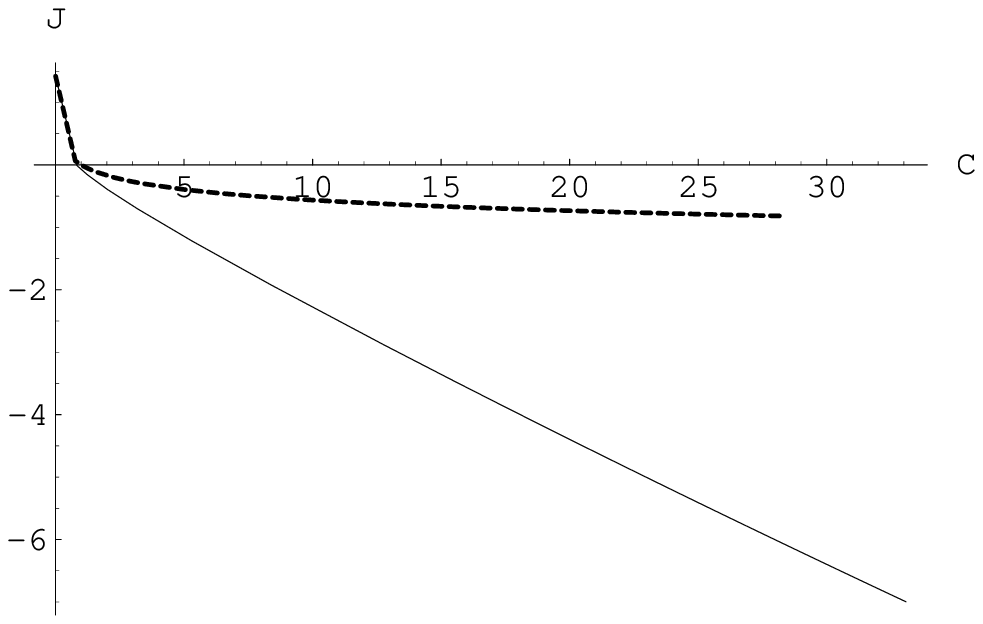}
\else
\end{center}
 \fi
\caption{\it
Dependence of the K\"ahler potential $J$ on the imaginary part $C$ of the complex coordinate $t$ for the $\gamma \, = \, - \ft 76$ model. The parameter $\lambda$ was set to one. At small values of $C$ the behavior of $J(C)$ is the same as $J_0(C) \, = \, -\frac{12}{49} \, \log[C] $, corresponding to an $\mathrm{SU(1,1)/U(1)}$ manifold with $q=\frac{12}{49}$. At larger values of $C$ the behavior of $J(C)$ changes significantly. The solid line in the graphs corresponds to  the exact function $J(C)$. The dashed line corresponds to the function $J_0(C)$ which approximates the small $C$ behavior.}
\label{calluspot}
 \iffigs
 \hskip 1cm \unitlength=1.1mm
 \end{center}
  \fi
\end{figure}

From eq.(\ref{celerus2}) we conclude that\footnote{Here we have reinstalled the parameter $\lambda$ since this helps us to fix the normalizations at $\lambda \, = \,0$. In the sequel we will put again $\lambda \, =\,1$.}:
\begin{eqnarray}\label{goebel}
    \frac{\mathrm{d}J}{\mathrm{d}\phi} & = & \frac{\mathrm{d}J}{\mathrm{d}C} \, \frac{\mathrm{d}C}{\mathrm{d}\phi}
    \, =\,\frac{\mathcal{P}(\phi)}{\frac{\mathrm{d} \mathcal{P}(\phi)}{\mathrm{d}\phi}} \, = \,
4 \sqrt{3}
   \left(\frac{13}{e^{\frac{13
   \phi }{2 \sqrt{3}}} \lambda
   +14}-1\right)
   \end{eqnarray}
   and by means of a simple integration we obtain
   \begin{equation}\label{cuperlo}
    J(\phi) \, = \, -\frac{2}{7} \left(\sqrt{3}
   \phi +6 \log
   \left(e^{\frac{13 \phi }{2
   \sqrt{3}}} \lambda
   +14\right)-6 \log
   (14)\right)
   \end{equation}
   where we have fixed the integration constant in such a way that at $\lambda\, = \,0$ we get:
   \begin{eqnarray}\label{deborah}
J(\phi) \, = \,    J(C) & \stackrel{\lambda \to 0}{\Longrightarrow} & - \frac{12}{49} \, \log \, [C] \nonumber\\
    C(\phi) & \stackrel{\lambda \to 0}{\Longrightarrow} & \exp\left[ \frac{7}{2\sqrt{3}}  \, \phi \right]
   \end{eqnarray}
 Performing the shift (\ref{glipsa})  on the $\phi$-field and replacing the shifted $\phi$ with the $U$ coordinate introduced in (\ref{gonone}) we obtain:
 \begin{equation}\label{carrettoagricolo}
    J(U) \, = \, \frac{12}{91} \left (-\log [U]-13 \log [U+14]+\log [\lambda]+13 \log [14] \right)
 \end{equation}
 By means of a parametric plot we can visualize the dependence of the K\"ahler potential with respect to the imaginary part $C$ of the complex coordinate $t$. Such behavior is displayed in fig.\ref{calluspot}.
Can we derive the analytic form of the function $J(C)$ that we have displayed graphically? The best we can do is to provide a series representation of $J(C)$ whose coefficients can be inductively derived from the coefficients in the series expansion of the Appel function. Hence let us first discuss the series expansion of the function $C_{-\ft 76}(U)$ defined in equation (\ref{C76funzia}).
\par
\begin{figure}[!hbt]
\begin{center}
\iffigs
 \includegraphics[height=60mm]{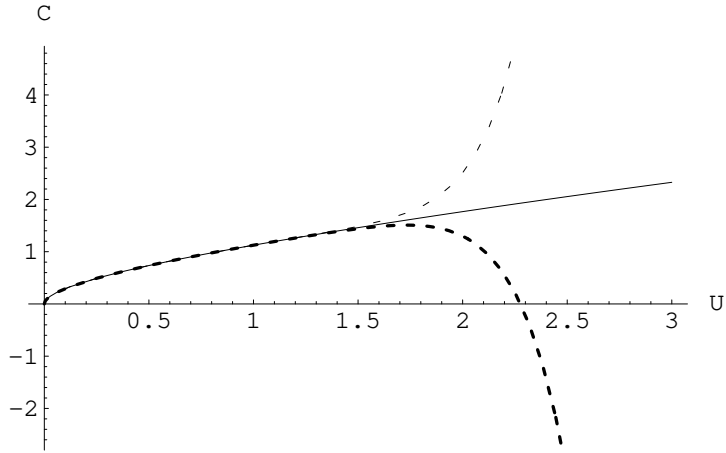}
\else
\end{center}
 \fi
\caption{\it
Comparison of the exact function $C_{-\ft 76}(U)$ with the truncation to a finite order of its series development. The solid line is the exact function, while the two dashed lines correspond to the  series expansion respectively truncated to $k_{\mbox{max}}=10$ (thicker line) and  $k_{\mbox{max}}=11$ (thinner line).}
\label{nogoodapprox}
 \iffigs
 \hskip 1cm \unitlength=1.1mm
 \end{center}
  \fi
\end{figure}

Inserting the parameter values (\ref{valliparami}) with $\gamma \, = \, - \, \ft 76$  into eq.(\ref{gingle}) we arrive at the final form of the series expansion for the $C_{-\ft 76}(U)$ function:
\begin{equation}\label{serialeCdiU}
    C_{-\ft 76}(U) \, = \, U^{\ft{7}{13}} \, \sum_{k=0}^\infty \, d_k \, U^k
\end{equation}
where:
\begin{equation}\label{dikappi}
  d_k \, = \,   \frac{7 (-1)^k \sqrt{\pi } \,
   _2\tilde{F}_1\left(1,-k;\frac{3}{2
   }-k;\frac{1}{14}\right) \sec (k
   \pi )}{2 (13 k+7) \Gamma (k+1)}
\end{equation}
In the above formula $ _2\tilde{F}_1$ denotes the regularized hypergeometric function of its arguments.
The  radius of convergence of the series representation (\ref{serialeCdiU}) is given by the condition $U < \ft 32$.
Within the convergence radius the approximation given by truncation to any finite order  $k_{\mbox{max}}$ is very good, but due to the alternating signs as soon as $U>\ft 32$ the truncation to any finite order provides no realistic approximation to the true function. This is illustrated in fig.\ref{nogoodapprox}.
Anyhow, within the convergence radius we can reconstruct the functional form of the K\"ahler potential introducing its series expansion according to the following equation:
\begin{equation}\label{molinodorino}
    J(C) \, = \, - \, \frac{12}{49} \, \log[C] \, + \, 1 + \, \sum_{k=1}^\infty c_k \, C^k
\end{equation}
The series expansion of $J(U)\, = \, J\left(C_{-\ft 76}(U)\right)$ is predetermined from eq.(\ref{carrettoagricolo}) and we have:
\begin{eqnarray}
  J(U)&=& -\frac{12 \log (U)}{91}\, + \,\sum _{k=1}^{\infty }  m_k \, U^k\nonumber\\
   &=& -\frac{12 \log
   (U)}{91}-\frac{6
   \, U}{49}+\frac{3
   \, U^2}{686}-\frac{\, U^3}{4802}+
   \frac{3
   \, U^4}{268912}-\frac{3
   \, U^5}{4705960}\nonumber\\
   &&+\frac{\, U^6}{26
   353376}-\frac{3
   \, U^7}{1291315424}+O\left(\, U^8
   \right) \label{felicetto}
\end{eqnarray}
On the other hand inserting the series (\ref{serialeCdiU}) into eq.(\ref{molinodorino}) we obtain another form of the same series:
\begin{equation}\label{ziocane}
    J(U)\,=\, -\frac{12 \log (U)}{91}\, + \,\sum _{k=1}^{\infty }  p_k \, U^k
\end{equation}
the coefficients $p_k$ being expressed in terms of the coefficients $c_k$ and $d_k$ according to the following pattern:
 \begin{eqnarray}
  p_1 &=& c_1-\frac{12 d_1}{49} \nonumber\\
  p_2 &=& \frac{1}{49} \left(49 c_2+91
   c_1 d_1+6 \left(d_1^2-2
   d_2\right)\right)\nonumber \\
  p_3 &=& \frac{1}{49} \left(49 c_3+182
   c_2 d_1+13 c_1 \left(3
   d_1^2+7 d_2\right)-4
   \left(d_1^3-3 d_2 d_1+3
   d_3\right)\right) \nonumber\\
  p_4 &=& \frac{1}{343} \left(21
   d_1^4-13 c_1 d_1^3-84 d_2
   d_1^2+1911 c_3 d_1+546 c_1
   d_2 d_1+42 d_2^2+343 c_4\right.\nonumber\\
   &&\left.+91
   c_2 \left(19 d_1^2+14
   d_2\right)+7 \left(91
   c_1+12 d_1\right) d_3-84
   d_4\right) \nonumber\\
  p_5 &=& \frac{1}{12005}  \, \left(-588 d_1^5+130 c_1
   d_1^4+34580 c_2 d_1^3+2940
   d_2 d_1^3-1365 c_1 d_2
   d_1^2-2940 d_3 d_1^2\right.\nonumber\\
   &&\left.-2940
   d_2^2 d_1+89180 c_4
   d_1+121030 c_2 d_2
   d_1\right.\nonumber\\
   &&\left.+19110 c_1 d_3 d_1+9555
   c_1 d_2^2+12005 c_5\right.\nonumber\\
   &&\left.+9555
   c_3 \left(16 d_1^2+7
   d_2\right)+44590 c_2
   d_3+2940 d_2 d_3+245
   \left(91 c_1+12 d_1\right)
   d_4-2940 d_5\right)\nonumber\\
   p_6 &=& \dots\nonumber\\
   \dots&=&\dots \label{triangolobermuda}
\end{eqnarray}
As we see the infinite system of equations $p_k \, = \, m_k$ is triangular in the coefficients $c_k$ and can be solved iteratively.
The result for the first seven coefficients are displayed in the following table, where we also mention the numerical values of the $d_k$ coefficients appearing in the series expansion of the $C_{-\ft 76}(U)$ function.
\begin{equation}\label{fascolono}
    \begin{array}{|c|l|l|}
    \hline
    \hline
      \mbox{k} & d_k & c_k \\
      \hline
      1 & \frac{3}{20} & -\frac{3}{35}\\
      2 & -\frac{61}{1848} & \frac{1877}{107800} \\
      3 & \frac{101}{9016}& -\frac{1630943}{173558000}\\
      4 & -\frac{13621}{2590336} &\frac{1338922439597}{189233758
   560000} \\
      5 & \frac{65635}{22127616} & \frac{209551731730673}{331159
   07748000000} \\
      6 & -\frac{2733169}{1462881280}&\frac{64501979213193313277}{10
   255069384138656000000} \\
      7 & \frac{3738761}{2951578112} & -\frac{38584080386830892078311
   67}{57452317046406130464000
   0000}\\
   \hline
    \end{array}
\end{equation}
As one sees the $d_k$ coefficient are all rational numbers and so are the $c_k$.
\par
\begin{figure}[!hbt]
\begin{center}
\iffigs
 \includegraphics[height=60mm]{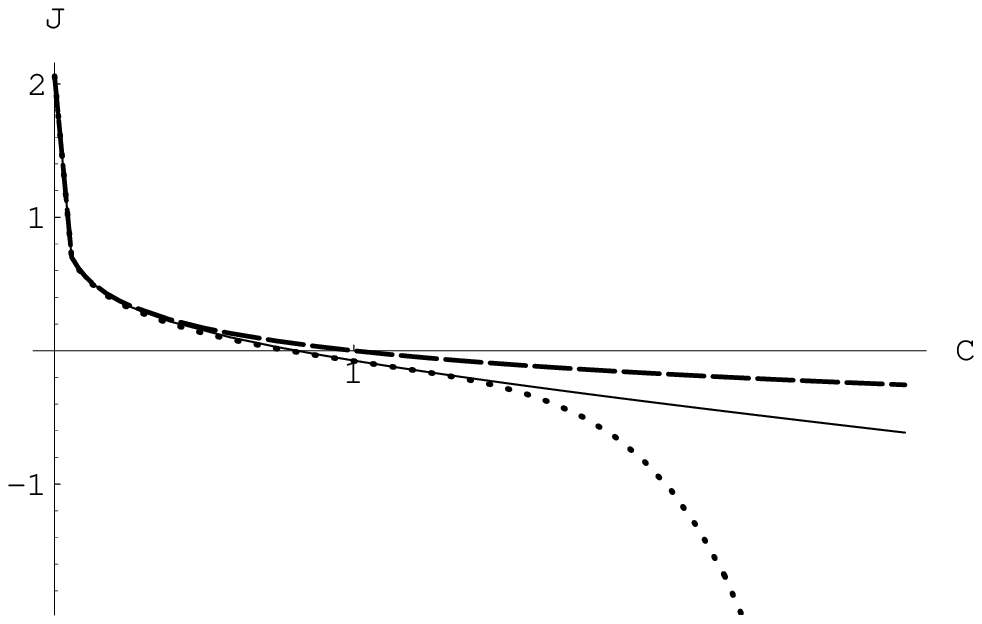}
 \includegraphics[height=60mm]{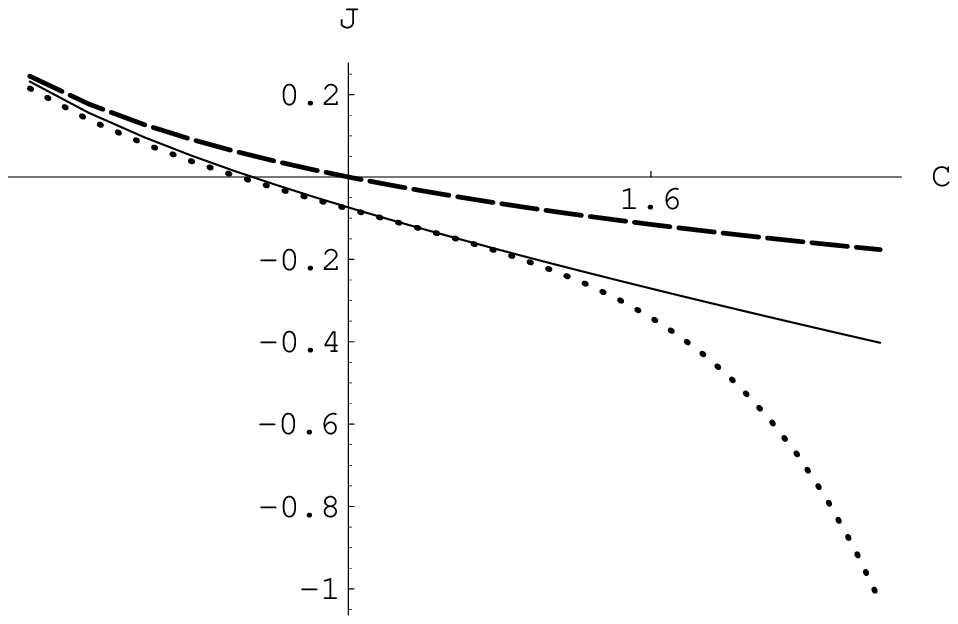}
\else
\end{center}
 \fi
\caption{\it
In this figure we compare the exact behavior of the K\"ahler potential $J(C)$ with both its series representation and with the leading logarithm which approximates it for small $C$.s The continuous line is the exact function, the dotted line is the series representation truncated to the order $k=7$ while the dashed line is the leading logarithm $J_0(C) \, = \, - \frac{12}{49} \, \log [C]$. For $C < 1.4$ the series representation is very accurate. For $C>1.4$ both the leading logarithm and the truncated series depart significantly from the behavior of the exact function.}
\label{seriokalero}
 \iffigs
 \hskip 1cm \unitlength=1.1mm
 \end{center}
  \fi
\end{figure}
In fig.\ref{seriokalero} we compare the behavior of the series representation of the K\"ahler potential with that of the exact function. It is suggested by the graph that the convergence radius of the new series should be $C<1.4$, yet this is very difficult to be established in want of exact recurrence relation for the series coefficients.
\par
All in all it remains the problem of defining intrinsically the function $J(C)$. This definition should emerge from its physical interpretation.
\section{Searching for Interpretations}
As we emphasized in the introduction, the most important conceptual challenge posed by the $D$-map that we have described in the previous sections is the need for a microscopic interpretation of the axial symmetric K\"ahler metric which,  after gauging of the axial symmetry,  leads to the inflaton potential. Without repeating the arguments put forward in the introductory section that we consider sufficiently detailed, we just stress that if such an interpretation were available, inflaton dynamics and, henceforth, inflation, the CMB spectrum  and the structure of the early Universe would be traced back to fundamental properties of a Superstring Vacuum encoded into some appropriate superconformal field theory. By this token Cosmology would shed light on the choice made by Nature among the plethora of available String Vacua in a cooperative way with the informations on the same choice that we have from Particle Physics. Without claiming that the here proposed scenario  is the only possible one, we have found that a quite convincing and well defined Ariadne's thread is provided by the mathematical notion of Axial symmetric descendants of special K\"ahler manifolds. By means of this token, that we deem might be physically justified in terms of type I or Heterotic String Theory duals of Closed Superstring models, one obtains a clearcut algorithm able to associate a restricted Picard Fuchs Equation\footnote{In this context restricted means that the linear system associated with the equation has a symplectic rather than special linear connection, as we shall explain.} to every positive-definite scalar potential $V(\phi)$. On its turn the Picard Fuchs equation is a sure bridge in the direction of singling out a microscopic interpretation of the special geometry ancestor manifold. Indeed the coefficient of the Picard Fuchs equation determine the three point function of the underlying topological $\sigma$-model or, equivalently, the three point correlator of a chiral ring.
\subsection{Axial symmetric K\"ahler descendants of Special K\"ahler Manifolds}
Let us consider a holomorphic section of the $\sym(4,\mathbb{R})$ symplectic bundle that defines  a one-dimensional special K\"ahler manifold. Following the notations and conventions of \cite{N2Wonder} we write:
\begin{equation}\label{chirollo}
  \mathbb{V}(\psi) \, \equiv \, \left\{X^0(\psi) \, , \, X^1(\psi) \, ,\, F_1(\psi) \, , \, - \, F_0(\psi) \right \}
\end{equation}
 where $\psi \in \mathbb{C}$ is any complex  coordinate labeling the points of the special K\"ahler manifold and where
\begin{equation}\label{ferdinandoprimo}
  F_\Lambda \, = \, \frac{\partial}{\partial X^\Lambda(\psi)}  \, F\left(X(\psi)\right)
\end{equation}
$F(X)$ being a homogeneous function of degree two of the variables $X^\Lambda(\psi)$, ($\Lambda \, = \, 0,1$). The K\"ahler potential $\mathcal{K}(\psi,\bar{\psi})$ which determines the metric and defines the geometry of the manifold is obtained as follows:
\begin{equation}\label{galstuck}
 \mathcal{K}(\psi,\bar{\psi}) \, = \, \log\left[- {\rm i} \overline{\mathbb{V}(\psi)} \, \mathbf{C} \, {\mathbb{V}}({\psi}) \right ]
\end{equation}
having denoted:
\begin{equation}\label{symplametrica}
    \mathbf{C} \, = \, \left(
                         \begin{array}{cccc}
                           0 & 0 & 1 & 0  \\
                           0 & 0 & 0 & 1 \\
                           -1 & 0 & 0 & 0 \\
                           0 & -1 & 0 & 0 \\
                         \end{array}
                       \right)
\end{equation}
the standard symplectic metric. The complex special coordinate $t=t(\psi)$ is defined by setting:
\begin{equation}\label{goliardusco}
    t (\psi) \, = \, \frac{X^1(\psi)}{X^0(\psi)} \quad \Rightarrow \quad F(X) \, = \, \left(X^0(\psi)\right)^2 \, \mathcal{F}(t)
\end{equation}
which implies:
\begin{equation}\label{corcullo}
    \mathbb{V}(\psi) \, = \, X^0(\psi) \, \mathbf{V}(t) \, = \, X^0(\psi)  \, \left\{1 \, , \, t \, ,\, 2 \,\mathcal{F}(t) \, - \, t \,  \mathcal{F}^\prime(t) \, , \, \mathcal{F}^\prime(t) \right \}
\end{equation}
where $\mathcal{F}(t)$ is a holomorphic function of its complex argument that is named the \textit{prepotential}. Correspondingly the K\"ahler potential becomes:
\begin{eqnarray}
  \mathcal{K}(\psi,\bar{\psi})  &=& \widehat{\mathcal{K}}(t(\psi),\bar{t}(\bar{\psi})) \, + \, \log |X^0(\psi)|^2 \label{grusco1}\\
  \widehat{\mathcal{K}}(t,\bar{t}) &=& \log \left[\, - \,{\rm i} \, \overline{\mathbf{V}({t})} \, \mathbf{C} \, \mathbf{V}(t) \right]\label{fontanacalda}
\end{eqnarray}
and the K\"ahler metric takes the form:
\begin{equation}\label{fischiaincurva}
    ds^2_{\mbox{K\"ahler}} \, = \, \partial_t \, \partial_{\bar{t}} \,\widehat{\mathcal{K}}(t,\bar{t})\, |\mathrm{d}t|^2
\end{equation}
The case of the constant curvature  coset manifold $\mathrm{SL(2,\mathbb{R})}/\mathrm{O(2)}$ is retrieved by choosing a cubic prepotential:
\begin{equation}\label{gonzetti}
    \mathcal{F}(t) \, = \, \ft 16 \, d_0 \, t^3
\end{equation}
so that the K\"ahler potential becomes:
\begin{equation}\label{cortiletto}
    \widehat{\mathcal{K}}(t,\bar{t}) \, = \, \log\left[ \left( t \, - \, \bar{t}\right)^3\right] \, + \, \mbox{const.}
\end{equation}
which depends only on the imaginary part of the field $t$:
\begin{equation}\label{fantabello}
    t \, = \, {\rm i} C \, + \, B
\end{equation}
and so does the K\"ahler metric which happens to be the Poincar\'e one:
\begin{equation}\label{poncarettus}
    g_{t\bar{t}}\, \propto \, \frac{1}{C^2}
\end{equation}
It is important to stress that the independence of the K\"ahler potential and of the K\"ahler metric from the real part $B$ of the field $t$, which is the very basis for the gauging of the $B$-translation and the generation of the $D$-type potential is a peculiarity of the power $3$ in the superpotential. For any other superpotential: $\mathcal{F}(t) \, = \, \ft {1}{q!} \, d_0 \, t^q$ this does not occur. Indeed with such a choice we obtain:
\begin{equation}\label{suriman}
   \widehat{\mathcal{K}}(t,\bar{t}) \, = \, \log \left(\frac{\left(t^q
   \bar{t} (q \bar{t}-(q-2)
   t)-t \bar{t}^q (q
   (t-\bar{t})+2
   \bar{t})\right) d_0}{t
   \bar{t} q!}\right)
\end{equation}
However let us note that we can easily invent a universal rule that associates an \textbf{axial symmetric K\"ahler manifold} to every \textbf{special K\"ahler manifold} whose holomorphic superpotential has the following reality property:
\begin{equation}\label{realita}
 \forall \, C \, \in \, \mathbb{R} \quad : \quad   \mathcal{F}({\rm i} \, C) \, = \, - \, \overline{\mathcal{F}({\rm i} \,C)} \quad \Rightarrow \quad  \mathcal{F}({\rm i} C) \, =  \,{\rm i} \, \mathfrak{H}(C) \quad ; \quad  \mathfrak{H}(C) \, \in \, \mathbb{R}
\end{equation}
where $\mathfrak{H}(C)$ is a real function of a real argument.
Indeed, if eq.(\ref{realita}) is satisfied, we can modify eq.(\ref{fontanacalda}) in the following way:
\begin{eqnarray}\label{soccalo}
    \widehat{\mathcal{K}}(t,\bar{t}) & = &  J(C)  \, \equiv \, - \, \log \left[ \, - \,{\rm i} \, \overline{\mathbf{V}(\,{\rm i} \,C)} \, \mathbf{C} \, \mathbf{V}({\rm i} C) \right] \nonumber\\
    & = & - \, \log \, \left[ 4 \, \mathfrak{H}(C)\,  \right]
\end{eqnarray}
and we obtain a K\"ahler metric of the form:
\begin{equation}\label{fartagliona}
    ds^2_{\mbox{K\"ahler}}\, = \,\ft 14 \,\frac{\mathrm{d}^2J}{\mathrm{d}C^2} \, \left(\mathrm{d}C^2\, + \, \mathrm{d}B^2\right)
\end{equation}
\begin{definizione}
\label{discendente}
Given a  one-dimensional special K\"ahler manifold that admits a prepotential $\mathcal{F}(t)$ with the reality property (\ref{realita}), we name its \textbf{axial symmetric descendant}  the one-dimensional K\"ahler manifold which is locally defined by the K\"ahler potential of eq.(\ref{soccalo}).
\end{definizione}
The name descendant is utilized because the above procedure is somehow reminiscent of the procedure by means of which one obtains open string theories from closed ones by properly identifying the superconformal algebras of the left and the right movers and by restricting holomorphic partition functions to functions of the real or imaginary part of their arguments.
\par
Leaving aside, for the moment, the physical justification of the  recipe encoded in definition \ref{discendente}, which probably might be sought for in some suitable projection of closed string theory that breaks $\mathcal{N}=2$ to $\mathcal{N}=1$ supersymmetry, let us observe
that the  condition (\ref{realita}) necessary to construct an axial symmetric descendant  is certainly satisfied when the prepotential has the form which emerges from instanton corrections to the moduli-space geometry of the K\"ahler class deformations in compactifications on Calabi-Yau manifolds, namely\footnote{See for instance eq.(8.5.82) in the book \cite{N2Wonder}.}:
\begin{equation}\label{cattus}
    \mathcal{F}(t) \, = \,{\rm i} A_0 \, + \, A_1 \, t \,+ \,  {\rm i} \, \ft 12 A_2 \, t^2 \, + \, \ft{1}{3!}\, d_0\, t^3 \, - \,{\rm i} \, \sum_{N=1} ^\infty \, \frac{d_N}{ \left(2\pi \, N\right)^3} \, \exp \left[2\pi\, {\rm i} \, N \, t \right]
\end{equation}
where $A_0,A_1,A_2,d_N$ are all real coefficients. In this case the function $\mathfrak{H}(C)$ advocated above is simply given by:
\begin{equation}\label{caruttone}
    \mathfrak{H}(C) \, = \, A_0 \, + \, A_1 \, C \, - \,   \, \ft 12 A_2 \, C^2\, - \, \ft{1}{3!}\, d_0\, C^3 \, \, - \, \sum_{N=1} ^\infty \, \frac{d_N}{ \left(2\pi \, N\right)^3} \, \exp \left[- 2\pi\, \, N \, C \right]
\end{equation}
Specifically for the quintic Calabi-Yau threefold $\mathbb{CP}_4[5]$, the coefficients $d_N$ are integer numbers that are related to the other integers $n_k$ expressing the number of rational curves of degree $k$ in $\mathbb{CP}_4[5]$ by means of the definition\footnote{See eq. (8.5.49) of the book \cite{N2Wonder}.}:
\begin{equation}\label{concollo}
    \sum_{N=0}^\infty \, d_N \, e^{2\pi\, {\rm i} \, N \, t } \, = \, 5 \, + \, \sum_{k=1}^\infty \, \frac{n_k \, k^3 \,e^{2\pi\, {\rm i} \, k \, t }}{1\, - \, e^{2\pi\, {\rm i} \,  t }}
\end{equation}
As it is well known, the miraculous mechanism that leads to the form (\ref{caruttone})  of the prepotential and to the celebrated prediction  by Candelas et al \cite{Candella1,Candella2} of the $n_k$ numbers is mirror symmetry which allows to calculate the  holomorphic symplectic section $\mathbb{V}(\psi)$ defined in eq.(\ref{chirollo}) by solving an appropriate
Picard-Fuchs equation. Assuming that the four entries of $\mathbb{V}(\psi)$ are the periods of the unique
holomorphic three-form $\Omega^{(3,0)}(\psi)$ on a basis of three cycles\footnote{We remind the reader that the existence and uniqueness of the holomorphic three-form $\Omega^{(3,0)}(\psi)$ is the very definition of CY three-folds, such a condition being equivalent to the vanishing of the first Chern class of the tangent bundle.}:
\begin{equation}\label{crili}
    \mathbb{V}(\psi) \, = \, \left \{ \int_{A_1} \,\Omega^{(3,0)}(\psi) \, , \, \int_{A_2} \,\Omega^{(3,0)}(\psi) \, , \, \int_{B_1} \,\Omega^{(3,0)}(\psi)\, , \, \int_{B_2} \,\Omega^{(3,0)}(\psi)\right\}
\end{equation}
we obtain that each entry satisfies a linear quartic differential equation (the Picard-Fuchs equation)\footnote{See eq.(8.3.15) of \cite{N2Wonder}.} which, in full generality we can write as follows:
\begin{eqnarray}\label{calossoloe}
  0 &= &   \left[\partial_\psi^4 \, + \, w_2 \, \partial_\psi^2 \, + \, \left(w_{3}\, + \, w_2^\prime\right) \, \partial_\psi  \, + \, \left(
    \ft 3{10} \, w_{2}^{\prime\prime}\, + \, \ft {9}{100} \, w_{2}^2 \, + \, \ft 12 \, w_3^\prime \, + \, w_4 \right) \right] \, \mathbb{V}(\psi)
\end{eqnarray}
In eq.(\ref{calossoloe}) the functions $w_{2,3,4}(\psi)$ are the so named tensor covariants of the differential equation. They are constructed as follows.
Given a generic linear differential equation of the $4th$ order:
\begin{equation}\label{codazzus}
 \mathcal{D}_\psi \mathbb{V}(\psi) \, \equiv \,   \left[\, \sum_{n=0}^4 \,a_n(\psi) \, \partial_\psi^n \, \right] \mathbb{V}(\psi) \, = \, 0
\end{equation}
where the coefficients $a_n(\psi)$ are functions of the parameter $\psi$, by means of the redefinition:
\begin{equation}\label{ridefinizia}
    \widetilde{\mathbb{V}}(\psi) \, = \, \exp \left[ - \, \ft 14 \, \int \frac{a_3(\psi)}{a_4(\psi)} d\psi\right] \, \mathbb{V}(\psi)
\end{equation}
the $3rd$ derivative term can  always be canceled and the differential equation (\ref{codazzus}) is put into the form:
\begin{equation}\label{corelli}
  0 \, = \,  \left[\,\partial_\psi^4 \, + \, c_2(\psi) \, \partial_\psi^2 \, + \, c_1(\psi) \partial_\psi \, + \, c_0(\psi) \right] \widetilde{\mathbb{V}}(\psi)
\end{equation}
Then the tensor covariants are defined by the following combinations:
\begin{eqnarray}
\label{vudoppioni}
  w_2 &=& c_2 \\
  w_3 &=& c_1 \, - \, c_2^\prime \\
  w_4 &=& c_0 \, - \, \ft 12 \, c_1^\prime \, + \, \ft 15 \, c_2^{\prime\prime} \, - \, \ft{9}{100}  c_2^2
\end{eqnarray}
Under a change of variable:
\begin{equation}\label{tildatus}
   \psi \, \rightarrow \, \tilde{\psi}(\psi) \quad ; \quad \partial_\psi \, \to \, \left(\frac{\partial \tilde{\psi}}{\partial\psi}\right)^{-1} \, \partial_{\tilde{\psi}}
\end{equation}
The objects $w_{3,4}(\psi)$ transform indeed as tensors, namely:
\begin{eqnarray}
\tilde{w}_{3}  & = & \left(\frac{\partial \tilde{\psi}}{\partial\psi}\right)^{-3} \, {w}_{3} \nonumber\\
    \tilde{w}_{4}  & = & \left(\frac{\partial \tilde{\psi}}{\partial\psi}\right)^{-4} \, {w}_{4} \label{cospetto}
\end{eqnarray}
while the object $w_{2}(\psi)$ transforms as follows:
\begin{equation}\label{canedatrifola}
    \tilde{w}_{2} \, = \, \left(\frac{\partial \tilde{\psi}}{\partial\psi}\right)^{-2} \left({w}_{2}\, - \, 5 \, \left\{\tilde{\psi},\, \psi\right\} \right)
\end{equation}
where
\begin{equation}\label{cagnesco}
    \left\{f,\, x\right\} \, \equiv \, \frac{f^{\prime\prime\prime}}{f^{\prime}} \, - \, \ft 32 \, \left(\frac{f^{\prime\prime}}{f^{\prime}}\right)^2
\end{equation}
is the schwarzian derivative.
\subsubsection{The setup of linear systems}
What we have so far recalled and which is instrumental to our further discussions is just part of the general theory of N-th order linear differential equations developed by Drinfel'd-Sokolov \cite{drinfeldo}, revisited from a mathematical-physicist's point of view in \cite{francescus} and finally, at the beginning of the nineties, applied to special K\"ahler geometry in many papers by many authors,  as reviewed in  the book \cite{N2Wonder}. In particular in the fundamental paper \cite{ceresolus}, it was clarified that a set of Picard Fuchs equations is intrinsically associated with any special K\"ahler geometry, independently from the geometrical algebraic origin of such a manifold  as moduli space of an algebraic $N$-fold. Furthermore, as we are going to recall below, in \cite{ceresolus} it was discovered the relation between the symplectic structure of the linear system associated with the Picard Fuchs equation and the vanishing of the $w_3$ tensor covariant which becomes, just because of this fact, the identity card of special geometry.
\par
 In view of this, the proper setup to understand the formal structure of  Picard Fuchs  equations and a very essential step in our quest for an interpretation of the axial symmetric K\"ahler manifolds advocated by inflaton potentials,  is provided by linear systems. We shortly summarize the notions of this theory that are necessary for our analysis.
\par
The starting point observation is that any linear
fourth order  differential equation (\ref{codazzus}) is,
for a suitable choice of the matrix $\IGa$,
equivalent to a first order matrix equation
\begin{equation}
\Big[\,{\bfone}\del\ -\ \IGa\,\Big]\cdot \mathcal{V}\ =\ 0
\label{matdeq}
\end{equation}
 where $\mathcal{V}$ is a
$4\times 4$
matrix whose first row is $\mathbb{V}$, namely the vector of
four independent solutions of the differential equation
(\ref{codazzus}).
 A matrix of the form
\begin{equation}
\IGa\ =\ \pmatrix{
\ast & 1 & 0 & 0 \cr
\ast & \ast  & 1 & 0 \cr
\ast & \ast & \ast & 1 \cr
\ast & \ast & \ast & \ast \cr}\ ,
\label{formga}
\end{equation}
corresponds to a fourth order operator ${\cal D}$
with $a_4=1$ whereas $tr \IGa
= 0$ leads to $a_3=0$.
Hence from our previous discussion of
the general form of a fourth order differential equation, we
conclude that the most general case is associated with a linear
system where the connection matrix $\IGa \, \in \, \slal(4,\mathbb{R})$ is
in the special linear Lie algebra and has the form (\ref{formga}).
However the corresponding fourth order differential operator
 $\cD$ is left invariant by local
gauge transformations acting as $\mathcal{V}\to S^{-1}\cdot \mathcal{V}$ and $\IGa\to
S^{-1}\IGa S-S^{-1}\del S$, if $S$ belongs to the nilpotent subgroup obtained by exponentiating the subalgebra generated by all negative roots, namely
\begin{equation}
S\ =\ \pmatrix{
1 & 0 & 0 & 0 \cr
\ast & 1  & 0 &0 \cr
\ast & \ast & 1& 0 \cr
\ast & \ast & \ast & 1\cr}\in \mathfrak{N} \subset \mathrm{SL(4,\mathbb{R})}\
\label{Strans}
\end{equation}
The top row of $\mathcal{V}$
corresponds to a highest weight of the $\slal(4,\mathbb{R})$ Lie algebra
and thus it is also $\mathfrak{N}$-invariant
(the other rows of $\mathcal{V}$ are gauge dependent). That is, the solutions
of (\ref{codazzus}) are completely invariant under
the local transformations (\ref{Strans}).
\par
Note also that the more general gauge transformations belonging
to the Borel subgroup $\mathcal{B} \,\subset \, \mathrm{SL(4,\mathbb{R})}$, where
\begin{equation}
S = \pmatrix{
\ast & 0 & 0 & 0 \cr
\ast & \ast  & 0 & 0 \cr
\ast & \ast & \ast & 0 \cr
\ast & \ast & \ast & \ast \cr}\ \in\ \mathcal{B}
\label{Stildetrans}
\end{equation}
do not leave $\mathcal{D}_\psi$ invariant but induce $a_3 \neq 0$ and $a_4 \neq
1$.
This corresponds to a rescaling of the solution $ \mathbb{V} \to f(\psi ) \mathbb{V}$
(and
corresponds to an irrelevant K\"ahler transformation in the context of special K\"ahler geometry).
\par
Using the gauge freedom we have discussed, we can put the
connection in the form
\begin{equation}
\IGa\ =\ \IGa_w\ \equiv\ \pmatrix{
0  & 1 &  0 & 0 \cr
-\ft 3{10}w_2 & 0  & 1 & 0 \cr
-\ft 12 w_3 & -\ft 4{10}w_2 & 0 & 1\cr
 -w_4 &-\ft 12 w_3 &-\ft 3{10}w_2 & 0\cr}\ \in \slal(4,\mathbb{R})
\label{wmat}
\end{equation}
 and the differential equation takes the form (\ref{calossoloe}).
\par
The reason why we spelled out the transcription of the differential equation (\ref{calossoloe}) in terms of a linear system is to put into evidence (see \cite{N2Wonder}) the meaning of the invariant condition $w_3\, = \,0$ which characterizes special geometry. Indeed as it is explained  in eq.(8.3.41) of the quoted book the vanishing of the covariant tensor $w_3$ is the necessary and sufficient condition for the connection $\IGa_w$ of the linear system to be in the symplectic subalgebra $\sym(4,\mathbb{R}) \subset \slal(4,\mathbb{R})$, namely:
\begin{equation}\label{chicchio}
    \IGa_w \, Q \, + \,Q \, \IGa_w^T \, = \, 0
\end{equation}
the invariant symplectic metric being, in the present conventions, given by:
 \begin{equation}
 Q\ =\ \pmatrix{
  & & & 1 \cr
  & & -1 & \cr
  & 1 & & \cr
  -1 & & &\cr}
 \label{Qsympdef}
\end{equation}
On the other hand that $\IGa_w$ be symplectic is a consistency condition for the Picard-Fuchs equation to be equivalent to the constraints of special K\"ahler geometry.
\subsubsection{The algorithm to reconstruct Special K\"ahler geometry descendance}
\label{reconstruo}
As it was stressed in \cite{N2Wonder} the final geometric structure of the special K\"ahlerian manifold is entirely encoded in the two holomorphic pseudo-tensors $w_{2,4}(\psi)$ which determine the Picard Fuchs equation and its four independent solutions. From such holomorphic data one retrieves the K\"ahler potential and the K\"ahler metric. On the other hand the holomorphic pseudo-tensors $w_{2,4}(\psi)$ are traced back to the chiral ring of the polynomial constraints that define the Calabi-Yau three-fold as an algebraic locus in $\mathbb{CP}_N$ or, more abstractly, to the chiral ring of the underlying $(2,2)$-superconformal field theory on which the superstring is supposed to be compactified.
\par
It is therefore quite challenging to answer the question whether the K\"ahler manifolds in the image of the $D$-map of interesting cosmological potentials (not necessarily integrable) might be identified with the axial symmetric descendants of appropriate special K\"ahler   manifolds which, on their turn, could be seen as moduli spaces of particular $(2,2)$ superconformal field theories. Ariadne's thread in this labyrinth is provided by the Picard Fuchs equation which, being holomorphic, can be reduced to the real numbers and applied to the real functions we deal with in our $D$-map constructions.
\par
The logical steps can be summarized as follows.
\begin{description}
  \item[a)] From the potential $V(\phi)$ we obtain first the momentum map $\mathcal{P}(\phi)$ and then the K\"ahler potential $J(\phi)$, by means of the integration mentioned in eq.s(\ref{celerus1},\ref{celerus2})
  \item[b)] With some art, but usually without too much effort, we derive a coordinate transformation
  \begin{equation}\label{coordabella}
    \phi \, = \, \phi(U)
  \end{equation}
  which eliminates all transcendental functions from the metric (\ref{metraxia}) leaving two coefficients $p(U)$ and $q(U)$ that are rational functions of the new variable. This step is not obligatory yet  very convenient in order to unwind Ariadne's thread.
  \item[c)] In the new coordinate $U$ the K\"ahler potential  has typically a fairly simple form $J(U)$.
  \item[d)] Consider next a candidate symplectic section of special geometry (\ref{chirollo}) which, reduced to a pure imaginary parameter $\psi \, = \, {\rm i}\,U$ has, by hypothesis, the following form:
  \begin{equation}\label{sezionata}
    \mathbb{V}({\rm i} \,U) \, =  \, \left\{1\, , \, {\rm i} \mathcal{A}(U)\, , \, {\rm i} \mathcal{B}(U) \, , \, {\rm i} \mathcal{C}(U)  \right\}
  \end{equation}
  $\mathcal{A}(U),\mathcal{B}(U),\mathcal{C}(U)$ being real functions of the real argument $U$. Constructing the K\"ahler potential according to the framework of special geometry (\ref{soccalo}) we obtain:
  \begin{equation}\label{formidabile}
    \widehat{\mathcal{K}} \, = \,  - \, \log \left[ \, - \,{\rm i} \, \overline{\mathbb{V}(\,{\rm i} \,U)} \, \mathbf{C} \, \mathbb{V}({\rm i} U) \right] \, = \, -\log \left[2 \mathcal{B}(U)\right]
  \end{equation}
  Hence the K\"ahler potential from special geometry coincides with the  K\"ahler potential $J(U)$ of the perspective descendant if we identify:
  \begin{equation}\label{corettodipifferi}
    \mathcal{B}(U) \, = \, \ft 12 \, \exp\left[ \, - \, J(U) \,\right]
  \end{equation}
  \item[e)] If special geometry holds the function $\mathcal{B}(U)$ defined in eq.(\ref{corettodipifferi}) must be a solution of a $4th$ order linear differential equation of the form (\ref{corelli}) characterized by $w_3\,=\,0$. This means:
      \begin{equation}\label{caralino}
        c_1(U)\, = \, \partial_U \, c_2(U)
      \end{equation}
  Furthermore the four entries of the symplectic section (\ref{sezionata}) must correspond to a basis of four independent solutions of the Picard-Fuchs equation. This means that the constant $w(U)\, = \, 1$ is necessarily a solution of the equation:
  \begin{equation}\label{aquata}
     0 \, = \,  \left[\,\partial_U^4 \, + \, c_2(U) \, \partial_U^2 \, + \, c_1(U) \partial_U \, + \, c_0(U) \right] \,w(U)
  \end{equation}
  This happens if and only if the coefficient $c_0(U) \, = \,0$ vanishes. Hence the entire structure of the    differential equation is reduced to the determination of a single function $c_2(U)$:
  \begin{equation}\label{cossobianco}
  0 \, = \,  w'(U)c_2 '(U)\, + \, c_2(U) w_2''(U)\, + \, w^{(4)}(U)
  \end{equation}
  \item[f)] Imposing that the known function $\mathcal{B}(U)$, as given in eq.(\ref{corettodipifferi}) should be a solution of the Picard-Fuchs differential equation (\ref{cossobianco}) yields a first order linear differential equation for $c_2(U)$ which can be always integrated.
   \item[g)] In this way from the form of the K\"ahler potential $J(U)$ as given in a non special coordinate $U$, under the hypothesis that our K\"ahler manifold should be the axial symmetric descendant of a special K\"ahler manifold we can reconstruct the intrinsic data of the Picard Fuchs equation, in particular the $w_2(U)$ pseudo tensor $\Phi(U)$ and  most important the covariant $w_4(U)$.
  \end{description}
  As explained in \cite{N2Wonder} the $w_4$ tensor is related to the three-point correlation function of the underlying $\sigma$-model or, in another interpretation, to the \textit{three-point correlator} of a topologically twisted $(2,2)$-superconformal field theory. In this way, by means of the above outlined algorithm we have constructed a bridge that relates any positive definite cosmological potential to the Picard Fuchs equation of a one dimensional special manifold. Whether reasonable Calabi-Yau threefolds or Landau-Ginzburg chiral rings do exist that yield such a Picard Fuchs equation is a question that we leave for future work.
\subsubsection{The example of the integrable series $I_1B$ as an illustration}
As a unique illustration of the algorithm presented in section \ref{reconstruo}  we reconsider the integrable series
$I_1B$ discussed in section \ref{serieUnoB}. Starting from the momentum-map (\ref{garlando2}) we reinstall the parameter
$\mu$  that we had reabsorbed by a shift of the field $\phi$ and we write:
\begin{equation}\label{gondalfano}
    \mathcal{P}(\phi) \, = \, \sqrt{\mu +e^{\sqrt{3} \phi }}
\end{equation}
Following the same steps as in section \ref{serieUnoB} and performing the change of variable (\ref{ciangio}) which corresponds to point b) of the algorithm outlined above we obtain the metric:
\begin{equation}\label{sifula}
  ds^2_{\mbox{K\"ahler}} \, = \,  \frac{3 \mathrm{dB}^2}{16 U^4 \left(\mu +\frac{1}{U^2}\right)}+\frac{\mathrm{dU}^2}{3 U^2}
\end{equation}
with curvature:
\begin{equation}\label{curvaRc}
 R(U) \, = \,   -\frac{9 \mu ^2}{\left(\mu +\frac{1}{U^2}\right)^2}-3
\end{equation}
and K\"ahler potential
\begin{equation}\label{ghirobernardo}
    J(U) \, = \, -\frac{2 \mu  U^2}{3}-\frac{4 \log(U)}{3}
\end{equation}
Eq.(\ref{ghirobernardo}) replaces eq.(\ref{gecolona}). Then according to point d) of the algorithm the function $\mathcal{B}(U)$ that should be a solution of the Picard Fuchs equation is:
\begin{equation}\label{gorinteo}
  \mathcal{B}(U) \, = \, \ft 12 \,   e^{\frac{2 U^2 \mu }{3}} U^{4/3}
\end{equation}
Replacing $w(U) \, \to \, \mathcal{B}(U)$ in eq.(\ref{cossobianco}) we obtain the following differential equation for $c_2(U)$:
\begin{eqnarray}\label{equaFi}
    0 & = & \frac{256 U^4 \mu ^4}{81}+\frac{2176 U^2
   \mu ^3}{81}+\frac{16}{9} U^2 c_2 (U)
   \mu ^2+\frac{1040 \mu
   ^2}{27}+\frac{44}{9} c_2 (U) \,\mu \nonumber\\
   &&+\frac{4}{3} U c_2 '(U) \mu
   +\frac{160 \mu }{81 U^2}+\frac{4 c_2
   (U)}{9 U^2}+\frac{4 c_2 '(U)}{3
   U}+\frac{40}{81 U^4}
\end{eqnarray}
which admits the following solution:
\begin{equation}\label{canedilatte}
    c_2(U) \, = \, \frac{-4 \mu  \left(\mu  \left(4 \mu
   U^2+21\right) U^2+12\right) U^2\,+\,2\,+\, 9 \,\lambda \,
   e^{-\frac{2 U^2 \mu }{3}}
   U^{5/3}}{9 \left(\mu
   U^4+U^2\right)}
\end{equation}
the real parameter $\lambda$ being the unique integration constant. It appears from its form that a nicer Picard-Fuchs equation is obtained by choosing $\lambda \, = \, 0$ which indeed is the value we consider.
If we also choose the reference value $\mu=1$ the Picard-Fuchs equation becomes:
\begin{eqnarray}\label{gerolinico}
    0 & = & U \left(U^2+1\right) \left(9 U^2
   \left(U^2+1\right) w^{(4)}(U)-2
   \left(8 U^6+42 U^4+24 U^2-1\right)
   w''(U)\right)\nonumber\\
   &&-4 \left(2 \left(4 U^8+8
   U^6+9 U^4+U^2\right)+1\right) w'(U)
\end{eqnarray}
By construction eq.(\ref{gerolinico}) admits the constant solution and the solution (\ref{gorinteo}), the other two solutions are more difficult to be found. However if in (\ref{canedilatte}) we set $\lambda = 0$ and we perform the limit $\mu \,\to \, 0$ we obtain the following Picard-Fuchs equation which corresponds to a pure exponential potential and therefore to an axial symmetric K\"ahler manifold that is a constant curvature coset manifold. This equation is:
\begin{equation}\label{catusrufus}
  0 \, = \,  -\frac{4 w'(U)}{9 U^3}+\frac{2 w''(U)}{9
   U^2}+w^{(4)}(U)
\end{equation}
whose four solutions are immediately worked out:
\begin{equation}\label{gurtillo}
    w(U) \, = \, \left\{ 1\, , \, U^{4/3} \, , \, U^{5/3} \, , \, U^3 \, \right \}
\end{equation}
Since the order in which we dispose the solutions into the symplectic section is irrelevant we see that from the same special manifold associated with the Picard-Fuchs equation (\ref{catusrufus}) we can extract three different axial symmetric descendants that are all $\mathrm{SL(2,\mathbb{R})/O(2)}$ coset manifolds but with different curvatures, one with $q=\ft 43$, one with $q=\ft 53 $, one with $q=3$. In none of the three cases the special geometry parent is a constant curvature manifold. This can happen only when $w_4 \, = \,0$.
\par
This is an important lesson to remember.
\section{Conclusions and Perspectives}
The conclusions of this rather long paper will be rather short since the conceptual set up has already been described in the introduction and the main results we  obtained have also been anticipated there.  Here we just mention the perspectives that, according to us, have been opened up by our present work. We think that they are three and of different nature:
\begin{description}
  \item[a)] From a more phenomenological point of view, the possibility that we have shown to exist of embedding (integrable) Starobinsky--like potentials into supergravity should be exploited in a full-fledged way. For those such potentials that descend from constant curvature K\"ahler manifolds the microscopic interpretation is readily available and not particularly challenging since an $\mathrm{SL(2,\mathbb{R})/O(2)}$ factor within the scalar manifold of supergravity can be retrieved in many different ways starting from superstring compactifications on torii and orbifolds. Yet some relevant information might be hidden in the particular values of the $\mathrm{SL(2,\mathbb{R})/O(2)}$ curvature that do the job. For the Starobinsky--like potentials that do not descend from constant curvature manifolds, but are integrable, integrability should be thoroughly exploited in relation with data analysis of the CMB spectrum. Probably new possibilities open up.  As for the microscopic interpretation what we are going to say in the next two items applies here as well.
  \item[b)] From the more conceptual point of view, the Ariadne's thread provided by the Picard Fuchs equation should be followed with determination and to do this is certainly the main point in our agenda for the next coming months. Working out the Picard Fuchs equation associated with interesting inflaton potentials, both integrable and not integrable is the first step to be  urgently done. Studying the properties of these equations and their monodromy group, possibly constructing their solutions is the next step. The final and most ambitious step is that of retrieving a chiral ring that determines such an equation unwielding in this way the possible microscopic interpretation of inflaton dynamics.
  \item[c)] The third point in the programme that naturally emerges from our results is also of conceptual type and hopefully provides the physical justification for the programme outlined in b). We refer to the superstring mechanism, to be explored that should implement the transition from the original special K\"ahler manifold to its axial symmetric descendant when we consider Open String descendant of closed ones.
\end{description}
\section*{Acknowledgments}
We express our gratitude to our friends and collaborators A. Sagnotti and M. Trigiante for very useful and inspiring comments and for many discussions during the elaboration of this work.
The work of A.S. was supported in part by the RFBR Grants No. 11-02-01335-a, No. 13-02-91330-NNIO-a and No. 13-02-90602-Arm-a.

\end{document}